\documentclass[sn-mathphys-num]{sn-jnl}% Math and Physical Sciences Numbered Reference Style
\usepackage{graphicx}%
\usepackage{multirow}%
\usepackage{amsmath,amssymb,amsfonts}%
\usepackage{amsthm}%
\usepackage{mathrsfs}%
\usepackage[title]{appendix}%
\usepackage{xcolor}%
\usepackage{textcomp}%
\usepackage{manyfoot}%
\usepackage{booktabs}%
\usepackage{algorithm}%
\usepackage{algorithmicx}%
\usepackage{algpseudocode}%
\usepackage{listings}%
\usepackage[sort&compress,numbers]{natbib}
\usepackage{doi}\usepackage{hyperref}
\usepackage{nicematrix}

\theoremstyle{thmstyleone}%
\theoremstyle{thmstyletwo}%

\theoremstyle{thmstylethree}%

\begin{document}

\title[XXX]{Theoretical Studies of $\alpha$ Clustering in Nuclei and Beyond}


\author*[1,2]{\fnm{Takaharu} \sur{Otsuka}}\email{otsuka@phys.s.u-tokyo.ac.jp}

\author[3]{\fnm{Alexander} \sur{Volya}}\email{avolya@fsu.edu}
%\equalcont{This author mainly contributed to Sect. 3 of this work.}

\author[4]{\fnm{Naoyuki} \sur{Itagaki}}\email{itagaki@omu.ac.jp}
%\equalcont{This author mainly contributed to Sect. 4 of this work.}

\affil*[1]{\orgdiv{Department of Physics}, \orgname{University of Tokyo}, \orgaddress{\street{Hongo 7-3-1}, \city{Bunkyo-ku}, \postcode{1130033}, \state{Tokyo}, \country{Japan}}}

\affil[2]{\orgname{RIKEN Nishina Center for Accelerator-Based Science}, \orgaddress{\street{Hirosawa 2-1}, \city{Wako-shi}, 
\postcode{3510198}, \state{Saitama}, \country{Japan}}}   %\orgdiv{Nuclear Structure Research Group}, 

\affil[3]{\orgdiv{Department of Physics}, \orgname{Florida State University}, \orgaddress{\street{311 Keen Building}, \city{Tallahassee}, \postcode{32306-4350}, \state{FL}, \country{USA}}}

\affil[4]{\orgdiv{Department of Physics}, \orgname{Osaka Metropolitan University}, \orgaddress{\street{3-3-138 Sugimoto}, \city{Osaka}, 
\postcode{558-8585}, \state{Osaka}, \country{Japan}}}


\abstract{This article comprises three sections after an introduction.  Section 2 starts with a quick review of the results of {\it ab initio} no-core shell model calculations by Monte Carlo Shell Model on light nuclei.  It is shown that $\alpha$ clustering arises in such first principles calculations for $^{8,10,12}$Be and $^{12}$C with the Daejeon16 and JISP16 interactions.  The $\alpha$ clustering occurs even in well bound states such as the ground state of  $^{12}$C.  The Hoyle state is shown to be dominated by $\alpha$ clustering in triangular configurations.  The crossover between clustering and nuclear matter is demonstrated.  As the ground and Hoyle states show strong deformations, they are also good cases to investigate rotational excitations.  As an original work, the recently proposed fully quantum (mechanical) formulation for deformation and rotation is extended to cluster or molecular states.  Dual rotational modes are proposed: compact-object rotation and distant-object rotation.  The former is found in many heavy nuclei, whereas the latter can be found for clustering states in Be and C isotopes.  While $^{8}$Be is a transparent example for the latter, $^{12}$C is a rare example that both modes appear in different states of the same nucleus, giving another novel significance to the Hoyle state.  The duality of rotation by compact-object and distant-object rotations is a visible outcome of the hierarchy by the cluster formation, placing $^{12}$C on the border.  Atomic molecules and hadrons can be viewed in terms of this duality.  Possible relevances to fission is mentioned.\\
Section 3 presents a general framework for an extended no-core shell model with cluster–nucleon configuration interaction, combining traditional shell-model–like configurations with explicit microscopic configurations representing cluster degrees of freedom. This approach offers a complementary perspective to the strategy discussed in the other sections. The section reviews the microscopic origins of cluster substructures in light nuclei, emphasizing how nucleonic degrees of freedom, nucleon–nucleon interactions, and continuum coupling naturally extend the traditional shell model into configuration-interaction frameworks that incorporate clustering and reaction dynamics. Both methodological developments and applications are discussed, including clustering in well-bound states as well as reaction processes involving alpha clusters.\\
Section 4 presents that although the cluster structure is robust in Be-C nuclei, some $jj$-coupling shell model components are mixed with clustering components in the ground state of $^{12}$C. This is a different feature than in the cases of the Be isotopes. Using the antisymmetrized quasi cluster model (AQCM), we can clearly model this competition between the cluster and shell components. The spin-orbit interaction is key to realizing the shell structure and contributes more to the $^{12}$C case than to the $^8$Be case due to closer $\alpha$-$\alpha$ distances in the former case, thanks to attraction among the clusters.\\
Section 5 presents remarks and prospects transcending the whole article, besides summarizing discussions within Sections 2-4.  
 }




\maketitle
%%%%%%%%%%%%%%%%%%%%%%%%\chapter{review on clustering}

\section{Introduction \label{introduction}}

%The atomic nucleus comprises $Z$ protons and $N$ neutrons, which are collectively called nucleons.   In the $\alpha$ clustering picture, as illustrated in Fig.~\ref{fig:classical}, the $\alpha$ particle ($Z$=$N$=2) is a building block, and some nuclei can be composed of $\alpha$ particles.  In such cases, $Z$=$N$=2$i$ holds with $i$ being an integer, and the mass number $A$=$Z$+$N$ becomes equal to 4, 8, 12, ...
%A given nucleus is labelled as $^A$X with X denoting the element name, e.g., $^8$Be 
%for beryllium-8.    Figures~\ref{fig:classical}{\bf b-c}  sketch intuitive pictures for possible $\alpha$ clustering in $^8$Be and $^{12}$C, respectively, where $\alpha$ particles are shown by mid-sized circles forming nuclei represented by the background green areas. 
%Such $\alpha$ cluster models have been developed since the 1930s \cite{wefelmeier,wheeler,morinaga_1956,brink_1966,ikeda_1968,arima_1973,freer_rmp}.
%It is, however, still difficult to {{\it experimentally observe} the $\alpha$ clustering inside nuclei. This is basically because the nucleus is not at rest (quantum mechanically) but we need its  snapshot (see Figure\ref{fig:classical}). 

\begin{figure*}[b]
  \centering
  \includegraphics[width=13cm]{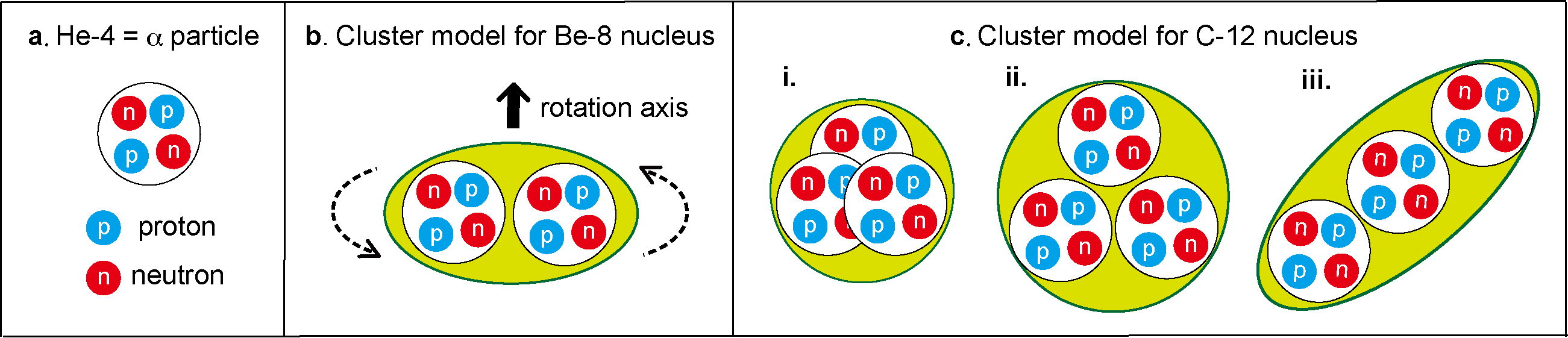}

    \caption{ {\bf Schematic illustrations of %classical picture of the 
    $\alpha$ clustering in atomic nuclei},  
    {\rm for} {\bf a} $^{4}${\rm He=}$\alpha$ {\rm particle,} {\bf b} $^{8}$Be, and {\bf c} $^{12}$C (three possible cases, i, ii and iii).  The green areas represent atomic nuclei allowing some movements of $\alpha$ clusters.  Taken from \cite{otsuka_2022} with permission.
} 
  \label{fig:classical}  
\end{figure*}

The atomic nucleus consists of $Z$ protons and $N$ neutrons, collectively referred to
as nucleons. In the $\alpha$-clustering picture, illustrated schematically in
Fig.~\ref{fig:classical}, the $\alpha$ particle ($Z = N = 2$) (see Fig.~\ref{fig:classical}{\bf a}) is regarded as a fundamental building
block, and certain nuclei may be described as aggregates of $\alpha$ particles, perhaps for main components of some states. In
such systems, the condition $Z = N = 2i$ holds, where $i$ is an integer, and the mass
number $A = Z + N$ takes the values $A = 4, 8, 12, \ldots$. A nucleus is denoted as
$^{A}X$, where $X$ represents the chemical element; for example, $^{8}$Be denotes
beryllium--8. Figures~\ref{fig:classical}{\bf b-c}  provide intuitive illustrations of possible
$\alpha$-cluster configurations in $^{8}$Be and $^{12}$C, respectively, in which
$\alpha$ particles are depicted as mid-sized circles forming a nucleus represented
by the surrounding shaded region.  

Models based on $\alpha$ clustering have been developed since the 1930s
\cite{wefelmeier,wheeler,morinaga_1956,brink_1966,ikeda_1968,arima_1973,freer_rmp}. Despite this long history, direct experimental observation of
$\alpha$ clustering within nuclei remains challenging. This is partly because
$\alpha$ clusters are not fundamental degrees of freedom but rather emergent
correlations within an interacting quantum many-body system of nucleons.
The Pauli principle and residual nucleon-nucleon correlations may  
constrain the allowed spatial configurations and/or may lead to substantial mixing 
between cluster-like and other components in some ways. 
As a result, geometric cluster pictures provide superb intuition but do not
correspond to directly observable structures; instead, experimental
signatures of clustering must be inferred indirectly through reaction dynamics
and spectroscopic observables. Part of this difficulty also arises from the
intrinsically quantum-mechanical nature of nuclear motion: nuclei are not static
objects, and cluster wave functions represent configurations defined in an underlying 
intrinsic frame rather than in the laboratory coordinate system
(see Fig.~\ref{fig:classical}).

Motivated by these challenges, a broad body of theoretical work has been
developed, several aspects of which are reviewed in the subsequent sections
from different theoretical perspectives. These sections are authored by
T.~Otsuka, A.~Volya, and N.~Itagaki, respectively.
In particular, a major research project headed by T.~Nakamura had been conducted in the recent past years \cite{nakamura_2025} focusing on hierarchy structures involving clustering.   A brief comment will be made in relations to Sect.~2.    

In Sect.~\ref{sec:otsuka}, a recent challenge involving super-large-scale {\it ab initio} no-core shell model calculations for Be and C isotopes is reviewed, depicting the emergence of $\alpha$(-like) clustering without assuming it {\it a priori}.   Contrary to the usual anticipation that $\alpha$(-like) clustering appears in the energy region around $\alpha$-particle emission threshold \cite{ikeda_1968}, $\alpha$(-like) four nucleon correlations, called $\alpha$ clustering for brevity hereafter, emerges even in well-bound states, and its mixing with normal nuclear states lowers the ground-state energy of $^{12}$C, for instance.  Similar $\alpha$ clustering arises more distinctly in $^8$Be and in the Hoyle state of $^{12}$C.  In other words, nuclear forces and nuclear many-body dynamics favor the $\alpha$ clustering, although details may vary.   This work suggests, from first principles viewpoint, that the hierarchy structure with the appearance of the $\alpha$ cluster seems to occur, but may not manifest as clearly as intuitively expected, because of nuclear forces and antisymmetrization constraints.   
Beyond this review, some new features are discussed.  Molecular(-like) configurations of $\alpha$ clusters can naturally result in rotational motion.  On the other hand, a new general formulation of rotational bands in atomic nuclei with ellipsoidal shapes has recently been presented \cite{otsuka_2025}, where not only the prevailing of triaxial (i.e., almond-like) shapes but also the excitation mechanism within a rotational band were major subjects.  This new picture can be confronted to the rotational excitation of molecular configurations, and we can indeed analyze their relationship by applying the new formulation.  This analysis leads to two basic modes of quantum mechanical rotations.  Such study goes beyonds the clarification of the $\alpha$ clustering, towards a new unified picture of the rotation of quantum many-body systems, possibly including hadrons and atomic molecules as future applications.  All these developments illuminate the importance of the $\alpha$ clustering in the global landscape of physics.  The transition between two rotational modes and the hierarchy boundary, likely found in $^{12}$C, may show a very interesting coincide \cite{nakamura_2025}.

In Sect.~\ref{sec:volya}, $\alpha$-clustering is reviewed from a microscopic
many-body perspective. The section focuses on extended
configuration-interaction frameworks that incorporate microscopically
constructed cluster configurations, in which full fermionic antisymmetrization
and proper treatment of center-of-mass dynamics are maintained. These
approaches illustrate how cluster substructures emerge naturally from
nucleonic degrees of freedom and realistic nucleon-nucleon interactions.
Emphasis is placed on the unified treatment of nuclear structure and reactions,
targeting both spectroscopic properties and scattering observables, with
representative applications spanning well-bound systems as well as reaction
processes involving $\alpha$ clusters.

Sect.~\ref{sec:itagaki} provides a complementary review of the competition and coexistence
between cluster and shell-model structures in light nuclei. Using the
antisymmetrized quasi-cluster model (AQCM), the section elucidates how
$\alpha$-cluster configurations mix with and can be continuously transformed
into $jj$-coupling shell-model states, thereby offering a unified description of
intermediate regimes. Applications to $^{8}$Be and $^{12}$C highlight the
decisive role of the spin-orbit interaction in governing cluster persistence
and breaking, clarifying the physical mechanisms that drive the evolution
from cluster-dominated to shell-model-like structures. 

In addition to summarizing discussions within Sects.~\ref{sec:otsuka} -~\ref{sec:itagaki}, Sect.~\ref{sec:end_summary} presents remarks and prospects transcending the discussions and results depicted in Sects.~\ref{sec:otsuka} -~\ref{sec:itagaki} focusing on $\alpha$ clustering in $^8$Be and $^{12}$C.

%%%%%%%%%%%%%%%%%%%%%%%%%%%%%%%%%%%%%%%%%%%%%%%%

%Motivated by these challenges, a broad body of theoretical work has been developed, several aspects of which are reviewed in the subsequent sections from different theoretical perspectives. These sections are authored by T.~Otsuka, A.~Volya, and N.~Itagaki, respectively.

%In Sect.~3, $\alpha$-clustering is reviewed from a microscopic
%many-body perspective. The section focuses on extended
%configuration-interaction frameworks that incorporate microscopically
%constructed cluster configurations, in which full fermionic antisymmetrization
%and proper treatment of center-of-mass dynamics are maintained. These
%approaches illustrate how cluster substructures emerge naturally from
%nucleonic degrees of freedom and realistic nucleon-nucleon interactions.
%Emphasis is placed on the unified treatment of nuclear structure and reactions,
%targeting both spectroscopic properties and scattering observables, with
%representative applications spanning well-bound systems as well as reaction
%processes involving $\alpha$ clusters.

%%%%%%%%%%%%  FIGURE 1  %%%%%%%%%%%%%
% Fig1: Classical cluster models

%%%%%%%   S E C T I O N     2    %%%%%%%%%%%%%%%%%%%%%%%%%%%%%%%%%%%%%%%%%%

\section{First-principle realization of $\alpha$ clustering and dual rotational modes in quantum many-body systems}
\label{sec:otsuka}

Quite a few theoretical approaches have been made for the understanding of $\alpha$ clustering in atomic nuclei.   Examples such as  \cite{Uegaki_1977,Kamimura_1981,tohsaki_2001,bijker_2002,itagaki_2004,Chermykh_2007_Hoyle,Kanada_2007,JISP16forBe,zhao_2015,dreyfuss_2017}   
were performed, up to around 2020, including limiting cases like   
linear chains\cite{morinaga_1956,zhao_2015}, equilateral triangles\cite{bijker_2002} and a Bose-Einstein condensate\cite{tohsaki_2001}.  In about the same and later periods, {\it ab initio} calculations were reported\cite{wiringa_2000,GFMC_2015,epelbaum_2012,12CBE2,shen_2023}.  
More references may be mentioned in Sect. 3.

The initial impactful outcome was the appearance of {\it di}-$\alpha$ clusters in the ground state of $^8$Be in a VMC calculation \cite{wiringa_2000,GFMC_2015}, similarly to Fig.~\ref{fig:classical}{\bf b}. 
The $\alpha$ clustering is more crucial but less clarified for the $^{12}$C nucleus: this nucleus can be formed by three $\alpha$ particles in configurations, triangular, linear, or others (see Figure~\ref{fig:classical}{\bf c}).   Its lowest spin/parity $J^{\pi}$=0$^+$ excited state, the famous Hoyle state\cite{hoyle_1954,dunbar_1953,freer_ppnp}, is a critical gateway in the nucleosynthesis to the present carbon-abundant world filled with living organisms \cite{Fynbo_2005,jin_2020}, but its structure remains to be clarified.  The clarification of these structures lead to a novel picture of dual rotational modes in quantum many-body systems.

\subsection{First-principles realization of $\alpha$ clustering in $^{8}$Be and $^{12}$C \label{first-principles realization}}

We now briefly review a set of computational simulations \cite{otsuka_2022} without assuming $\alpha$ clustering {\it a priori}, which exhibited that $\alpha$ clustering indeed occurs for the ground and excited states of $^{8,10,12}$Be and $^{12}$C isotopes, including the Hoyle state, in varying formation patterns. 
The simulations are performed by full Configuration Interaction (CI) calculations from first principles with the Daejeon16 interaction \cite{shirokov_2016} for $^{12}$C and with JISP16 interaction \cite{shirokov_2007} for the Be isotopes, and their validity is further examined for some observables by comparing with experimental data \cite{ensdf} (see \cite{otsuka_2022}).  In the computational side, the Monte Carlo Shell Model \cite{mcsm1995,mcsm1998,mcsm2001,mcsm2012} has been used for the study of Sect. 2, and relevant results for light nuclei are also shown in \cite{Abe_2012,Abe_2021}.  
 %, also with substantial differences from traditional views.   
 
%%%%%%%%%%%%  FIGURE 2  %%%%%%%%%%%%%
% Fig2: levels of 12C

\begin{figure}[tb]
  \centering
    \includegraphics[width=4.4cm]{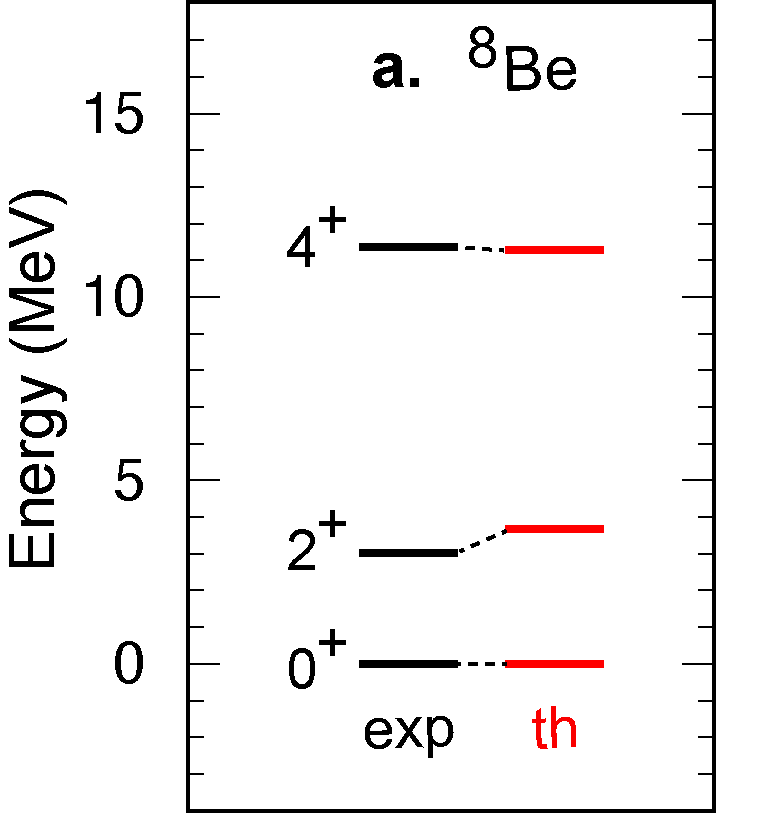}\includegraphics[width=8.4cm]{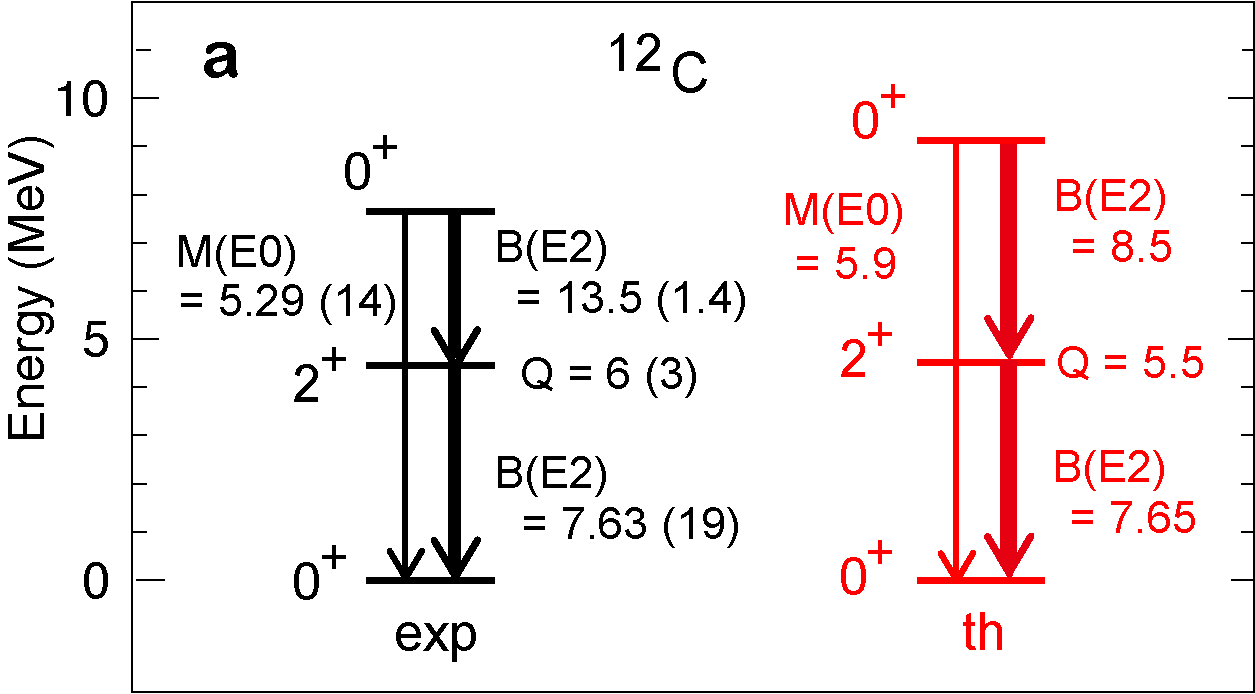}
    \caption{ {\bf Level energies of {\bf a} $^{8}$Be and {\bf b} $^{12}$C }  
    (left) Experimental data\cite{ensdf,12CBE2}, and (right) theoretical results in each panel.
  Taken from \cite{otsuka_2022} with minor changes with permission.
} 
  \label{fig:12C_levels}  
\end{figure}  
%%%%%%%%%%%%%%%%%%%%%%%%%%%%%%%%%%%

Figure~\ref{fig:12C_levels}{\bf a,b} display, respectively, the level energies of $^{8}$Be and $^{12}$C, both experimental and theoretical, with good experiment-theory agreement.  Note that the theoretical calculations were performed in the bound state approximation, which is considered to be sensible for the states to be discussed.  
We point out that theoretical 0$^+_1$ and $2^+_1$ states can be also obtained up to 99\% probability by simply projecting the same intrinsic state ($K^P$=0$^+$) extracted from the shell model wave functions.  Considering large E2 matrix elements related to these states, the 0$^+_1$ and $2^+_1$ states are identified as members of a strongly deformed rotational band with a prolate (an oblate) shape of the deformation parameter $\beta_2 \sim$ 1 (0.6) for $^{8}$Be ($^{12}$C) \cite{otsuka_2022}.  This is an important point, and we shall use this feature later.  The $B(E2;2^+_1 \rightarrow 0^+_1)$ value has been also measured experimentally for $^{12}$C \cite{12CBE2} in a high precision, and it has been reproduced well by the present calculation with free charges as displayed in Figure~\ref{fig:12C_levels}{\bf b}. 

The Hoyle state in the theoretical calculation is still too high as compared to the experimental one.  This is probably due to the size of the model space (seven harmonic oscillator shells).  This model space appears to suffice for the description of the ground and 2$^+_1$ state, but a somewhat wider space may improve the quality of the description of the Hoyle state. We, however, assume that the present computational setup is sufficient for the discussions below, where conceptual features are main subjects and extrapolations towards better descriptions are also possible.  

\begin{figure}[tb]
  \centering
    \includegraphics[width=13cm]{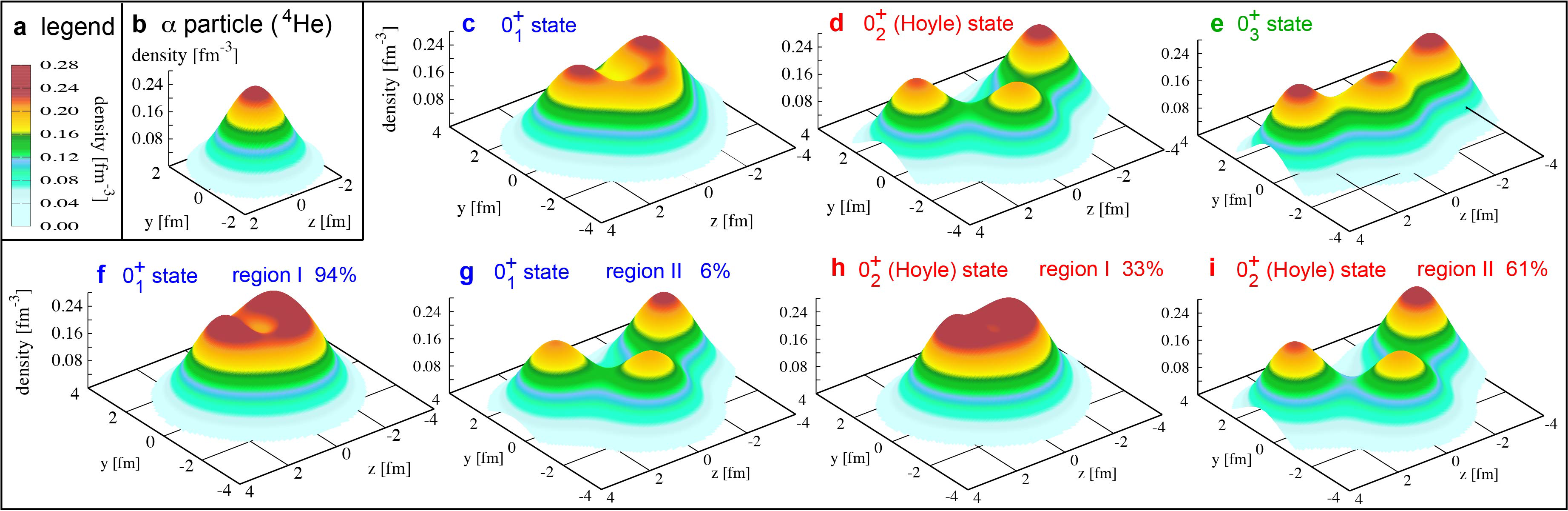}
    \caption{ {\bf Density profiles of $^{12}$C with that of $^{4}${\rm He=}$\alpha$ particle }
    The density profiles for $^{12}$C were obtained from intrinsic states, and in panels ${\bf i} - {\bf i}$,  eigenstates are  decomposed according to shapes using T-plot analysis.
  Taken from \cite{otsuka_2022} with permission.
} 
  \label{fig:12C_density}  
\end{figure}

Figure~\ref{fig:12C_density} shows density profiles of the ground, Hoyle and 0$^+_3$ states of $^{12}$C, and their decompositions according to their structures. The density profile of $^{4}$He, or $\alpha$ particle, is shown for comparisons.  We will later present a beautiful {\it di}-$\alpha$ structure of $^8$Be emerging also from first principles.  It is mentioned that the present no-core calculation is very suitable for the density profile, because in-medium corrections, including couplings to Giant Resonances, are basically treated explicitly, in contrast to usual shell-model calculations with a one or at most two valence shells.    

The $\alpha$ clustering can then be identified by three separate high peaks of the nucleon density, particularly clearly in panels {\bf d},   {\bf e}, {\bf g}, and {\bf i}.  We here decomposed MCSM basis vectors into three groups, (i) medium $\beta_2$ ($<$0.7), (ii) large $\beta_2$ ($>$0.7) not too close to prolate shape (6$^{\circ}\leq\gamma\leq 60^{\circ}$), (iii) large $\beta_2$ ($>$0.7) near prolate shape ($\gamma< 6^{\circ}$).  This classification was supported independently by a statistical learning technique in data science \cite{learning_book} (see \cite{otsuka_2022}).

Eigenstates of MCSM calculations are superpositions of these basis vectors.  
The ground state is composed of group (i) by 94 \% (panel {\bf f}), but contains group (ii) by 6 \% (panel {\bf g}), which is not negligible.
The group (ii) of the ground state (panel {\bf g}) exhibits a clear $\alpha$ clustering.
We stress that this occurs in the ground state which is well bound.  So, this is not an effect of loose binding, in contrast to some beliefs that $\alpha$ clustering arises as a consequence of weak or no binding near $\alpha$ threshold ~\cite{ikeda_1968}.  In other words, the $\alpha$ clustering can occur without the threshold effect of Ikeda {\it et al.}~\cite{ikeda_1968}, but this work does not deny possible appearance of this threshold effect.  The $\alpha$ clustering in well-bound states may be one of the very important answer to the $\alpha$-clustering question from the first principles.

%%%%%%%%%%%%  FIGURE 3  %%%%%%%%%%%%%
% Fig3: 3-dim. density of 12C

%%%%%%%%%%%%%%%%%%%%%%%%%%%%%%%%% 

Furthermore, the total density of the Hoyle state (panel {\bf d}) also manifests an even clearer $\alpha$ clustering.  As this calculation is still a shell model calculation, all single-particle basis vectors are provided by the eigenfunctions of the harmonic oscillator potential and no explicit continuum components are involved.  We still see a good picture of $\alpha$ clustering, probably because the $\alpha$ clustering is also a correlation effect.  The above decomposition can be performed for the Hoyle state as well.  Panel {\bf h} shows the density profile formed by the basis vectors in group (i), which constitutes 33 \% of the eigen wave function.  Likewise, Panel {\bf i} shows the density profile by group (ii), which constitutes 61 \% of the eigen wave function.  A strong mixing is found between a nuclear matter-type density (panel {\bf h}) and a molecular-type density (panel {\bf i}).   Because the present 0$^+_2$ state is formed primarily of the $\alpha$ clustering state and its energy is close to the experimental value of Hoyle state, it is implied that the Hoyle state has thus been reproduced from first principles, apart form possible improvement of the precision.  Note that Daejeon 16 interaction was not tuned for this calculation at all.
 
Thus, $\alpha$ clustering structures emerge from the first principles without assuming them {\it a priori}, as emphasized in \cite{otsuka_2022}.  This can be an important achievement of no-core MCSM calculation powered by Daejeon 16 interaction.
 
\begin{figure}[tb]
  \centering
  \includegraphics[width=13cm]{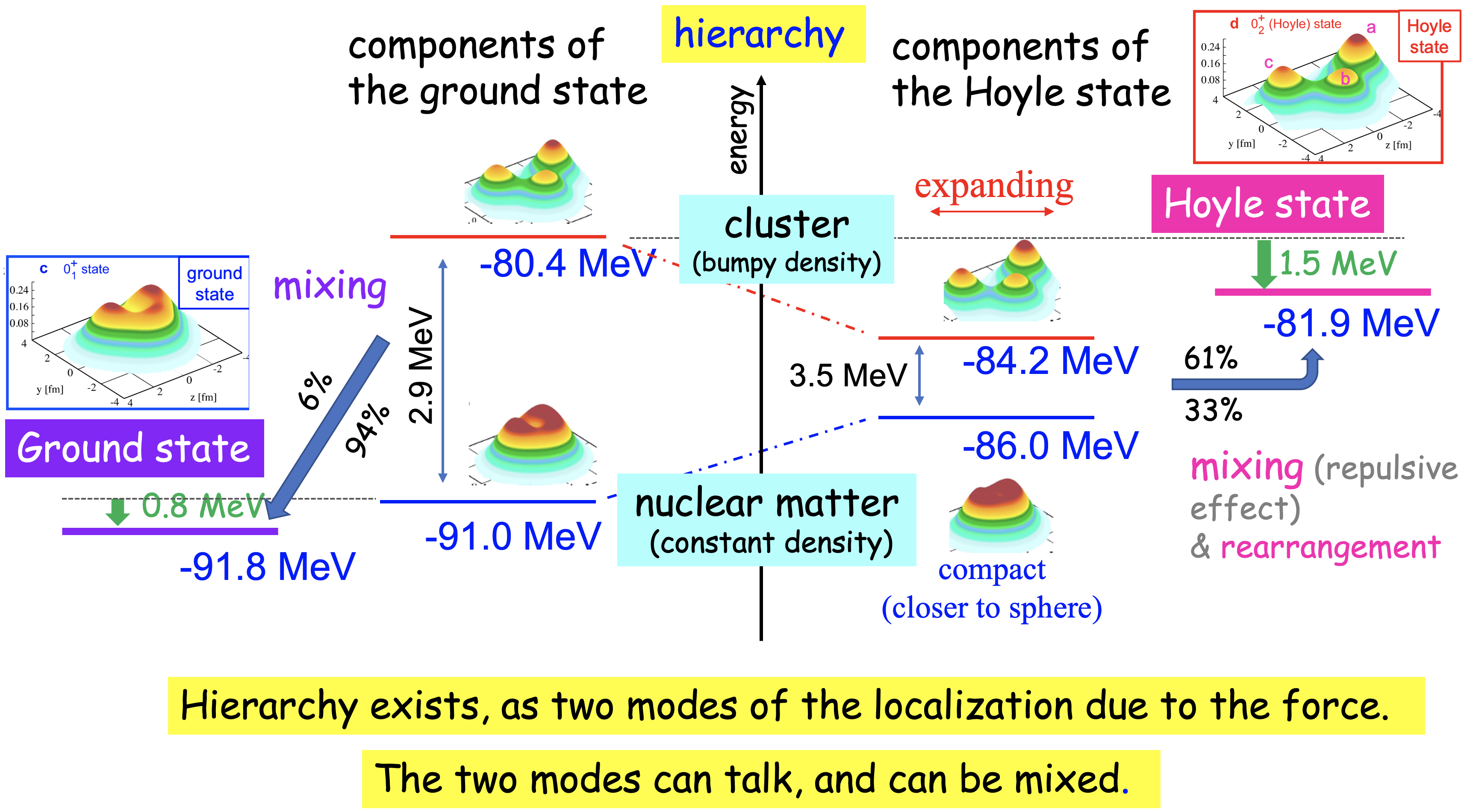}
    \caption{ {\bf Overall picture of ground and Hoyle states of $^{12}$C }
    } 
  \label{fig:12C_overall}  
\end{figure}

The nuclear-matter-type density profile, characterized by a constant density over a certain region, emerges definitely in the ground state, and rather modestly in the Hoyle states.   At the same time, $\alpha$-clustering-type density profile also emerges in these states, certainly to different extents.  It is of interest how building-block states producing these density profiles arise and are mixed.  Figure~\ref{fig:12C_overall} exhibits a sketch for this point.   

%%%%%%%%%%%%  FIGURE 4  %%%%%%%%%%%%%
% Fig4: overall picture of 12C

%%%%%%%%%%%%%%%%%%%%%%%%%%%%%%%%%%%%%%%%

We begin with the ground state.  The two components corresponding to groups (i) and (ii) are located at the energies $E=$ -91.0 and -80.4 MeV, as shown in the left half of Fig.~\ref{fig:12C_overall}.   Due to the nuclear forces (Daejeon16 interaction in this case), they are mixed, and the resulting state appears at the energy lower by 0.8 MeV, with the mixing probabilities 94\% and 6\%, respectively, for nuclear matter (group (i)) and cluster (group (ii)) components. 
Note that the classification was made according to ellipsoidal shapes of basis vector states, but the two components were named after their density-profile characters. 
The mixing of 6\% gives the ground state additional binding energy of 0.8 MeV.   Thus, the nuclear forces favor the $\alpha$ clustering even in the middle of nuclear matter, although it may occur mainly around the surface.  This surface enhancement is another interesting feature to be further investigated.

%%%%%%%%%%%%  FIGURE 5  %%%%%%%%%%%%%
% Fig5: 2-dim. density of 12C and 8Be

\begin{figure}[tb]
  \centering
    \includegraphics[width=13cm]{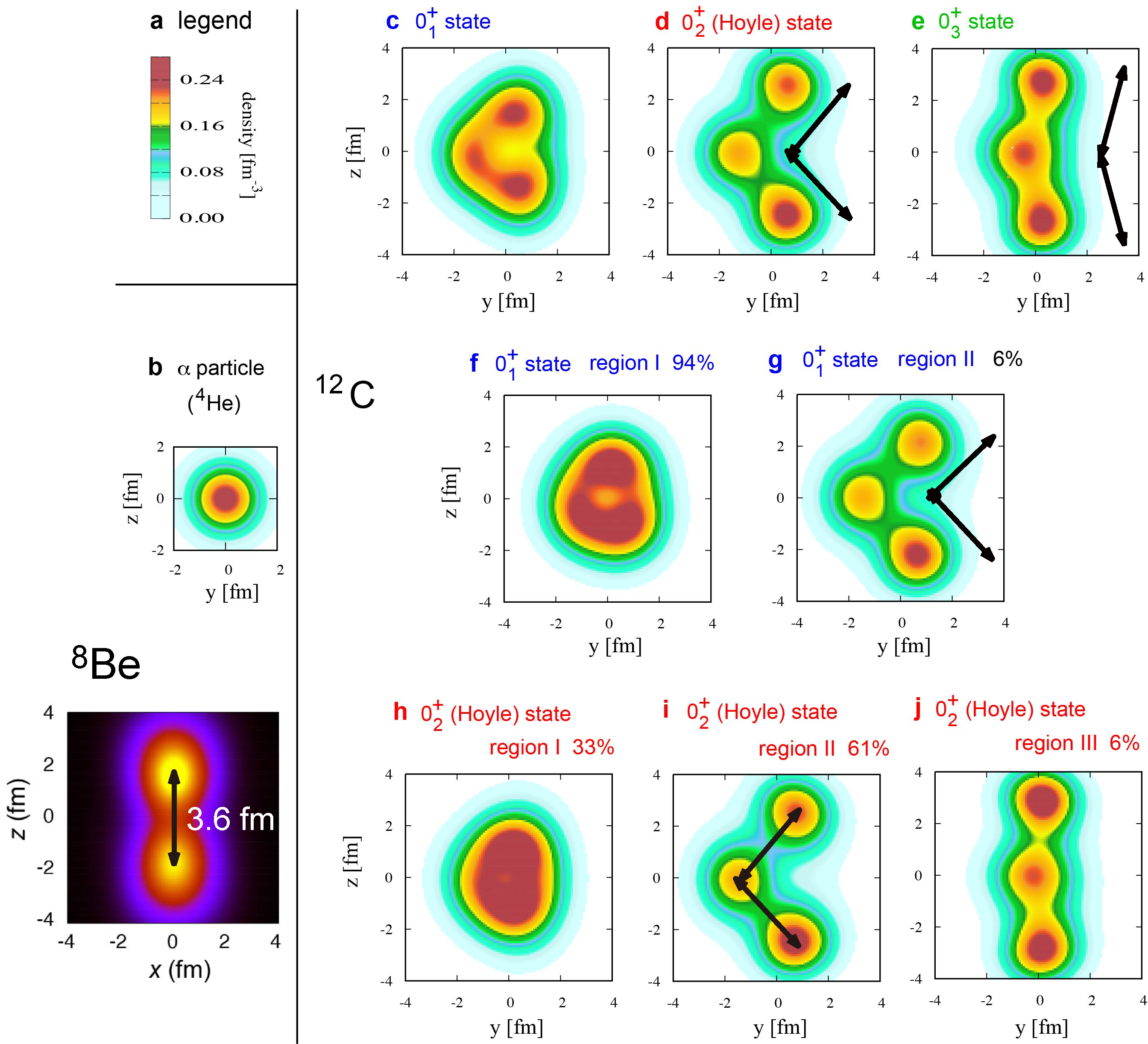}
    \caption{ {\bf Two-dimensional representation of matter density profile of $^{12}$C compared to $^{4}$He and $^{8}$Be}
    Densities of $^{4}$He and $^{8}$Be are shown in the far left part.  
    Panels {\bf c} - {\bf i} correspond, respectively, to Panels {\bf c} - {\bf i} of Fig.~\ref{fig:12C_density}.  
    Two-way arrow indicates the distance between two peaks of $^{8}$Be density with the length $\sim$3.6 fm, and is also 
    shown for some panels for $^{12}$C.
    Modified from figures in \cite{otsuka_2022} with permissions.} 
  \label{fig:12C_8Be}  
\end{figure}  

We now move on to the Hoyle state.  The nuclear-matter component formed by group (i) is located at -86.0 MeV.  It is closer to a sphere than the one for the ground state.  The cluster component is located at -84.2 MeV, lower by 3.8 MeV than its counterpart for the ground state.  This is 7.6 MeV above the ground-state energy, which is closer to the experimental Hoyle-state energy.   However, nature is not so nice in this case, and pushes up the mixed state to -81.0 MeV.  This repulsive mixing occurs basically because of the orthogonality to the ground state.  The mixed state is still below the unperturbed cluster component of the ground state, so there is still some binding-energy gain.

This is a sketch how the ground and Hoyle state are formed.  The relation between clustering and normal nuclear matter is considered to be a crossover\cite{crossover} as pointed out in \cite{otsuka_2022}, and appears to be rather complicated as each structure also has variations. Indeed, the mixing between such two structures plays significant roles, especially in the ground state, and may lead us to further understanding of $\alpha$ decay of the states well-bound in general.  The cluster-component wave function of the Hoyle state may be somewhat improved by including more single-particle states of harmonic oscillator potential, making individual ``clusters'' more $\alpha$-particle-like as a possibility.   The positions of those clusters may not substantially change, however.  In order to see this point, we show the distances between clusters in Fig.~\ref{fig:12C_8Be}.   The density profile for $^8$Be is shown in the left lower corner of Fig.~\ref{fig:12C_8Be}, and the distance between two peaks appears to be about 3.6 fm. This value is obtained by human eyes, but other manners will lead to similar values.
The distance is displayed by two-way arrow. We can put the same arrow with angles tilted in panels for $^{12}$C in Fig.~\ref{fig:12C_8Be}, where panels {\bf c} - {\bf i} correspond, respectively, to panels {\bf c} - {\bf i} in Fig.~\ref{fig:12C_density}.   We find remarkable similarities between the $^8$Be distance and the distances in those $^{12}$C cases.  The distance $\sim$3.6 fm has a special meaning for keeping $\alpha$-cluster-like structures, as a balance between kinetic energy and binding effects by nuclear forces.  For $^{12}$C, additional single-particle states of harmonic oscillator potential may shape up wave functions of individual clusters, but may not change their positions much.

Summarizing this subsection, $\alpha$ clustering emerges as a consequence of nuclear forces in the present first principles calculation.   The Hoyle state is primarily made up of three $\alpha$ clusters as discussed in quite a few earlier works of various types, but its triangular configuration is presently not equilateral.  
The ground and 2$^+_1$ states are members of an oblate rotational band with $\beta_2 \sim$ 0.6 \cite{otsuka_2022}.   The Hoyle state is also strongly deformed in terms of quadrupole deformation with $\beta_2 \gtrsim$ 1.0 \cite{otsuka_2022}.  We now turn to discussions on such strongly deformed state and their rotations.

\subsection{General description of rotational excitations within quantum many-body theory \label{general}}

\begin{figure}[tb]
  \centering
    \includegraphics[width=13cm]{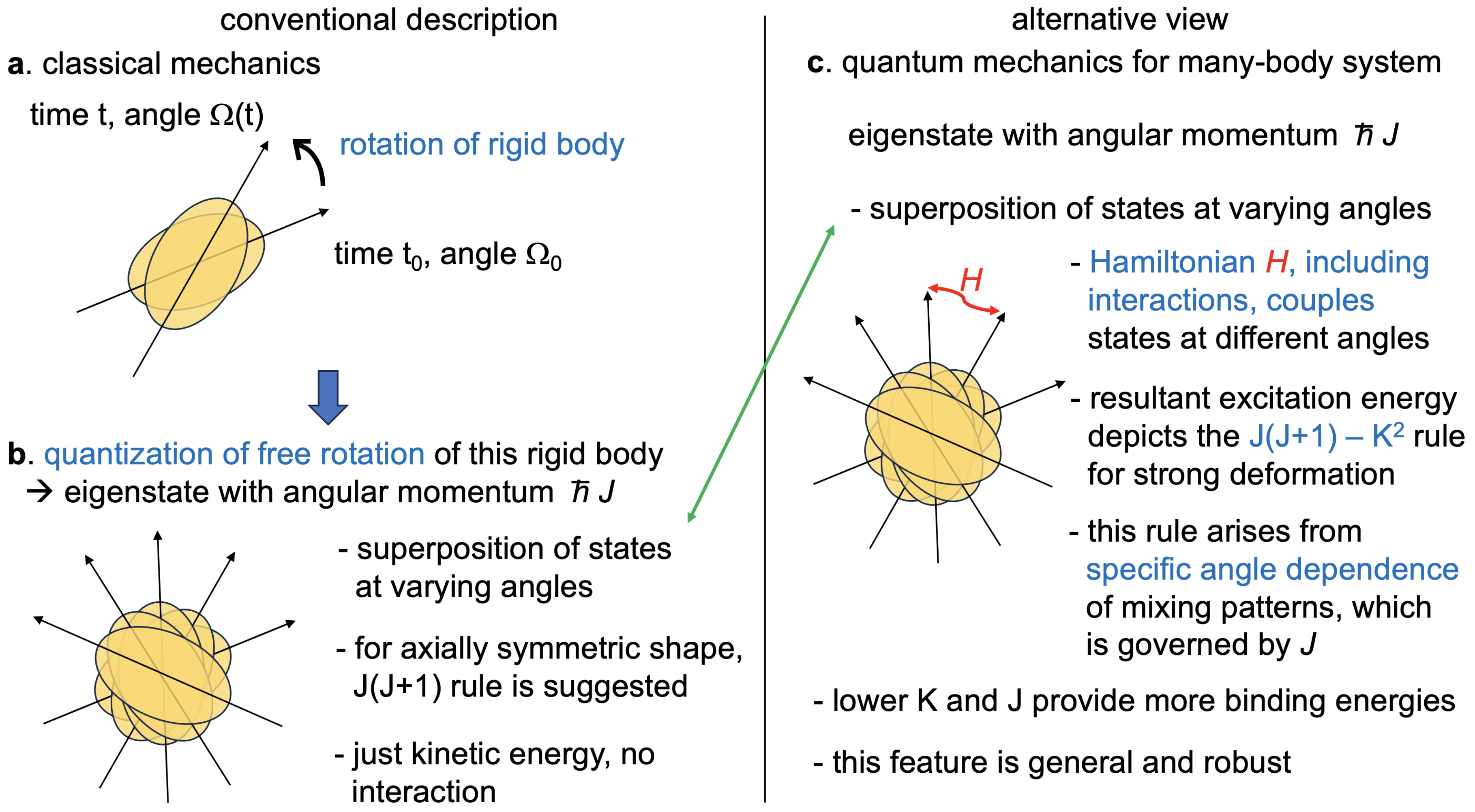}
    \caption{ {\bf Schematic illustrations of the rotation in classical and quantum mechanics.}    
    {\bf a.} Classical mechanical view.  {\bf b.} Quantization of freely rotating rigid body. 
    {\bf c.} View of the quantum mechanical system composed of many constituents with angular momentum $\hbar J$.  The $J(J+1)$-$K^2$ rule arises. 
     The green arrow indicates similarity in the wave function, but the energy comes from different 
     origins.     Taken from Fig. 24 of \cite{otsuka_2025} with kind permission of The European Physical Journal (EPJ).     
    } 
  \label{fig:rotation}  
\end{figure}

In the classical mechanics, a rigid body rotates as sketched in Fig.~\ref{fig:rotation}{\bf a}.
This motion evolves as time goes by, following the Newtonian equation.  

We then consider the free rotation of a rigid body in the quantum mechanics. The angular momentum has to be quantized.
Provided that the rigid body is of axially symmetry, as illustrated in Fig.~\ref{fig:rotation}{\bf b}, the rotational kinetic energy is proportional to $(\vec{J}\cdot \vec{J})$, where the angular momentum of the rigid body is denoted by $\hbar J$.   The eigenvalues of the rotational kinetic energy are then proportional to $J(J+1)$.

We now move on to nuclei.  In the picture of deformed nuclei proposed by Aage Bohr \cite{bohr1952,bohr_mottelson1953,aage_bohr_nobel,bohr_mottelson_book2}, the nucleus is described as a deformed object with a fixed shape filled with a uniform-density matter.  It was still a rigid body.  The argument in the previous paragraph is applied to the nuclear case.  The so-called Bohr Hamiltonian contains a kinetic term representing the rotational motion of this rigid body about three principal axes (see e.g.,  \cite{ring_schuck_book}).  By assuming the axial symmetry of this object, the $J(J+1)$ rule of excitation energies within a rotational band arises, exhibiting an account for the origin of observed level-energy regularity ($\propto J(J+1)$) in many nuclei.   This axially-symmetric rigid-body picture for deformed nuclei seems to be one of the major elements of the Nobel prize of physics in 1975 \cite{aage_bohr_award_speech}, and has remained as a paradigm of nuclear rotational bands for (the majority of) the community.

We now turn to another formulation which is free from the classical (or semi-classical) picture/interpretation where a deformed nucleus is regarded as an axially-symmetric rigid-body.
In quantum many-body theory, a rotational band can be defined as a set of many-nucleon states, where the member of angular momentum $J$ is generated by projecting a common intrinsic state onto this angular momentum $J$.  This definition itself may not be new, but we start from it and re-formulate the whole description of rotational bands, staying inside quantum mechanics (without resorting to the quantization of the free rotation of an axially-symmetric rigid-body). We will come back to differences from conventional formulations later.

We somewhat elaborate on the actual theoretical process as pedagogically as possible, largely because it matters to major discussions later.  The angular-momentum projection method is formulated with Wigner's D function (see eq.(9) of \cite{otsuka_2025} where \cite{ring_schuck_book} is cited for it), and we begin with its concise sketch.  First, the intrinsic state is denoted by $\phi$.  The state $\phi$ can be a sophisticated state containing full of correlations by nuclear forces.  So, it does not have to be a simple state.  From this $\phi$, we obtain the state of definite $J$ and $M$, the total angular momentum and its z-projection in the laboratory frame.  This projection can be performed by rotating $\phi$ in the three-dimensional space with three Euler angles $\alpha, \beta$ and $\gamma$, and by integrating it with an appropriate weighting factor, Wigner's D function.  The obtained state is written as,
\begin{eqnarray}
&\Psi \bigl[\phi, J, M, K \bigr] \,=\,&(2J+1)/(8 \pi^2)    \int_0^{2\pi}   d\alpha \int_0^{\pi} d\beta \, {\rm sin}\beta \, \int_0^{2\pi} d\gamma \,\,\,\,\, \nonumber \\
&& \big\{ D^J_{M,K} (\alpha, \beta, \gamma) \big\}^*\, e^{i\alpha \hat{J}_z} \, e^{i\beta \hat{J}_y} \, e^{i\gamma \hat{J}_z} \, | \, \phi \rangle,
\label{eq:rot_phi}
\end{eqnarray}
where $D$ is the Wigner's function.  
%Note that $\gamma$ here is different from the deformation parameter $\gamma$, but no confusion is expected in this traditional usage of the character. Likewise, $\alpha$ here has nothing to do with the $\alpha$ particle. 

Equation~(\ref{eq:rot_phi}) implies that the three-fold rotation of $\phi$ generates states with good ($J$, $M$) pairs.
One notices an additional index of $K$.  In fact, there can be different and independent states from the same $\phi$ for a given pair ($J$, $M$), and $K$ specifies them.
By doing the $J_z$ rotation ($e^{i\gamma \hat{J}_z}$) with proper weighting factor, 
we project $\phi$ onto a specific value of $K$, the $z$-component of $\vec{J}$ of $\phi$.  
In this section, $K$=0 is assumed for clarity, while more general cases can be discussed similarly \cite{otsuka_2025}.  

This $K=0$ state, denoted by $\phi_0$, is obtained by extracting the relevant parts from eq.~(\ref{eq:rot_phi}), 
\begin{equation}
\phi_0  \,=\, \frac{1}{2\pi}  \int_0^{2\pi} d\gamma \, e^{i\gamma \hat{J}_z} \, \phi  . 
\label{eq:phi_0}
\end{equation}
It is clear that all relevant orientations are superposed with the same amplitude.  

We also take $M$=0 without losing generality, because the Hamiltonian is rotationally invariant.  The relation
\begin{equation}
D^J_{M,K} (\alpha, \beta, \gamma) \,=\, e^{iM\alpha} \, d^J_{M,K} (\beta) \, e^{iK\gamma } ,
\label{eq:Wigner}
\end{equation}
is used with $d^J_{M,K} (\beta)$ being the (small) $d$ function.     With $M$=$K$=0, eq.~(\ref{eq:Wigner}) becomes 
\begin{equation}
D^J_{M=0,K=0} (\alpha, \beta, \gamma) \,=\, d^J_{0,0} (\beta).
\label{eq:Wigner00}
\end{equation}
With this, we consider $J$-projected norms and Hamiltonian matrix elements. 
Because of $M$=0, the $\phi_0$ state appears for the bra state. 
The norm of the $J$-state component contained in $\phi_0$ is now given by,
\begin{eqnarray}
&|{\mathcal N_J}|^2  \,&=\,\frac{2J+1}{8 \pi^2}\, \int_0^{2\pi} d\alpha \,  \int_0^{\pi} d({\rm cos}\beta) \, \int_0^{2\pi} d\gamma \,  \nonumber \\
& &   \,\,\,\,\,\,\,\,\,\,\, \langle \phi \,|\, \big\{ D^J_{0,0} (\alpha, \beta, \gamma) \big\}^*
 e^{i\alpha \hat{J}_z} \, e^{i\beta \hat{J}_y} \, e^{i\gamma \hat{J}_z} \,  | \, \phi \rangle \nonumber \\
& &= \, \frac{2J+1}{2} \, \int_0^{\pi} d({\rm cos}\beta) \, d^J_{0,0} (\beta) \, \langle \phi_0 \,|\, e^{i\beta \,\hat{J}_y} | \, \phi_0 \rangle . 
\label{eq:Nj}
\end{eqnarray}  
The corresponding quantity for the Hamiltonian, $H$,  is obtained by inserting $H$ after $\langle \phi \,|$ or $\langle \phi_0 \,|$.

The normalized expectation value of the Hamiltonian $H$ for the projected state is then given by 
\begin{eqnarray}
& E_J \,&=\, \frac{\, \int_0^{\pi} d({\rm cos}\beta) \, d^J_{0,0} (\beta) \, \langle \phi_0 \, |  \, H \, e^{i\beta \hat{J}_y} \, | \, \phi_0 \,\rangle}  {\, \int_0^{\pi} d({\rm cos}\beta) \, d^J_{0,0} (\beta) \, \langle \phi_0 \, |  \, e^{i\beta \hat{J}_y} \, | \, \phi_0 \,\rangle} \nonumber \\
& &=\, \frac{ \, \int_0^{\pi} d({\rm cos}\beta) \, d^J_{0,0} (\beta) \, h_y(\beta)}
                  { \, \int_0^{\pi} d({\rm cos}\beta) \, d^J_{0,0} (\beta) \, n_y(\beta)},
\label{eq:H yn}
\end{eqnarray}
where the energy and norm kernels are introduced as, 
\begin{equation}
h_y(\beta)  \,=\,  \langle \, \phi_0 \,|\, H \, e^{i\beta \hat{J}_y} \,  | \, \phi_0 \rangle ,
\label{eq:EJkernel}
\end{equation}
and 
\begin{equation}
n_y(\beta)  \,=\,  \langle \, \phi_0 \,|\, e^{i\beta \hat{J}_y} \,  | \, \phi_0 \rangle .
\vspace{0.5cm}
\label{eq:NJkernel}
\end{equation} 
The following identity is recalled, 
\begin{equation}
d^J_{0,0} (\beta)  \,=\,  P_J ({\rm cos}\beta), 
\label{eq:dfunc}
\end{equation}
where $P_J ({\rm cos}\beta)$ stands for a Legendre polynomial. 
By expanding it in terms of $({\rm cos}\beta - 1)^k$, with $k$=0, 1, 2, ...,
the first two terms of the expansion are written as, 
\begin{equation} 
P_{J} ({\rm cos}\beta) \, = \, 1 \,+\, J(J+1)/2 \, ({\rm cos}\beta - 1)  \,+\,  ... \,.
\label{eq:Legendre3}
\end{equation}
The two terms in eq.~(\ref{eq:Legendre3}) give a good approximation if ${\rm cos}\beta$ is close enough to unity. 

The values of $n_y(\beta)$ and $h_y(\beta)$ are reduced quickly as $\beta$ moves away from 0, as a consequence of strong deformation.  
With this situation, the $d$ function is approximated as 
\begin{equation} 
d^J_{0,0} (\beta)  \,=\,  P_J ({\rm cos}\beta) \, \approx \, 1 \,+\, F_J \, ({\rm cos}\beta - 1)\,\,\,\, {\rm for} \,\, \beta \approx 0 ,
\label{eq:PJeq3}
\end{equation}
with
\begin{equation} 
 F_J = J(J+1)/2.  
\label{eq:FJ0}
\end{equation}
As the integral is carried out with the variable ${\rm cos}\,\beta$ in eqs.~(\ref{eq:Nj}) and (\ref{eq:H yn}), 
the $d$-function in eqs.~(\ref{eq:Nj}) and (\ref{eq:H yn}) is naturally replaced by the function in eq.~(\ref{eq:PJeq3}), a polynomial of $({\rm cos}\,\beta - 1)$.

The range of $\beta$ runs from 0 to $\pi$.  Sizable contributions to the quantities are expected also for $\beta$ close to $\pi$, 
as the overlap is generally restored. For $\beta \sim \pi$, the linear and other approximations starting from $\beta = \pi$ back to smaller values work well also.  Although these contributions can be evaluated in the same way as those from $\beta \sim 0$, their concrete description is omitted for brevity.

We define
\begin{equation} 
n_k \, = \, \int d({\rm cos}\beta) \, n_y(\beta) \, ({\rm cos}\beta - 1)^k , \,\,\,\,  {\rm for } \, k=0, 1, 2, \dots ,
\label{eq:ni}
\end{equation}
%and
%\begin{equation} 
%n_1 \, = \, \int d({\rm cos}\beta) \, n_y(\beta) \, ({\rm cos}\beta - 1).
%\label{eq:n1}
%\end{equation}
and
\begin{equation} 
e_k \, = \, \int d({\rm cos}\beta) \, h_y(\beta)  \, ({\rm cos}\beta - 1)^k , \,\,\,\, {\rm for } \, k=0, 1, 2, \dots \,.
\label{eq:hi}
\end{equation}
%and
%\begin{equation} 
%e_1 \, = \, \int d({\rm cos}\beta) \, h_y(\beta) \, ({\rm cos}\beta - 1).
%\label{eq:h1}
%\end{equation}
The projected energy of the state of $J$ is given by
\begin{equation} 
E_J \, \approx \, \frac{e_0 \,+\, F_J e_1}{n_0 \,+\, F_J n_1}.
\label{eq:EJ}
\end{equation}
As the inequalities, $n_1/n_0, e_1/e_0 \ll$ 1, hold for strongly deformed states, the energy eigenvalues are then given by,
\begin{equation} 
E_J \, \approx \, E_0 \, + \, J(J+1) \,\, \frac{1}{2} \, \frac{e_0}{n_0} \,\{ \, \frac{e_1}{e_0} \,-\, \frac{n_1}{n_0} \}, \,\,\, {\rm with} \,\, E_0 \, = \, \frac{e_0}{n_0}.
\label{eq:E}
\end{equation}
The $J (J+1)$ rule of the rotational excitation energy thus emerges within quantum many-body theory for strongly deformed states. We stress that the quantization of free rotation of axially symmetric rigid-body is not used.  
Some details about the polynomial expansion of the $d$ function are found in \cite{otsuka_2025}, including some history.  For general $K>0$ values, more general $\{ J(J+1) - K^2 \}$ rule has also been given in \cite{otsuka_2025,otsuka_2025A}.  
   
%%%%%%%%%%%%%%%%%%%%%%%%%%%%%%%%%%%%%%%%%%%%%%%%
\begin{figure}[tb]
  \centering
    \includegraphics[width=6cm]{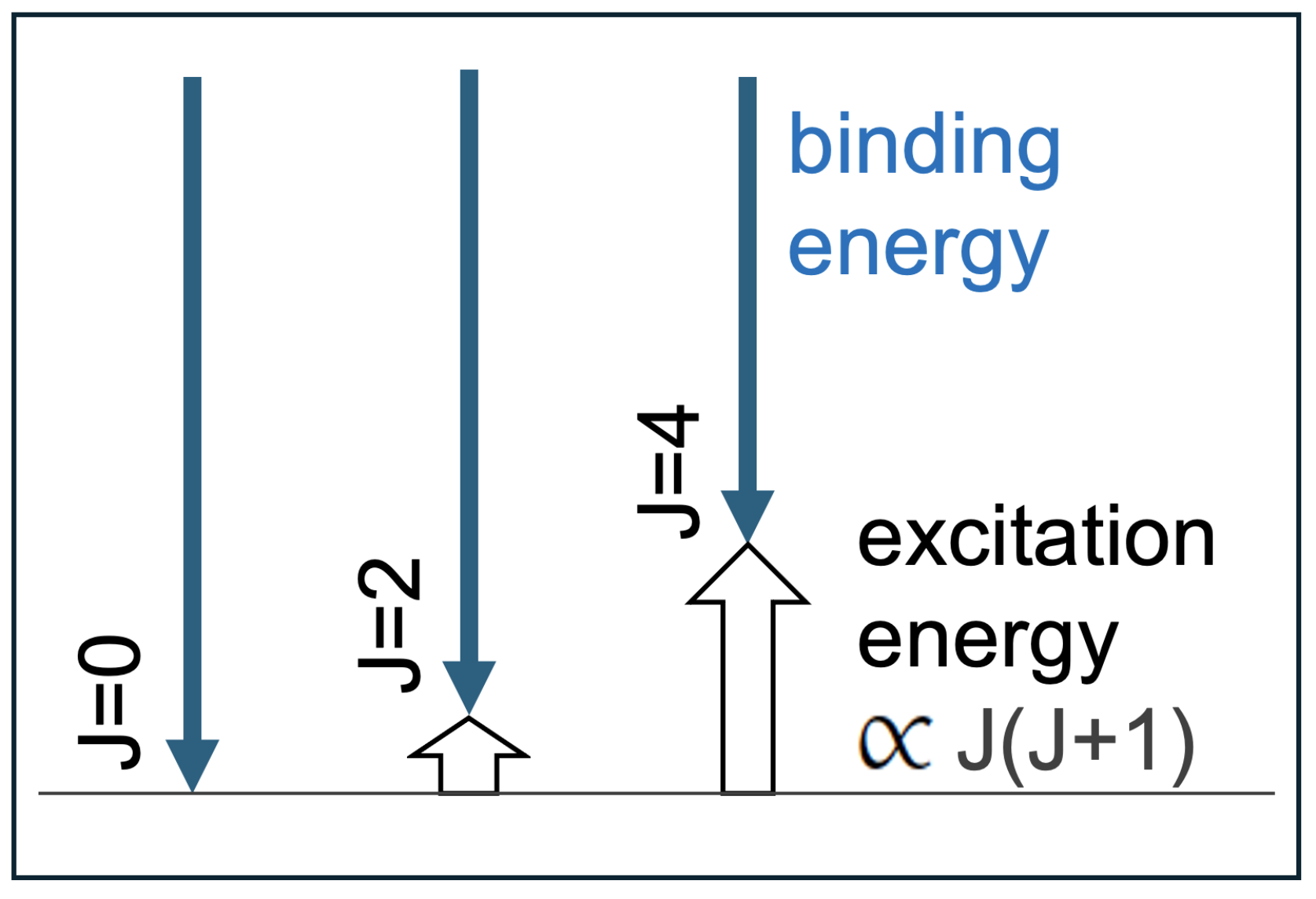}
    \caption{ {\bf A graphical illustration of the origin of the $J(J+1)$ rule of excitation energies within a rotational band.}
    Blue downward arrows mean binding energies for the states with the angular momentum $J$=0, 2, 4.  The differences from the energy of the $J$=0 state are exhibited by open upward arrows.  The present fully quantum mechanical study indicates their height being proportional to $J(J+1)$, if strong ellipsoidal deformation occurs, irrespectively of triaxiality.  For $K>$0, the $\{ J(J+1)-K^2 \}$ rule is obtained similarly, on top of the $J$=$K$ state.
    Taken from figures in \cite{otsuka_2025A}.}   
    \label{fig:binding}  
\end{figure}  
%%%%%%%%%%%%%%%%%%%%%%%%%%%%%%%%%%%%%%%%%%%%%%%%

The most important property suggested by eq.~(\ref{eq:E}) may be the feature displayed in Fig.~\ref{fig:binding}: the $J(J+1)$ dependence of the excitation energy originates in the $J$-dependent reduction (i.e., decrease) of the binding energy provided by the Hamiltonian, $H$.  This reduction occurs primarily in the contributions of various parts of the nuclear forces, including single-particle energies, two-nucleon forces, three-nucleon forces, {\it etc}.  All parts give the $J(J+1)$ dependence for strong deformation.
% if one looks at appropriate differences from the values for the band head ($J$=0 value in Fig.~\ref{fig:binding} and $J$=$K$ value in general).  
In many practical calculations made so far, the kinetic energy appears to be a tiny fraction, or even can work against (lowering rather than raising).  
%Note that in Fig.~\ref{fig:binding}, practical conservation of $K$ quantum number is already incorporated, as it has been demonstrated in \cite{otsuka_2025}.   

%%%%%%%%% Rotational excitations built on clustering states   %%%%%%%%%%

\subsection{Rotational excitations built on clustering states \label{distant}}

The quantum many-body formulation of the rotational mode presented in the previous subsection can obviously be applied to the ground and 2$^+_1$ states of $^{12}$C, as they are nuclear-matter-type states.  The properties shown in Fig.~\ref{fig:binding} should basically hold for them.

A question arises here as to what picture is appropriate for the states dominated by cluster formation, which differ in structure from nuclear-matter states as discussed in Subsect.~\ref{first-principles realization}.    
The most straightforward example may be the ground and 2$^+_1$ states of $^8$Be, where the ${\it di}$-$\alpha$ cluster appears to be a full description of these states as demonstrated in the left lower corner of Fig.~\ref{fig:12C_8Be}.

\begin{figure}[tb]
  \centering
  \includegraphics[width=13cm]{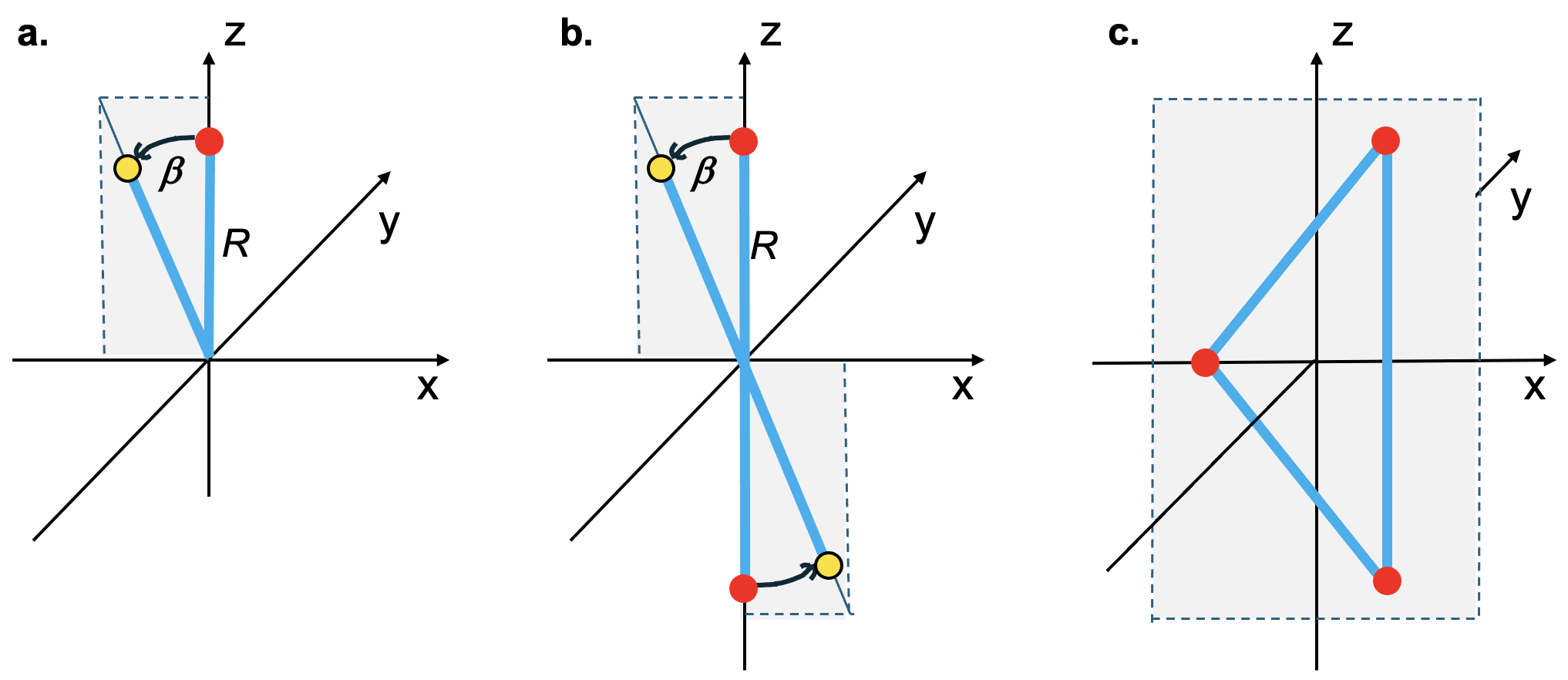}
    \caption{ {\bf A schematic illustration of the rotational mode of clustering states.} \\
      Red and yellow circles imply clusters for systems with 
     {\bf a} single cluster,  {\bf b} two clusters just opposite over the origin, and 
     {\bf c} a possible configuration of three clusters.   Yellow circles in {\bf a,b} denote rotation of red ones by angle $\beta$ in the integral (see the text). In {\bf c}, such rotation angles are too complicated to be included.  Gray rectangular boards indicate the planes where clusters are in relevant senses.
$R$ stands for the radius of cluster's moving surface.Bold blue lines are to guide the eye.}   
    \label{fig:cluster_rot}  
\end{figure}

We therefore extend the general formulation described in Subsect.~\ref{general} so as to be applicable to the case of one cluster, as shown in Fig.~\ref{fig:cluster_rot}{\bf a}, where a small circle stands for a cluster.  This is certainly a simplified treatment of more realistic case shown in Fig.~\ref{fig:cluster_rot}{\bf b}, which corresponds to $^8$Be case.
%It is straightforward to extend it to {\it di-}cluster cases such as $^8$Be.
As the cluster is supposed to be stable internally and have spin zero, its motion as a whole is described in terms of its center of gravity.  
%We implicitly assume that there are more than one cluster keeping the center of gravity of the whole nucleus separated from the present discussion.
The intrinsic wave function of this cluster is denoted by $\phi_0$, following the convention of Subsect.~\ref{general}.  By definition, it is a $K^P$=0$^+$ state.  We assume that the internal state of the cluster is of angular momentum $J_c$=0 with positive parity, for simplicity.

%%%%%%%%%%%%%%%%%%%%%%%%%%%%%%%%%%%%%%%%%%%%%%%%
%%%%%%%%%%%%%%%%%%%%%%%%%%%%%%%%%%%%%%%%%%%%%%%%

The overlap between intrinsic state $\phi_0$ and its rotated state $e^{i\beta \hat{J}_y} \,  | \, \phi_0 \rangle$ is called norm kernel as mentioned earlier, and is shown in eq.~(\ref{eq:NJkernel}).
Likewise, the energy kernel is shown in eq.~(\ref{eq:EJkernel}).  
Both quantities should have large magnitudes at and near $\beta$=0, and damp quickly as $\beta$ moves away from $\beta$=0. 

The Hamiltonian $H$ of the present system contains two parts: the kinetic energy of the center of gravity of the cluster, $H_g$, and the internal part $H_i$ of the cluster.  
The cluster is considered to be self-contained, implying that the internal state is the lowest eigen state of  $H_i$ with the eigenvalue $\epsilon$.  We assume this for obtaining the basic picture to start with, and obtain 
\begin{eqnarray}
h_y(\beta)  \,&=&\,  \langle \, \phi_0 \,|\, ( H_g \,+\, H_i ) \, e^{i\beta \hat{J}_y} \,  | \, \phi_0 \rangle  \nonumber \\
   &=&\,  \langle \, \phi_0 \,|\, ( H_g \,+\, \epsilon ) \, e^{i\beta \hat{J}_y} \,  | \, \phi_0 \rangle  \nonumber \\
   &=&\,  \langle \, \phi_0 \,|\, H_g \, e^{i\beta \hat{J}_y} \,  | \, \phi_0 \rangle  + \epsilon \, n_y(\beta) .
   \label{eq:E_ker_1}
\end{eqnarray} 
Because the cluster is in the $J _c=0^+$ ground state with its internal eigen energy denoted by $\epsilon$,  the second term on the right-hand-side produces a constant shift of $\epsilon=E_{J_c=0^+}$ (see eq.~(\ref{eq:H yn})).  Keeping this in mind, we shall not consider contributions of this term hereafter.
Likewise, besides $H_g$ and $H_i$, there is Hamiltonian representing the radial motion of the cluster.  It is not considered here either, as the radius of the center of gravity of the cluster is fixed at a given length.  This is taken as a reasonable modeling for the present basic picture.  Note that various properties of the radial motion, {\it e.g.} radial smearing of wave function, can be included in more elaborate calculations at varying levels of sophistication, keeping basic features to be shown here.

In the present case, as shown in Fig.~\ref{fig:cluster_rot}{\bf b}, the center of gravity of the cluster is rotated by the Euler angle $\beta$, with a fixed radius, $R$, from the center of the gravity of the whole nucleus.    The center of gravity of the cluster can be treated as a point mass.  As the radius is fixed, the relevant part of the Hamiltonian is its zenith-angle dependent part of the kinetic energy, which can be found in any textbook of quantum mechanics:
\begin{equation}
H_g  \,=\,-\,\frac{\hbar^2}{2 \, \mathcal{I} } \,  \frac{1}{{\rm sin}\beta} \frac{\partial}{\partial \beta} ({\rm sin}\beta  \frac{\partial}{\partial \beta} ),   
\label{eq:Hg}
\end{equation} 
where $\mathcal{I}$ is moment of inertia with $\mathcal{I}  
=m R^2$ with $m$ being the mass of cluster. 
The energy kernel is 
\begin{eqnarray}
h_y(\beta)  \,&=&\, - \,\frac{\hbar^2}{2 \, \mathcal{I} }  \langle \, \phi_0 \,|\, \frac{1}{{\rm sin}\beta} \frac{\partial}{\partial \beta} ({\rm sin}\beta  \frac{\partial}{\partial \beta} ) \, e^{i\beta \hat{J}_y} \,  | \, \phi_0 \rangle, \nonumber \\
&=&\, - \,\frac{\hbar^2}{2 \, \mathcal{I} } \frac{1}{{\rm sin}\beta} \frac{\partial}{\partial \beta} ({\rm sin}\beta  \frac{\partial}{\partial \beta} ) \, n_y(\beta) .
\label{eq:EK2}
\end{eqnarray} 

By performing partial integration twice, we obtain 
\begin{eqnarray} 
e_0 \, &=& \, - \,\frac{\hbar^2}{2 \, \mathcal{I} } \, \int d{\beta} \, {\rm sin}\beta \, \frac{1}{{\rm sin}\beta} \frac{\partial}{\partial \beta} ({\rm sin}\beta  \frac{\partial}{\partial \beta} ) \, n_y(\beta), \nonumber \\
&=& \, - \,\frac{\hbar^2}{2 \, \mathcal{I} } \, [ ({\rm sin}\beta  \frac{\partial}{\partial \beta} ) \, n_y(\beta) ]^{\pi}_0 = 0.
\label{eq:e0i}
\end{eqnarray}
Here $({\rm sin}\beta  \frac{\partial}{\partial \beta} ) \, n_y(\beta)$ is assumed to vanish at $\beta$=0 and $\pi$, as it is very likely.  

Similarly, we obtain 
\begin{eqnarray} 
e_1 \, &= \, - \,&\frac{\hbar^2}{2 \, \mathcal{I} } \, \int d{\beta}({\rm sin}\beta) \, ({\rm cos}\beta - 1) \, \frac{1}{{\rm sin}\beta} \frac{\partial}{\partial \beta} ({\rm sin}\beta  \frac{\partial}{\partial \beta} ) \, n_y(\beta) \nonumber \\
&=    \textcolor{blue}{+ \,}   &\frac{\hbar^2}{2 \, \mathcal{I} } \, \int d{\beta} ( 2 \,{\rm cos}\beta \,{\rm sin}\beta) \, n_y(\beta),
\label{eq:e1a}
\end{eqnarray}
If $n_y(\beta)$ is non-vanishing only for $\beta$ very close to 0, this quantity becomes
\begin{equation} 
e_1 \, \approx \, \textcolor{blue}{+ \,} \frac{\hbar^2}{2 \, \mathcal{I} } \, 2 \int d{\beta} \, ({\rm sin}\beta) \, n_y(\beta)   
       = \, \textcolor{blue}{+ \,} \frac{\hbar^2}{2 \, \mathcal{I} } \, 2 \, n_0 .
\label{eq:e1b}
\end{equation}

Because of eq.~(\ref{eq:e0i}), $\frac{e_0}{n_0} \frac{n_1}{n_0}$ term in eq.~(\ref{eq:E})  vanishes.
Equation~(\ref{eq:E}) then becomes 
\begin{equation} 
E_J \,  \approx  \, E_0 \, + \, J(J+1) \,\, \frac{1}{2} \, \frac{e_1}{n_0} \,
           \approx  \,\, \frac{\hbar^2}{2 \,\, \mathcal{I} } \,\,J(J+1) ,
\label{eq:E_J_final}            
\end{equation}
with
\begin{equation}
E_0 \, = \, \frac{e_0}{n_0} \,= \,0.
\end{equation}
We thus obtain the energy formula for strongly localized cluster contents.  Note that the approximation in eq.~(\ref{eq:e1b}) does not affect the appearance of the $J(J+1)$ rule, and changes emerge in the coefficient depending on the deviation from this approximation.  The deviation is expected to be small for visible clustering cases.  The formula looks the same as  the equation for the quantized energies of free rotation of a point mass, while corrections to it are also in the scope.  

The excitation energy can be similarly derived for {\it di}-cluster systems like the one shown in Fig.~\ref{fig:cluster_rot}{\bf b} with proper value of the moment of inertia.
Furthermore, the axially symmetric rigid-body follows the same equation.
We stress that details of the localization are irrelevant, as $n_0$ cancel each other between the numerator and the denominator in eq.~(\ref{eq:E_J_final}), bringing about a kind of beauty.

%%%%%%%%%%%%%%%%%%%%%%%%%%%%%%%%%%%%%%%%%%%%%%%%
\begin{figure}[tb]
  \centering
  \includegraphics[width=13cm]{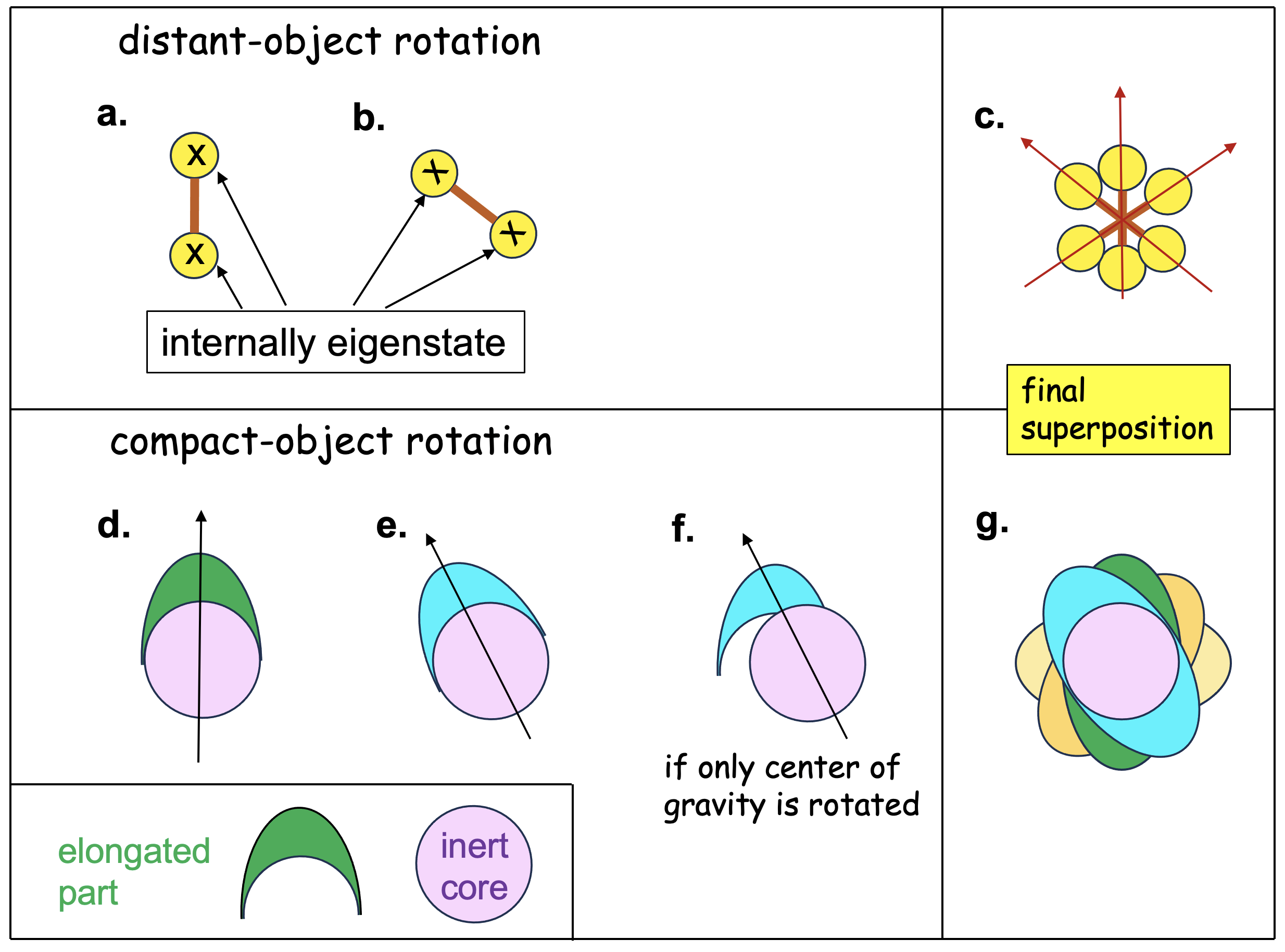}
  \caption{ {\bf Conceptual comparison between (upper part) distant-object rotation and (lower part) compact-object rotation.}
  Panels {\bf a} and {\bf b} represent {\it di}-cluster system at different orientations, where each cluster is an eigenstate of internal Hamiltonian.  Panel {\bf c} schematically displays superposition of the {\it di}-cluster system at various orientations, yielding distant-object rotation.    Panels {\bf d} and {\bf e} represent the ellipsoid at different orientations.  The elongated parts (green and blue) are coupled by the Hamiltonian including interactions among constituents.  Panel {\bf f} shows what would happen if only the center of gravity of blue part were to be rotated to another orientations.  Panel {\bf g} schematically displays superposition of the same ellipsoid at various orientations, yielding compact-object rotation.  
  }
  \label{fig:cluster_matter}  
\end{figure}  
%%%%%%%%%%%%%%%%%%%%%%%%%%%%%%%%%%%%%%%%%%%%%%%%

It is very interesting that the excitation energy now originates in rotational kinetic energy in contrast to the situation discussed in Subsect.~\ref{general}, the outcome of which is graphically depicted in Fig.~\ref{fig:binding}.
A comparison between these two situations is intuitively illustrated in Fig.~\ref{fig:cluster_matter}, where Panels {\bf a}-{\bf c} exhibit the rotational mode of clustering system and Panels {\bf d}-{\bf g} are about the rotational mode of ellipsoidal matter system.
The ``free'' rotation of clusters within quantum many-body picture is characterized here: while the wave function of good $J$ is a superposition over all possible orientations, the excitation energy is provided by the total rotational kinetic energy of the centers of gravity of the clusters.  Such a natural picture arises from the present formulation.
%This outcome illuminates the distinct difference between the cluster rotation and the picture presented in Subsect.~\ref{general}.  

For the sake of transparency, we now characterize these two types of situations by calling the former ``distant-object'' rotation (see Figs.~\ref{fig:cluster_matter}{\bf a}-{\bf c}) and the latter ``compact-object'' rotation (see Figs.~\ref{fig:cluster_matter}{\bf d}, {\bf e} and {\bf g}).  Both are quantum mechanical in the sense that the wave functions are superpositions over all orientations with specified amplitudes.  However, the dynamical origin differs completely between the two.  In the distant-object rotation, each cluster is basically in an eigenstate of its internal Hamiltonian and the excitation energy within a band represents rotational kinetic energy.  On the contrary, the compact-object rotation occurs, for instance, with ellipsoids, where the elongated parts are not eigenstates at all (see Panels {\bf d} and {\bf e}), and those in different orientations are coupled by the Hamiltonian including interactions among constituents, which are nuclear forces presently.  The change from Panels {\bf d} to {\bf e} involves proper tilting of the elongated part.  This cannot be made by simply rotating the center of gravity of the elongated part, as intuitively shown from Panels {\bf d} to {\bf f}.  The rotational kinetic term of this center of gravity is considered to yield quite minor contributions because of reduced overlap due to the tilting between Panels {\bf e} and {\bf f}.  Note that this overlap remains unity for the distant-object rotation.  A  classical counterpart probably does not exist for the compact-object rotation. 

We note that in both compact-object and distant-object rotations, the $J$=0 state (or state of no spinning) implies that the phase remains the same in all orientations for the same intrinsic state, instead of pointing to one direction.

\subsection{Rotational mode in $^{8}$Be \label{8Be}} 

The formulation of Subsect.~\ref{distant} is applied straightforwardly to the structure of $^{8}$Be.
For simplicity, we identify the peak position of the matter density as the center of gravity of cluster, and the density distribution is assumed to be localized enough, consistently with the picture of Subsect.~\ref{distant}.  The radius of the center-of-gravity movement is estimated as $R= 1.8$ fm as shown in Fig.~\ref{fig:12C_8Be}.  From this value, we can obtain
\begin{equation}
E_J \, \approx \, 0.80 \, J(J+1) \, {\rm MeV},
\label{eq:8Be_rot}
\end{equation}
where $J$ stands for the angular momentum of $^{8}$Be nucleus and $E_J$ is the excitation energy of the corresponding state in the rotational band.
This gives us $E_{2^+} \, \approx \,$ 4.8 MeV.
This value compares well with the result ($\approx$ 4 MeV) of the {\it ab initio} no-core MCSM calculation shown in Fig.~\ref{fig:12C_levels}. 
The moment of inertia parameter should be larger, as the density distribution is somewhat more spread (see Fig.~\ref{fig:12C_8Be}).  From this viewpoint, the agreement here is considered to be quite good.

The present feature arises from the localization of the cluster (i.e., quick damping of norm kernel), but is independent of details of cluster wave functions, because the same norm kernel appears both in the numerator and the denominator ($\frac{e_1}{n_0}$ in eq.~(\ref{eq:E_J_final})).
So, this beautiful feature is robust in this respect.  

\subsection{Coexistence of two rotational modes in $^{12}$C \label{dual_12C}} 

We now come back to the nucleus $^{12}$C, which is not as simple as $^8$Be. 

The structure of the Hoyle state is of great interests and still attracts attentions of comparatively recent works \cite{Chernykh_2007,Zimmerman_2011,Zimmerman_2013,Ogloblin_2014,Marin_Lambarri_2014,Funaki_2015,Marevic_2019}.   It is characterized by cluster configurations as suggested in Figs.~\ref{fig:12C_density} and \ref{fig:12C_8Be}.   For this subsection, the latter figure is more suitable.  Figure~\ref{fig:12C_8Be}{\bf d} indicates the emergence of three $\alpha$-like clusters.  We assume that each cluster is rather self-contained and stable, and behaves like an $\alpha$ particle.  In fact, by enlarging the single-particle model space in the MCSM calculation, each cluster may better resemble a free $\alpha$ particle.  
With this expectation, certain properties of the Hoyle state may be described, in a reasonable approximation, as a system of three clusters with individual center of gravity located in the peak positions in Fig.~\ref{fig:12C_8Be}{\bf d}.   These positions seem to be useful in further studies, because experimental data of E0 decay \cite{ensdf} and root-mean-square radius \cite{danilov_2009} compare well with the calculated values \cite{otsuka_2022}.  

%%%%%%%%%%%%%%%%%%%%%%%%%%%%%%%%%%%%%%%%%%%%%%%%
\begin{figure}[tb]
  \centering
  \includegraphics[width=13cm]{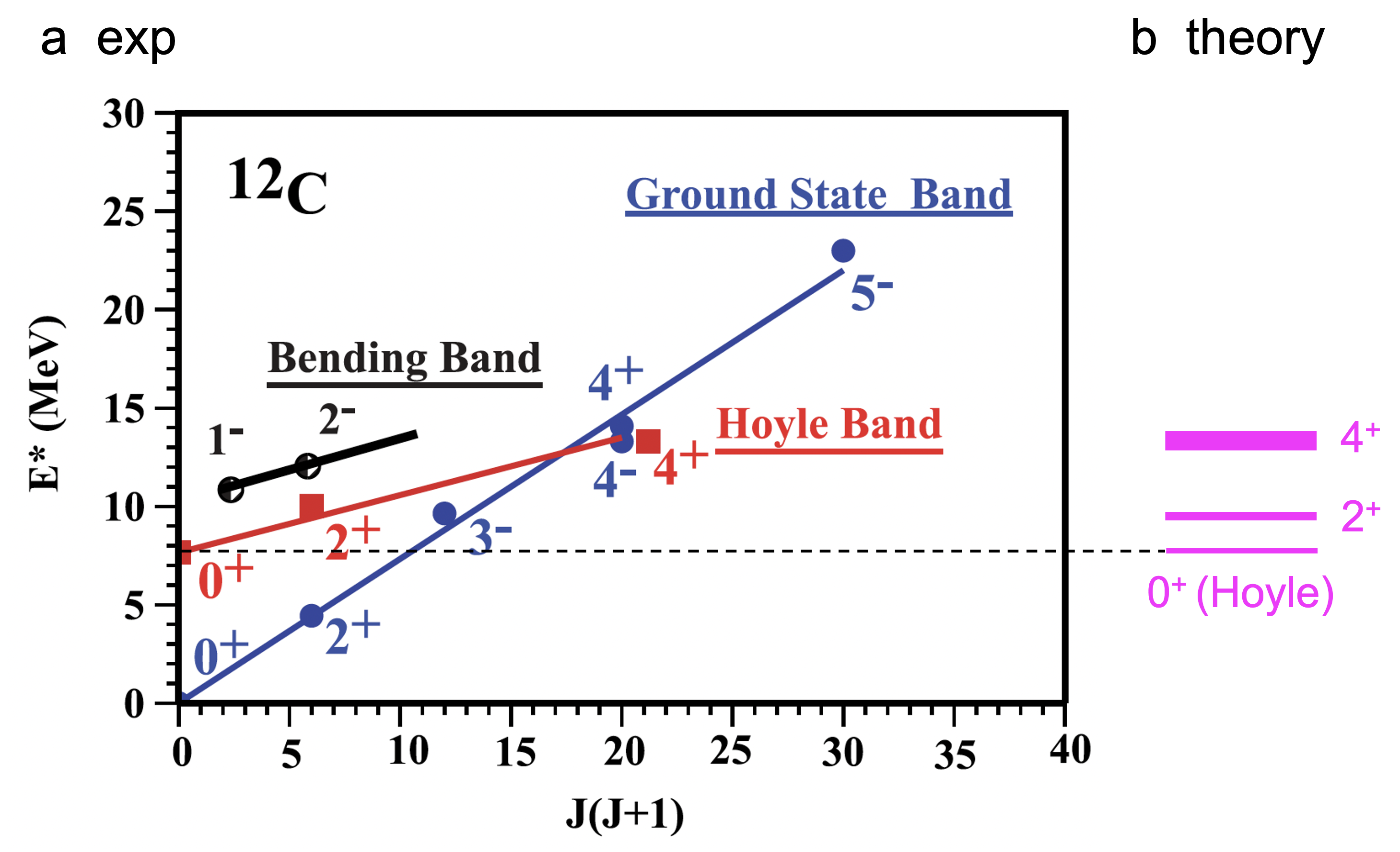}
  \caption{ {\bf Possible rotational band built on Hoyle state, a, experimental and b, theoretical.}
  Panel {\bf a} is taken from \cite{Marin_Lambarri_2014} with permission.  Panel {\bf b} displays the result of the present theoretical estimate with uncertainty depicted by the thickness of the bars due to simple approximation to precise calculation (see text). 
  }
  \label{fig:Hoyle_band}  
\end{figure}  
%%%%%%%%%%%%%%%%%%%%%%%%%%%%%%%%%%%%%%%%%%%%%%%%

The rotational band built on the ground state has been discussed in Subsects.~\ref{first-principles realization} and \ref{general}.  
This band is indicated in Fig.~\ref{fig:12C_levels}.

We now discuss another possible rotational band built on the Hoyle state.  Its experimental observation has been reported in \cite{Ogloblin_2014,Marin_Lambarri_2014}, as displayed in Fig.~\ref{fig:Hoyle_band} partly taken from \cite{Marin_Lambarri_2014}.
Although the establishment of this band may remain open experimentally, we assume it exists in the way proposed.  While this band appears above $\alpha$ threshold, we suppose that we discuss doorway states before decays.
The Hoyle state and the rotational band are then described by the distant-object rotation of a {\it tri}-$\alpha$ intrinsic state.  However, as depicted in Fig.~\ref{fig:12C_8Be}, the {\it tri}-$\alpha$ configuration is not of axial symmetry, and is indeed triaxial like the one shown in Fig.~\ref{fig:cluster_rot}{\bf c}.   
The structure is then not as simple as the one for $^8$Be. 

The intrinsic state, $\phi$, represents a {\it tri}-$\alpha$ state like the one shown in Fig.~\ref{fig:cluster_rot}{\bf c},  putting distortions from the pure (or free) $\alpha$ structure aside.  
We first look into the $K$ quantum number, which is the $z$ component of the angular momentum.
For triaxial configurations like Fig.~\ref{fig:cluster_rot}{\bf c}, three moments of inertia emerge in the classical mechanics, with different values depending on the rotation axis.  
The moment of inertia can be similarly calculated from the nucleon density distribution of the intrinsic state, within quantum mechanics.  We then define the $z$ axis so that the moment of inertia about the $z$ axis becomes the smallest, for the reason stated later.  
In this way, the $K$ quantum number gains a certain physical meaning rather than an arbitrary index, also in the cases being discussed.  This definition is surely consistent with the one for compact-object rotation, where the $z$ axis is usually taken to be along the longest axis of the ellipsoid. 
We here mention that this $K$ value is practically conserved by the same argument as in \cite{otsuka_2025}, provided that in $\phi$, strong mutual localization of nucleons into small volume of a specified cluster occurs and the center of gravity of each cluster is also localized in the intrinsic state, as expected in the usual $\alpha$ clustering. These arguments are partly based on \cite{otsuka_utsuno_2026}.

Following eq.~(14) of \cite{otsuka_2025}, the energy of the state with a good $K$ value, denoted by $\Phi \bigl[\phi, K\bigr]$, is given by
\begin{eqnarray}
& h_{K} \,&=\langle \Phi \bigl[\phi, K\bigr] \, | H \, | \Phi \bigl[\phi, K\bigr] \rangle 
\nonumber \\
 & &\propto %\frac{1}{|{\mathcal N}|^2} \, \frac{1}{{2\pi}}
      \, \int_0^{2\pi} d\gamma \, {\rm cos}(K\,\gamma) \, \langle \phi \, | H \, | \,     
     e^{i\gamma \hat{J}_z} \,  \phi \rangle.
\label{eq:H Kx}
\end{eqnarray}
The relevant part of Hamiltonian is the kinetic energy for the rotation about the $z$ axis of the centers of gravity for three $\alpha$ clusters in a fixed configuration, like the one shown in Fig.~\ref{fig:cluster_rot}{\bf c}.  The angle of this $z$-axis rotation is denoted by $\gamma$.  We then obtain
\begin{equation}
H_{\gamma}  \,=\,-\,\frac{\hbar^2}{2 \, \mathcal{I} } \,  \frac{\partial^2}{\partial \gamma^2} ,   
\label{eq:Hg_gamma1}
\end{equation} 
where $\mathcal{I}$ is the corresponding moment of inertia for this rotation.
% with $\mathcal{I} =m R^2$ with $m$ being the mass of cluster. 
By performing partial integration twice for eq.~(\ref{eq:H Kx}), $h_{K}$ is expressed by the product of the norm part and the $\frac{\hbar^2}{2 \, \mathcal{I} } \,  K^2$ term.  One then obtains the normalized contribution to the energy,
\begin{equation}
E_{K}  \,=\,\frac{\hbar^2}{2 \, \mathcal{I} } \,  K^2 .
\label{eq:Hg_gamma2}
\end{equation} 
This is nothing but the kinetic rotational energy due to the rotation about the $z$ axis.
As this energy increases as $K$, the lowest state is of $K$=0.
It is of interest that $K$=0 is favored both in the distant-object rotation and in the compact-object rotation \cite{otsuka_2025}.  In the latter, it is a consequence of the maximization of the binding energy, but in the former it arises  in order to avoid rotational kinetic energy with $K >$ 0.  In fact, by having $K$=0, all orientations about the $z$ axis are superposed with equal amplitudes, and the $K^2$ expectation value vanishes.  This is a quantum realization of ``stopped'' ($z$ axis) spinning. It is pointed out that the value of $\mathcal{I}$($>$0) does not matter for the realization of $K$=0 lowest band in the present scheme.

The rotational kinetic energy due to $K >$ 0 (see eq.~(\ref{eq:Hg_gamma2})) becomes higher with the present assignment of the $z$ axis than the corresponding energies with the $z$-axes assigned otherwise, which implies that the lowest set of $K$=0 states are better separated in energy from other states with $K >$ 0.

%%%%%%%%%%%%%%%%%%%%%%%%%%%%%%%%%%%%%%%%%%%%%%%%
\begin{figure}[tb]
  \centering
  \includegraphics[width=11cm]{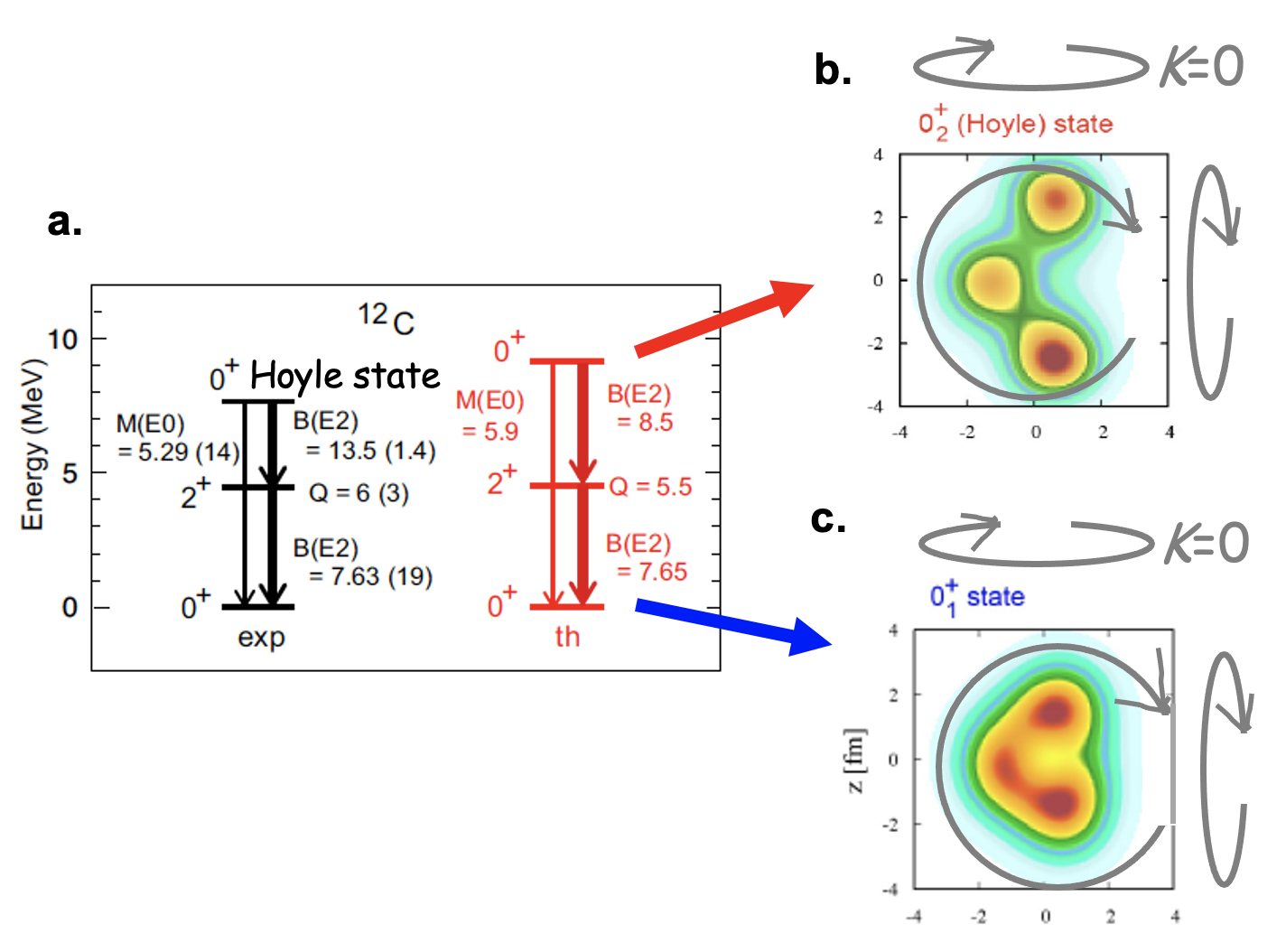}
  \caption{ {\bf Schematic picture of proposed rotational bands built on the ground and Hoyle states of $^{12}$C.}
  Panel {\bf a} is taken from \cite{otsuka_2022} with permission.  Panels {\bf b} and {\bf c} display, respectively, the rotation of  the Hoyle and ground states (see text). 
  }
  \label{fig:Hoyle_rotation}  
\end{figure}  
%%%%%%%%%%%%%%%%%%%%%%%%%%%%%%%%%%%%%%%%%%%%%%%%

Figure~\ref{fig:Hoyle_rotation} shows how rotational modes are created for the ground and Hoyle states.
We start with the Hoyle band.
Panel~{\bf b} indicates that the $K$=0 projected intrinsic state enters the process, as discussed above.  There are two other axes, $x$ and $y$ (see Fig.~\ref{fig:cluster_rot}{\bf c}).  In the classical mechanics, the moment of inertia takes different values for the rotations about these axes.  The motion is then considered to be very complicated.    In quantum mechanics, the $K$=0 projection comes in, which brings about a kind of averaging over the $x$-axis and $y$-axis rotations.  This effect can be precisely incorporated by assessing energy kernels following the arguments in Subsect.~\ref{general} (or eventually \cite{otsuka_2025}), but will be an elaborate work.  Anyway, a simple formula like the one $\propto J(J+1)$ arises in the quantum mechanical treatment, if the cluster localization is strong enough. 

Putting such an elaborate calculation aside for the time being, we look into physics cases in a simpler manner.  We here estimate the moment-of-inertia for the $y$ axis rotation and the $x$ axis rotation in Fig.~\ref{fig:cluster_rot}{\bf c} in a similar approximation used for $^8$Be in Subsect.~\ref{8Be}.  
For the $y$-axis rotation axis, the axis perpendicular to the paper plane of Fig.~\ref{fig:12C_8Be} is taken,  and the positions of the center of gravity of each cluster are read from the figure.   This is precise enough for the present purpose, similarly to the $^8$Be case.  Fig.~\ref{fig:12C_8Be}{\bf d} is used for this purpose, and the axis goes through the center of gravity of the whole nucleus.  The calculated coefficient in front of $J(J+1)$ is about 0.25 MeV.
The other value of the moment of inertia is for the $x$ axis rotation, which corresponds to the rotation about the horizontal axis in the paper plane of Fig.~\ref{fig:12C_8Be}.  This axis goes through the cluster on the $x$ axis, yielding no contribution from this cluster.  The calculated coefficient in front of $J(J+1)$ is about 0.30 MeV.

The precise calculation can be performed with an appropriate Hamiltonian with a proper $| \phi  \rangle$ or $| \Phi \bigl[\phi, K\bigr] \rangle$.  Such calculations are expected to provide results somewhere in between the corresponding values given by the present coefficient equal to 0.25 MeV or 0.30 MeV.  This range generates the excitation energies of the 2$^+$ and 4$^+$ states in the Hoyle rotational band at 1.50-1.80 MeV and 5.0-6.0 MeV, respectively.  The $J(J+1)$ rule will be maintained well, but other mechanisms like accidental mixing with a nearby state may disturb this regularity.  
Far right part of Fig.~\ref{fig:Hoyle_band} indicates that these values already depict quite interesting agreement with (possible) rotational levels observed in experiments \cite{Ogloblin_2014,Marin_Lambarri_2014}.
%It is rather fortunate that the triaxiality of the Hoyle state is adequate so as to enable us to perform such a simple estimate.   

The ground band presents basically the same property (see Panel~{\bf c}).  Because of faster convergence, the energies of the 0$^+_1$ and 2$^+_1$ states are calculated precisely enough by the MCSM calculations, which give a proper treatment of the $y$- and $x$-axis rotations. Namely, the elaborate calculation was feasible for these states, but infeasible for the Hoyle band due to the current computer resources.  

In a comparison to the $^8$Be case in Subsect.~\ref{8Be}, 
the difference of the observed 2$^+$ excitation energies between the $^8$Be band and the Hoyle band is as large as a factor of three.  This agrees, to a good extent, with the present theoretical values: 0.80/0.25=3.2 and 0.80/0.30=2.67.  

The Hoyle state is of great importance in the nucleosynthesis.  The present discussion adds another fundamental significance to it as a showcase of the coexistence of dual rotational modes within the same nucleus: distant-object and compact-object rotations.  This is so rare, and further investigations are of extreme importance.

\subsection{Summary}

In summary, the $\alpha$-clustering is shown, in First-Principles no-core shell model calculations, to emerge, without assuming it {\it a priori}, as an effect of nuclear forces.  This calculation, performed by the Monte Carlo Shell Model, reproduces various observed quantities.  The $\alpha$-clustering is shown to appear even in the well-bound ground state of $^{12}$C, as a non-negligible component.  Although, it is not a major component, this mixture lowers the energy of the ground state.
In contrast to this, the $\alpha$-clustering is virtually solo mechanism for the formation of the $^8$Be ground state.  The $\alpha$-clustering remains the dominant mechanism for the Hoyle state, where about 2/3 probability of the wave function, with proper superposition of basis vectors, is composed of $\alpha$-like clustering.  There are nuclear-matter components also in the Hoyle state, and it is of interest that the cluster and matter components are mixed repulsively, partly due to the orthogonality to the fully correlated ground state.   This is another very interesting point, partly because the ground-state wave function is believed to be almost converged.  

The clustering state is deformed, by definition; the density profile is not spherical in its intrinsic structure.
As the deformed shape is connected to the rotational excitation, the clustering states are a very good testing ground of the theory of rotational excitations.  The recently presented formulation comprising (i) fully quantum-mechanical derivation of ``rotational'' excitation energies, (ii) full inclusion of triaxiality as consequences of rotational symmetry and nuclear forces, (iii)  practical conservation of $K$ quantum number in contrast to traditional belief, can be applied to clustering states.  The same fundamental equation (eq.~(\ref{eq:E})) is used for the same picture that the intrinsic state pointing to all orientations are superposed properly according to the angular momentum of the state.  The difference from the rotational excitation of ellipsoidal matter lies in the relevant parts of the Hamiltonian.  For ellipsoidal matter cases, the nucleon-nucleon interactions produce the major contributions to the so-called rotational excitation energy.  But, in the case of clustering states, each cluster is basically in eigenstates internally.  What matters is the kinetic energy of the center of gravity of each cluster, which can be treated as a point mass.  If the relative configuration of these centers of gravity is fixed (allowing quantum fluctuations), the whole system can be described as an intrinsic state to be projected on a given angular momentum (and parity).  In this case, the origin of the ``rotational'' excitation energy is kinetic, and the same fundamental equation gives us the same energy formula as the one obtained from the quantization of free rigid-body with axial symmetry imposed, even for systems without axial symmetry thanks to the $K$ restoration.  
This $K$ quantum number is defined with the $z$ axis producing the smallest value of moment of inertia among three possible axes.
Rigorously speaking, some of the discussions here may hold only in ideal cases such as perfect $\alpha$ cluster, but such approaches provide us with simple fundamental pictures, from which further understandings of more complicated cases may be developed.    
It is noted that the present formulation is completely different from and independent of pictures based on moving  wave packet, which are quite problematical especially for rotational mode, in view of actual wave functions and also from standpoint of finite range of angles and periodic boundary condition.  We definitely do not need them.

We presented the concept of dual rotational modes: the rotation of nuclear-matter ellipsoid is called compact-object rotation, whereas the rotation of clusters with fixed configuration is called distant-object rotation.   We point out that the former occurs, if the range of interactions between constituents (e.g. nucleons for nucleus) and the size of the whole system (e.g. nuclear radius) are comparable.  On the other hand, the latter occurs, if the clusters are separated enough and the clusters behave as eigenstates internally.  The duality of rotational modes can be interpreted as an outcome of the clustering hierarchy \cite{nakamura_2025}, and remains visible despite sizable mixing of two hierarchies, as seen in $^{12}$C.  It is of great interest to explore, experimentally or theoretically, any physical systems exhibiting these features, including mixed/intermediate situations.  In fact, the compact-object rotation is obviously rather special from the global viewpoint of entire physics, and might be one of the treasures of nuclear physics, unrecognized, at least openly, so far.  The hadron spectroscopy can be a good candidate, where quark-gluon system can be compact but also can be spread like deuteron or neutron nuggets.

The $K$=0 dominance in the lowest states of deformed systems can be applied to more complex systems.  Although our arguments have been limited to positive-parity states, negative-parity states with clustering structure are very likely of $K$=0, which may be consistent with algebraic models for $^{12}$C \cite{bijker_2002}.   This $K$=0 dominance seems to hold also for the lowest states with three-dimensional configurations/shapes like tetrahedron or $\alpha$ quartet, for instance, \cite{bijker_2014}.

We further add two prospect subsections.

\subsection{Prospects 1:  Atomic molecules} 

It is evident that the present formulation extended to distant-object rotation can be applied to atomic molecules \cite{atkins_book}.
%The use of classical values of moment of inertia appears to be reasonable, which can fit quantum mechanical treatment if each clusters are individually eigenstates.  Quantum fluctuations do not matter, if sufficient localizations take place.
The discussions of $^8$Be may be applied to the linearly configured structures of the O$_2$ and CO$_2$ molecules, for example.  In the O$_2$ case, by putting O atoms at the red circles in Fig.~\ref{fig:cluster_rot}{\bf b}, the arguments for the $^8$Be nucleus can generally be extended to the O$_2$ molecule.

The structure of the Hoyle state shows certain similarity to the structure of the H$_2$O molecule; 
 in Fig.~\ref{fig:cluster_rot}{\bf c}, H atoms can be put at the red circles at the top and the bottom, and O atom at the red circle in the middle.  
After the $K$=0 projection, the rotational mode can be described with one rotational band on top of the ground state with the moment of inertia with the value between two classical ones obtained for two axes.   
It is really interesting how we can apply some of the arguments here to molecular structures, while certain approximations will be needed if norm and energy kernels are actually calculated.

\subsection{Prospects 2: Fission} 

The dual rotational modes may have another completely different relevance.  That is the nuclear fission \cite{fission_book}.
Before fission, the nucleus may be deformed, and shows a compact-object rotation like many other heavy deformed nuclei.     
In the case of fast neutron capture, for instance, the nucleus gains a certain value of angular momentum.
This angular momentum is conserved.  
After passing around scission point during the fission process, two fragments are formed, and their mutual rotation likely occurs.   This mutual rotation should belong to distant-object rotation.
For this distant-object rotation, excitation energies must be very low compared to compact-object rotation at the same angular momentum, and states may be quasi-degenerate, which is a favorable situation for a linear motion.  
If the nucleus gains a high value of the angular momentum at an initial stage of fission, this angular momentum is maintained.
However, after scission point, the two fragments may start certain distant-object rotations, taking certain amount of angular momentum.  As this mode has very low excitation energies  as compared to compact-object rotation, the nucleus may decay from its finite-angular-momentum state populated by fast neutron capture to a low-excitation-energy states of similar angular momenta.
This means that a sizable energy may be released to neutron emission, with larger phase space.  This mechanism may facilitate fission processes with neutron emission(s).   

On the other hand, individual fission fragments likely generate own rotational modes, which are compact-object rotations.  The angular momenta of distant-object rotation is then coupled with or transferred to those of such compact-object rotations.  The interplay between the distant-object rotation and the compact-object rotation may thus occur in certain types of fission.  As recently explored in \cite{wilson_2021}, the time evolution of the angular-momentum distribution contains exciting open questions, and in such studies, the interplay between the compact-objet and the distant-object rotational modes can be an interesting aspect.   It is mentioned that simultaneous emergence of compact-object and distant-object modes may result in spontaneous fission, as another interesting subject.
\\
 
%%%%%%%   S E C T I O N     3    %%%%%%%%%%%%%%%%%%%%%%%%%%%%%%%%%%%%%%

\section{A review on cluster model approaches \label{sec:volya}}  

\subsection{Introduction}
\begin{figure}[h] % Placement options: here, top, bottom, page
    \centering
    \includegraphics[width=1.0\textwidth]{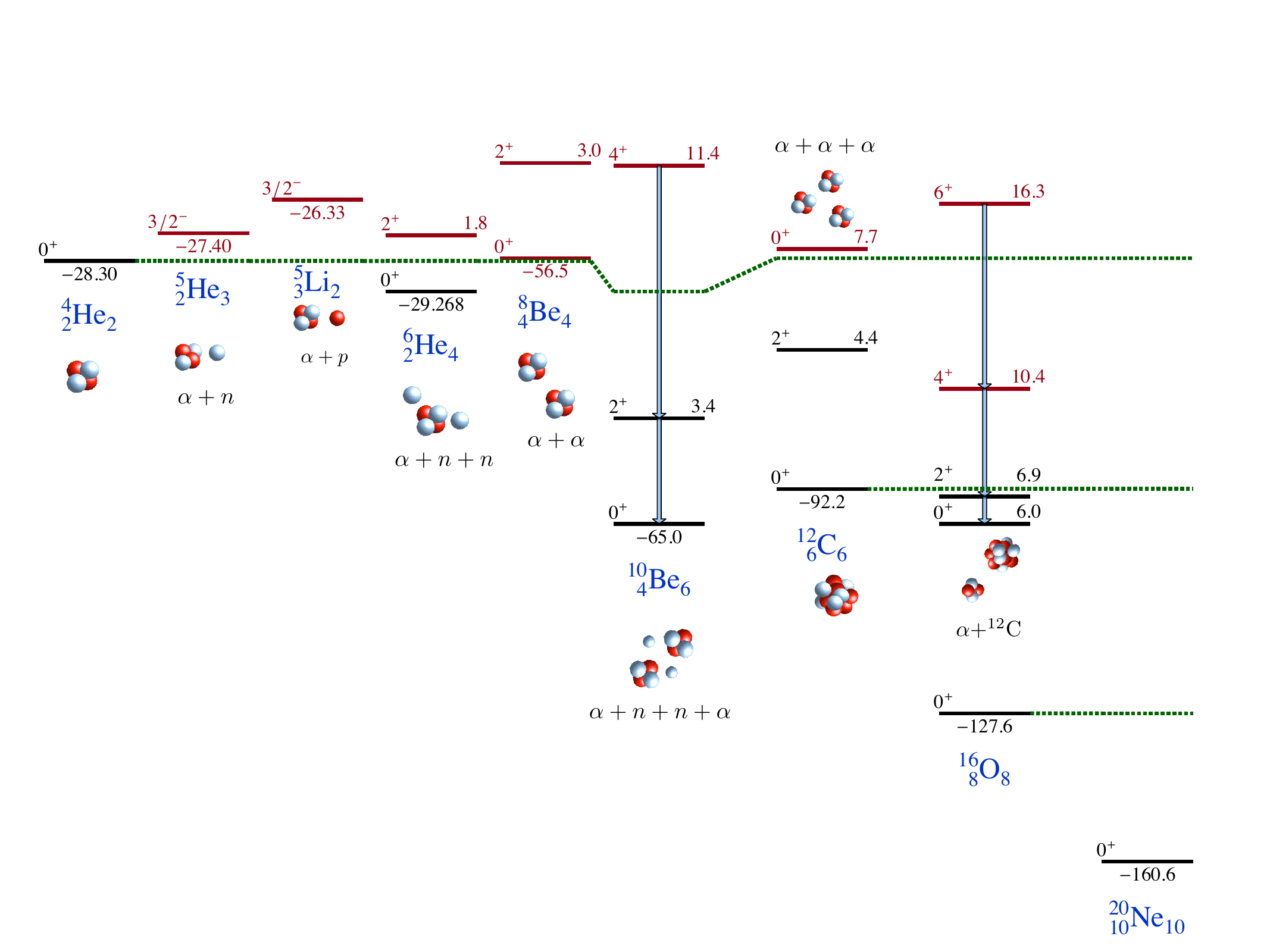} % Replace with your image file name
    \caption{Ikeda-style diagram illustrating clustering phenomena in light nuclei. The spectra of several light nuclei are shown along with illustrations of their potential clustering structures. Cluster decay thresholds are indicated with dashed lines, and decaying states are highlighted in red. For each state, the spin and parity are shown on the left, and the excitation energy is shown on the right. The total binding energy is indicated for the ground state. All energies are in units of MeV.}

    \label{fig:fig1} % Label to reference the figure
\end{figure}

One of the enduring challenges of nuclear many-body physics is to elucidate how structures---such as clusters, phonon excitations, and shape changes---emerge from fundamental interactions among nucleons. 
As was discussed earlier, while considerable progress has been made by describing nuclei in terms of individual protons and neutrons, the phenomenon of nuclear clustering, emergence of cluster degrees of freedom, most notably $\alpha$-like correlations, and their role in nuclear structure and dynamics remains not fully understood. 
Indeed, $\alpha$-particle formation has been suggested since the early days of nuclear science as one of the guiding ideas behind observed decay modes and the unusual stability of certain nuclear configurations~\cite{Gamow:1930,Hafstad1938,Blatt:Book}. 
Many multi-cluster, molecular-like states in light nuclei appear near their respective cluster-decay thresholds—a feature succinctly illustrated in the well-known Ikeda diagram~\cite{ikeda_1968}—supporting the idea that alpha clusters serve as important building blocks for such states.
A version of this diagram is shown in Fig.~\ref{fig:fig1}, highlighting the thresholds and pictorially illustrating cluster structures. Although threshold effects are known to restructure states, often helping to align structures to favor decay into corresponding thresholds \cite{Volya2022Superradiance}, as shown in examples in Sec.~2 and further illustrated in this section and Sec.~3, clustering is not just a threshold phenomenon.
Explicit $\alpha$-particle clustering is closely related to pairing and quartet correlations in nuclei~\cite{Sandulescu2024} and even in randomly interacting quantum many-body systems clusterization seems to naturally emerge~\cite{white2023Structured}, the mechanisms behind this remain to be fully explained. 

Within this paradigm, the connection between the microscopic description of nuclear structure—based on nucleonic degrees of freedom and methods such as the nuclear shell model and, more broadly, configuration interaction—and clustering correlations, preformation, dynamics, and decay remains a central question. 
Historically, the two descriptions—one based on nucleonic degrees of freedom and the other on cluster degrees of freedom—have largely evolved in parallel. Nevertheless, the overlap between these parallel lines of thought has been growing. For example, antisymmetrized molecular dynamics has been extensively applied to the study of molecular-type states in clustered nuclei~\cite{Kanada-Enyo2012Antisymmetrized}, incorporating both clusters and valence nucleons in a molecular-like structure, where valence nucleons mediate bonds between clusters. 

Present-day advanced models employ fully microscopic descriptions based on nucleon-nucleon interactions and treat cluster dynamics microscopically using methods such as the Resonating Group Method (RGM)~\cite{Wheeler1937mathematical,Wildermuth:1979}, as well as modern configuration-interaction approaches like the Monte Carlo Shell Model discussed in Sec.~2.
Furthermore, shell model calculations inspired by certain symmetry-based approaches, such as symplectic or SU(3) symmetries, have increasingly been able to capture key aspects of clustering~\cite{Draayer:1975,Smirnov1977Cluster,Chung:1978,Anantaraman:1975}. Such symmetry-based methods are among the most promising present-day techniques~\cite{Dreyfuss2020, Dytrych2020}.

Microscopic calculations, such as those using Green’s Function Monte Carlo, have demonstrated the emergence of clustering in $^{8}$Be directly from nucleon-nucleon (NN) interactions~\cite{wiringa_2000}. This fundamental result has reignited modern, \textit{ab initio}, theoretical interest in clustering. A variety of approaches—including mean-field models~\cite{Ebran:2012}, lattice simulations~\cite{Epelbaum2011initio,Elhatisari:2015}, large-scale {\it ab initio} no-core shell model calculations~\cite{dreyfuss_2017}, and Bose-Einstein condensate wave studies functions~\cite{tohsaki_2001,Schuck:2016,tohsaki_2017}—have been employed to better understand how clusters form and which features of nuclear interactions and many-body dynamics favor the development of correlated substructures. 
Fully microscopic calculations that integrate nuclear structure and reactions from a configuration interaction perspective have also made significant progress~\cite{Navratil:2012, Navratil2016, Zhang:2020}.
A comprehensive list of references and historical discussions on the subject can be found in review \cite{freer_rmp}.

The study of nuclear clustering is not only about emergent phenomena and the appearance of clustering degrees of freedom; it has become increasingly clear—supported by growing experimental evidence—that clustering plays a fundamental role in connecting nuclear structure and reactions. 
Clusters are predominantly observed near their corresponding decay thresholds, a pattern first noted by Ikeda~\cite{ikeda_1968} and subsequently confirmed by numerous experiments. 
The physics of open quantum systems have introduced a new perspective in clustering studies~\cite{Johnson2020}. As mentioned earlier, the threshold for a given decay channel significantly impacts the restructuring of the many-body wave function along this channel. This is evident in the observation of cluster states near corresponding thresholds, as well as in the convenient placement of broad states near these thresholds. The effect can manifest in different ways—either enhancing clustering or, conversely, blocking certain decay channels, as seen in proton-decaying $^{11}$B, which inhibits the $\alpha$-decay pathway~\cite{Okolowicz2020, Volya2020a}.
The near-threshold character of clustering makes many of these questions of paramount importance in astrophysics. The astrophysical significance of clustering is epitomized by the Hoyle state in $^{12}$C~\cite{dunbar_1953,hoyle_1954}, whose triple-$\alpha$ nature is crucial for the nucleosynthetic pathway to heavier elements.

Furthermore, recent studies indicate that clustering can persist at high excitation energies and large angular momenta, sometimes in conjunction with rotational phenomena, all while embedded in the continuum of open decay channels~\cite{Avila2014,Kuchera2013Clustering,Okolowicz2003Dynamics,Auerbach2011,Kravvaris2017Constructing}. This behavior may be driven by superradiance effects, leading to a separation of states into broad, strongly coupled, and structurally organized configurations that facilitate cluster decay, as well as narrow, trapped states~\cite{Auerbach2011}. Recent experimental results leveraging isospin symmetry have highlighted the distinct impact of the continuum on the realignment of wave functions toward clustering~\cite{Volya2022Superradiance}.

Advances in experimental techniques have significantly expanded the scope of clustering studies. The development of rare isotope beams has enabled novel studies of resonant reactions induced by radioactive nuclei. In particular, the Thick Target Inverse Kinematics (TTIK) approach~\cite{Koshchiy2020} has proven to be a powerful tool for rare-beam experiments, facilitating the search for $\alpha$-cluster states and systematic analysis of clustering strength distribution~\cite{Barbui2022, Goldberg2022, Nauruzbayev2017, Nauruzbayev2017, Nurmukhanbetova2019, Nurmukhanbetova2024, Nurmukhanbetova2024b, Upadhyayula2020, Volya2022Superradiance, Volya2022a}.

Against this backdrop, this section reviews how microscopic principles give rise to cluster substructures in light nuclei, with particular emphasis on the essential rosectionle of continuum coupling and reaction theory in shaping spectroscopic properties. In parallel with the discussion of the MCSM in Sec.~2, the framework presented here represents an alternative strategy within the broader configuration-interaction approach. We approach clustering from the perspective of nucleonic degrees of freedom and nucleon-nucleon interactions, tracing its development from the traditional shell model into configuration interaction techniques that incorporate cluster configurations. In fact, the approach discussed here and that in Sec.~2 are fundamentally equivalent in their microscopic foundations, and in some examples employ identical Hamiltonians (e.g., JISP16), differing primarily in the strategy of configuration selection. They also offer complementary physical perspectives: a rotating intrinsic (body-fixed) frame in which clustering appears as spatially localized structures, versus the laboratory-frame picture adopted here, where clustering is characterized through overlaps and spectroscopic factors. The direction outlined here serves as a bridge between various other theoretical limits, including algebraic and symmetry-based methods, the resonating group method, the traditional shell model, the no-core shell model, and its continuum extensions.

In the following subsections of this section, we outline an alternative technique to that discussed in Sec.~2. Rather than identifying clusters from large-scale shell model solutions, we proceed in the opposite direction: we construct configurations starting from clusters and then mix them via configuration-interaction techniques with, albeit much smaller, traditional shell-model configurations of Slater determinants. In this way, we build what we call Cluster-Nucleon CI.
We first present the details of this approach and then show applications, including the study of $^{12}$C, which reproduces the results discussed earlier while simultaneously bridging to the discussion in the final section, highlighting the competition between traditional shell-model configurations and cluster structures. We examine clustering in deeply bound and ground states, and briefly discuss the impact of the Pauli principle, as well as how four-nucleon correlations connect to physical $\alpha$ particles at larger distances. We conclude with the discussion of reactions, cluster resonances, and near-threshold clustering, which are addressed at the end of this section.

\subsection{Cluster configurations}
\subsubsection{Center of mass and boosting}
As in the standard shell model and no-core shell model approaches, we use the single-particle harmonic oscillator (HO) basis as the foundation of our configuration interaction (CI) framework:
\begin{equation}
\langle {\bf r} | n \ell m \rangle = \phi_{n\ell m}(r,\theta,\phi) = \frac{\phi_{n\ell}(r)}{r} Y_{\ell m}(\theta,\phi),
\label{eq:howf}
\end{equation}
where \( n \) is the radial quantum number, \( \ell \) the orbital angular momentum, and \( m \) its projection. The HO potential is defined by frequency \( \omega \), we use \( m \) for nucleon mass, the oscillator length \( b = \sqrt{\hbar / m \omega} \). 
The single-particle energy eigenvalues are
\begin{equation}
E = \hbar\omega(N + 3/2), \quad N = 2n + \ell.
\end{equation}
For explicit expressions and discussion of properties of HO wave functions, see e.g.~\cite{de-shalit:2004}.

Many-body wave functions are expressed as linear combinations of Slater determinants:
\begin{equation}
|\Psi \rangle = \Psi^\dagger |0\rangle = \sum_{\{1,\dots,A\}} \langle 1,\dots,A | \Psi \rangle\, a^\dagger_1 \dots a^\dagger_A |0\rangle,
\label{eq:sup}
\end{equation}
where \( a^\dagger_i \) creates a nucleon in a single-particle state labeled by HO quantum numbers and spin. The polymorphism between operators and states allows us to treat \( \Psi^\dagger \) as a many-body creation operator acting on the vacuum. Pauli antisymmetry is ensured by fermionic commutation relations and operator ordering. Products of antisymmetrized states are written as
\begin{equation}
|\mathcal{A}\{\Psi_\alpha \Psi_\beta\}\rangle = \Psi^\dagger_\alpha \Psi^\dagger_\beta |0\rangle.
\label{eq:second}
\end{equation}

Unlike the traditional shell model, which relies on a predetermined basis, CI methods allow for flexible, on-demand construction of configurations. This generality enables efficient incorporation of physically relevant degrees of freedom, including symmetry-adapted or cluster-like configurations, the construction of those we discuss next. 

The center-of-mass (CM) coordinate plays a crucial role in studies of clustering, where identifying and controlling the motion of cluster centers is essential. In traditional shell model calculations, the treatment of the CM has long been recognized as an important issue, particularly in ensuring translational invariance and in interpreting reaction channels. The use of the harmonic oscillator (HO) basis provides a powerful framework for addressing this problem, as it enables exact factorization of the CM degree of freedom due to the rich symmetry structure of many-body HO wave functions.

Following the no-core shell model approach~\cite{Barrett2013}, we exploit the symmetry of the many-body HO Hamiltonian, where energy eigenstates are degenerate with respect to the total number of oscillator quanta \( N \). States with fixed \( N \) form representations of symmetry groups, including SU(3) and O($A$), the latter of which enables exact CM separation in a truncated configuration space defined by a maximum excitation \( N_{\text{max}} \). For states with a given number of quanta
\begin{equation}
N = N_{\text{CM}} + N'.
\end{equation}
where $N_{\text{CM}}$ is the number of quanta in the CM excitation and $N'$ is the number of quanta in the intrinsic wave function. 
In traditional applications, states of interest are those with \( N_{\text{CM}} = 0 \), corresponding to a CM in the ground-state oscillator mode:
\begin{equation}
\Psi = \phi_{000}(\mathbf{R}) \Psi',
\label{eq:cm}
\end{equation}
where \( \mathbf{R} \) is the CM coordinate and \( \Psi' \) is the intrinsic wave function~\cite{Palumbo:1968,Gloeckner:1974}.

For clustering studies, however, we require more general configurations in which the CM component can take any form. To achieve this, we construct CM-boosted states where the CM motion is expanded in terms of HO eigenfunctions with arbitrary quantum numbers.

\begin{equation}
\Psi_{n\ell m} = \phi_{n\ell m}(\mathbf{R}) \Psi',
\label{eq:boost}
\end{equation}
where \( N_{\text{CM}} = 2n + \ell \) defines the CM excitation. 
Although such CM-excited states naturally appear in full shell model diagonalizations—where they are typically regarded as spurious—our approach constructs them directly by acting on the CM coordinate. This CM boost procedure is significantly simpler and requires no additional diagonalization beyond the initial shell model solution \cite{Kravvaris2017Constructing}.

To manipulate the CM motion, we use the standard CM creation and annihilation operators:
\begin{align}
\mathcal{B}_\mu^\dagger &= \frac{1}{\sqrt{2Am\omega\hbar}} (Am\omega R_\mu - i P_\mu), \\
\mathcal{B}_\mu &= \frac{1}{\sqrt{2Am\omega\hbar}} (Am\omega R_\mu + i P_\mu),
\end{align}
which relate to the isoscalar E1 operator:
\begin{equation}
D_\mu = \sqrt{\frac{4\pi}{3}} \sqrt{\frac{\hbar}{2Am\omega}} (\mathcal{B}_\mu^\dagger + \mathcal{B}_\mu).
\end{equation}
Here, \( R_\mu \) represents the spherical component \( \mu \) of the center-of-mass (CM) radius vector, and \( P_\mu \) is the corresponding component of the CM momentum.

The operator \( \mathcal{B}_m^\dagger \) increases \( N_{\text{CM}} \) by one and transforms as a vector. 
States with CM in a given state \( (n,\ell,m) \) can be constructed recursively. Node number is increased via the scalar product:
%%%%%%%   FIRST  ERROR -> solved Taka
\begin{align}   
{\bf \mathcal{B}}^\dagger \cdot {\bf \mathcal{B}}^\dagger 
&= \mathcal{B}_{+1}^\dagger \mathcal{B}_{-1}^\dagger + \mathcal{B}_{-1}^\dagger \mathcal{B}_{+1}^\dagger - \mathcal{B}_{0}^\dagger \mathcal{B}_{0}^\dagger, \\
{\bf \mathcal{B}}^\dagger \cdot {\bf \mathcal{B}}^\dagger \Psi_{n\ell m} 
 &= \frac{1}{4} \sqrt{(2n+2)(2n+2\ell+3)}\, \Psi_{n+1,\ell,m}.
\end{align}  
Angular momentum \( \ell \) is increased by acting on aligned states:
\begin{equation}
\mathcal{B}_{+1}^\dagger \Psi_{n\ell\ell} = \sqrt{\frac{(\ell+1)(2n+2\ell+3)}{4(2\ell+3)}} \Psi_{n,\ell+1,\ell+1}.
\end{equation}
The CM angular momentum operator is:
\begin{equation}
\mathcal{L}_{\pm} = \pm 4\sqrt{2} \left( \mathcal{B}_0^\dagger \mathcal{B}_{\pm1} - \mathcal{B}_{\pm1}^\dagger \mathcal{B}_0 \right),
\end{equation}
with action:
\begin{equation}
\mathcal{L}_{\pm} \Psi_{n\ell m} = \sqrt{(\ell \mp m)(\ell \pm m + 1)} \Psi_{n\ell, m \pm 1}.
\end{equation}
The boosted basis \eqref{eq:boost} generalizes the non-spurious form \eqref{eq:cm} and preserves translational invariance, as operations on the CM coordinate do not affect the intrinsic structure of \( \Psi' \). A related discussion can be found in Refs.~\cite{Smirnov1977Cluster,moshinksy:1996,Kravvaris2018Clustering}.

Next we comment on the structure of CM-boosted states and their connection to SU(3)-based models widely used in the literature~\cite{Volya2015,Smirnov1977Cluster,Ichimura1973Alphaparticle,Draayer:1975}. For simplicity, we focus on CM-boosted wave functions of $\alpha$ particles.
The ground state of an $\alpha$ particle is dominated by the fully symmetric $s^4$ configuration, which accounts for more than 90\% of the wave function across a wide range of oscillator frequencies $\hbar\omega$. Under the approximation that the $\alpha$ particle has this $s^4$ structure (equivalent to an $N_{\rm max} = 0$ truncation), we recover the algebraic limit~\cite{Draayer:1975,Smirnov1977Cluster,Ichimura1973Alphaparticle}.

In this limit, where the internal excitation is zero ($N' = 0$), the CM-boosted wave function \(\Psi_{n\ell m}\) carries all oscillator quanta \( N = 2n + \ell \) in the CM motion. The spatial part is fully symmetric under particle exchange, and the SU(3) symmetry of the system is restricted to irreducible representations with \((\lambda, \mu) = (N, 0)\). The wave function can be expanded as:
\begin{equation}
\Psi_{n\ell m} = \sum_\eta X^\eta_N\, \Phi^\eta_{(N,0):\ell m},
\label{eq:expand}
\end{equation}
where each term corresponds to a partition \(\eta = \{\alpha_i, N_i\}\) satisfying:
\begin{equation}
A = \sum_i \alpha_i, \qquad N = \sum_i \alpha_i N_i.
\end{equation}
Here, $A$ is the total number of nucleons, \(\alpha_i\) is the number of particles in oscillator shell \(i\), and \(N_i\) the number of quanta associated with that shell.

The expansion coefficients \(X^\eta_N\), known as cluster coefficients~\cite{Volya2015,Smirnov1977Cluster,Ichimura1973Alphaparticle}, are given analytically by:
\begin{equation}
X^\eta_N = \sqrt{ \frac{1}{4^N} \cdot \frac{N!}{\prod_i (N_i!)^{\alpha_i}} \cdot \frac{A!}{\prod_i \alpha_i!} }.
\label{eq:CC1}
\end{equation}
The states \(\Phi^\eta_{(N,0):\ell m}\) are SU(3)-symmetry states with SU(3) quantum numbers \((\lambda, \mu) = (N, 0)\); such states are unique for each configuration.

This algebraic framework forms the basis of many SU(3)-based shell model studies of $\alpha$ clustering in nuclei~\cite{Draayer:1975,Smirnov1977Cluster,Ichimura1973Alphaparticle}. The construction of CM-boosted wave functions via direct CM operations on intrinsic states, as employed in this work, reduces to the SU(3) expansion in the algebraic limit, thereby establishing a direct connection to these earlier models. However, since any intrinsic configuration \(\Psi'\) can serve as a starting point, the CM boosting approach is considerably more general applicable to clusters of any size and with realistic internal structure.

One notable limitation of the algebraic framework is the requirement to use the same HO frequency for all nuclei involved—such as the parent, daughter, and $\alpha$ particle in $\alpha$ decay—which can be problematic, as the optimal frequencies for these systems typically differ. While the CM boosting approach retains this frequency-matching constraint, the ability to represent the $\alpha$ particle in a more realistic basis, going beyond the $s^4$ configuration, helps to alleviate this issue.

\begin{figure}[htbp]
    \centering
    \includegraphics[width=0.8\textwidth]{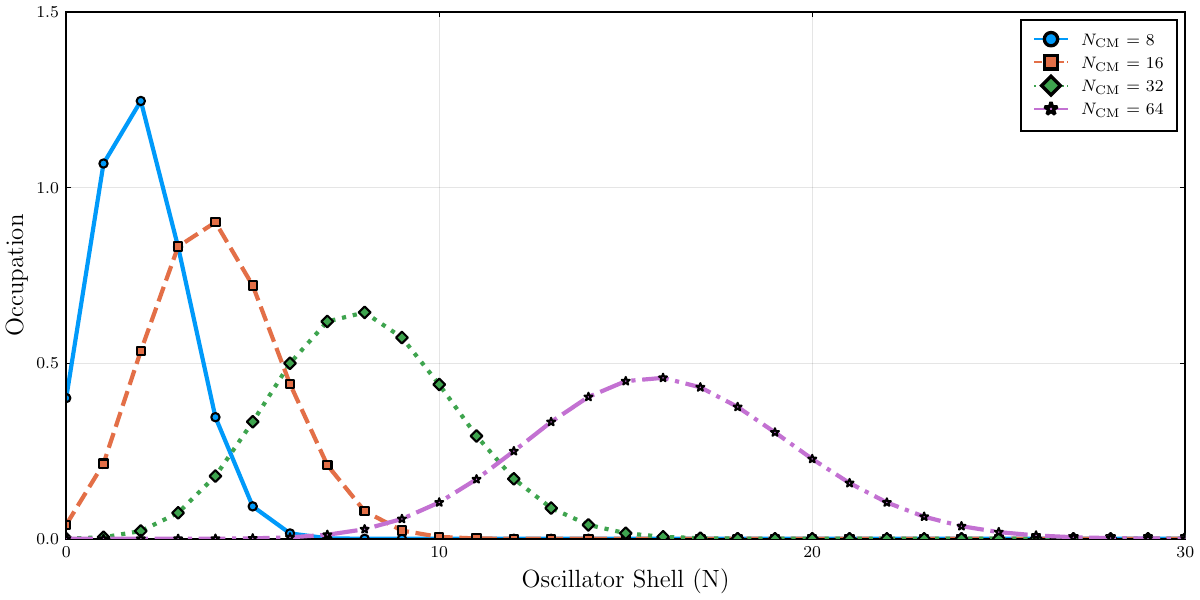}
    \caption{Nucleonic occupation numbers across oscillator shells in a boosted wave function of an $\alpha$ particle, shown for different values of the center-of-mass (CM) excitation quanta. See ref. \cite{Kravvaris2019}}
    \label{fig:alpha_occupancy_plot}
\end{figure}

In Fig.~\ref{fig:alpha_occupancy_plot}, to highlight the structure of the boosted wave function we illustrate the distribution of nucleons across oscillator shells in a CM-boosted wave function of an $s^4$ alpha particle, shown for different numbers of center-of-mass (CM) quanta. The structural content of a boosted state is presented in Tab.~\ref{tab:ncsm-config} two illustrative examples of an alpha-particle state boosted by 8 oscillator quanta, denoted as $\Psi_{n=4,\ell=0}^{(\alpha)}$. The left column corresponds to the alpha state described as $s^4$ ($N_{\rm max}=0$) and agrees with the result in \eqref{eq:CC1}. The right column shows the case with $N_{\rm max}=4$ wave function for the alpha particle calculated using JISP16  nucleon-nucleon interaction Hamiltonian \cite{Maris:2013} with oscillator basis frequency of $\hbar\omega =20$ MeV.

\begin{table}[htbp]
\centering
\caption{Select configuration content of NCSM wave functions for $^4$He with $\hbar \Omega = 20$ MeV boosted by 8 quanta ($L = 0$). This would correspond to a minimal number of quanta creating an alpha particle configuration within the $sd$ valence space. }
\begin{tabular}{lcc}
\toprule
\textbf{Configuration} & \boldmath$N_{\text{max}} = 0$ & \boldmath$N_{\text{max}} = 4$ \\
\midrule
$(sd)^4$                     & 0.038 & 0.035 \\
$(p)(sd)^2(pf)^1$            & 0.308 & 0.282 \\
$(p)^2(pf)^2$                & 0.103 & 0.094 \\
$(p)^2(sd)^1(sdg)^1$         & 0.154 & 0.141 \\
$(p)(sd)(sdg)(pfh)^1$        & - & 0.005 \\
$(p)(sd)(pf)^1(sdg)^1$       & - & 0.009 \\
\bottomrule
\end{tabular}
\label{tab:ncsm-config}
\end{table}

\subsubsection{Cluster channels}
While the formalism is general and allows construction of configurations with multiple clusters it is instructive to 
concentrate on a two-body problem by considering clusters with $A_1$ and $A_2$ nucleons, which combine to form a system of $A = A_1 + A_2$. A \emph{reaction channel} is defined as the asymptotic state of this two-cluster system, composed of wave functions $\Psi^{(1)}$ and $\Psi^{(2)}$ (obtained, for instance, from shell model or NCSM) and their relative motion specified by partial wave $\ell$. In line with the standard Resonating Group Method (RGM) or Generator Coordinate Method~~\cite{Wheeler1937mathematical,Wildermuth:1979}, we construct these channels from center-of-mass (CM) boosted fragments, Eq.~\eqref{eq:boost}. We construct the basis states to expand the channel wave functions as follows:
\begin{equation}
\Phi_{n\ell m}
= \mathcal{A}\,\Bigl\{
\phi_{000}(\mathbf{R})\,\phi_{n\ell m}(\boldsymbol{\rho})\,\Psi^{\prime(1)}\,\Psi^{\prime(2)}
\Bigr\},
\label{eq:wfcouple}
\end{equation}
where \(\mathcal{A}\) ensures proper antisymmetrization among all nucleons. In this construction, we recouple the center-of-mass (CM) coordinates of the two clusters into an overall CM coordinate vector \(\mathbf{R}\) and a relative coordinate \(\boldsymbol{\rho}\):
\begin{equation}
\mathbf{R}=\frac{A_1 \mathbf{R}_1 + A_2 \mathbf{R}_2}{A_1 + A_2},
\quad
\boldsymbol{\rho}=\mathbf{R}_1 - \mathbf{R}_2.
\label{eq::cm}
\end{equation}
To incorporate these channel basis states within the configuration interaction (CI) approach, we focus exclusively on states whose overall CM motion is represented by the ground-state harmonic oscillator (HO) wave function $\phi_{000}(\mathbf{R})$, thus ensuring the states are non-spurious. The relative motion of the two clusters is described by an HO wave function characterized by the chosen quantum numbers \(n, \ell, m\). 

\begin{figure}[h] % Placement options: here, top, bottom, page
    \centering
    \includegraphics[width=0.5\textwidth]{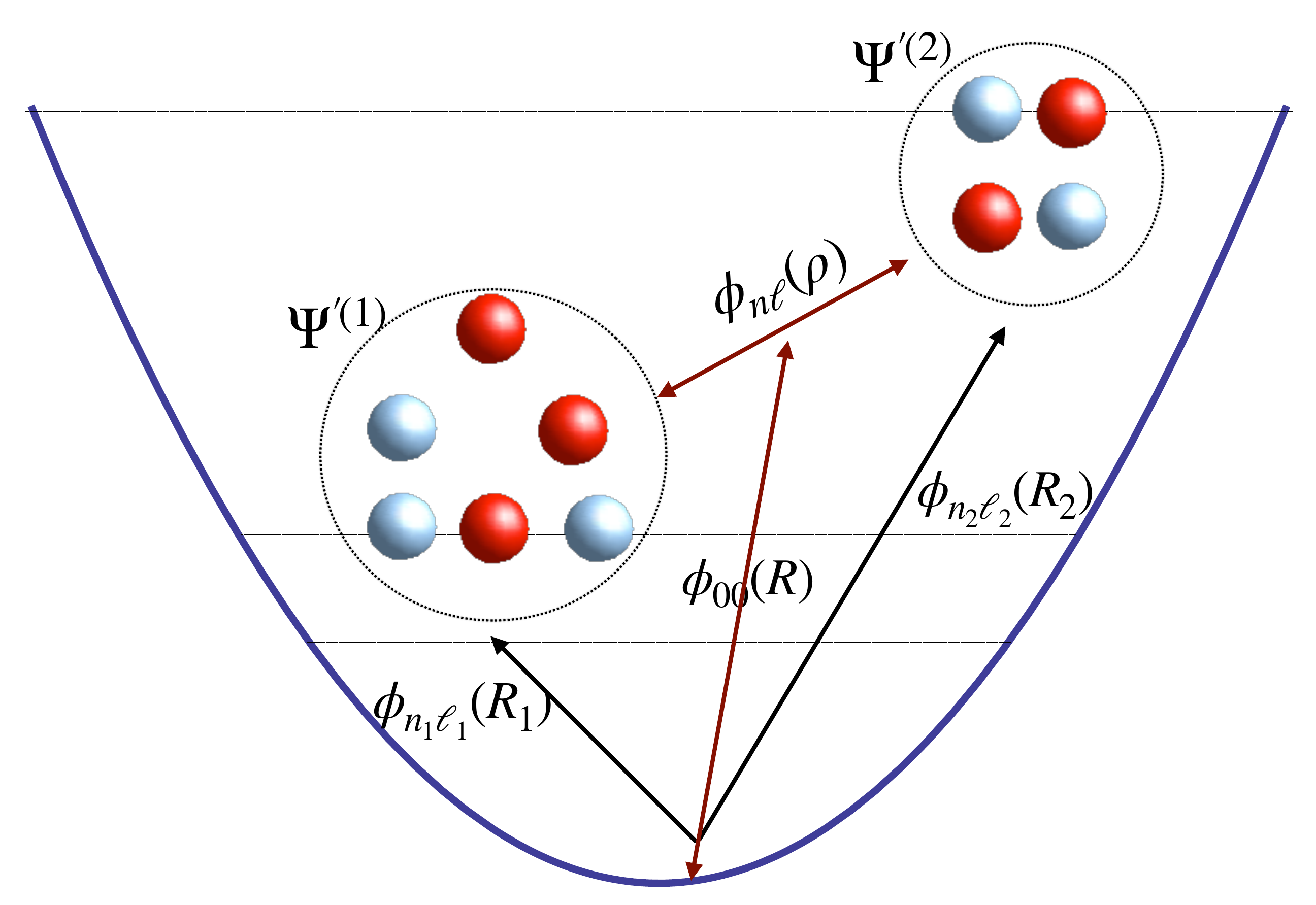} % Replace with your image file name
    \caption{Illustration showing the construction of the channel basis wave function $\Phi_{n\ell m}$ in \eqref{eq:wfcouple}}

    \label{fig:fig3} % Label to reference the figure
\end{figure}

Although expressed in cluster form, Eq.~\eqref{eq:wfcouple} remains a full many-body state that can be represented through Slater determinants (see Eq.~\eqref{eq:sup}). The recoupling of the individual HO wave functions of fragments \(A_1\) and \(A_2\) into combined CM and relative HO wave functions is accomplished using Talmi--Moshinsky--Smirnov coefficients~\cite{Trlifaj:1972}:
\begin{equation}
\Phi^\dagger_{n\ell}
= \sum_{n_1 \ell_1,\,n_2 \ell_2}\!
\mathcal{M}_{n_1 \ell_1\,n_2\ell_2}^{n \ell\,00;\ell}
\Bigl[\Psi^\dagger_{n_1\ell_1} \times \Psi^\dagger_{n_2\ell_2}\Bigr]_{\ell}.
\label{eq:Recoupling}
\end{equation}
Here we omit the magnetic quantum number \(m\) and interpret \(\ell\) as a combined set of asymptotic quantum numbers. The recouping procedure of constructing a channel basis wave function is illustrated in Fig. \ref{fig:fig3}

The configurations described in Eq.~\eqref{eq:wfcouple} are obtained directly from existing solutions for the systems $A_1$ and $A_2$ and do not require additional diagonalizations. However, these states significantly enlarge the configuration interaction space, making this a powerful cluster-nucleon configuration interaction approach. The variety of available configurations can be further increased by considering channels constructed from excited states of the nuclei. Extensions to multi-cluster systems are also feasible, although analogs of analytic Talmi--Moshinsky coefficients are generally unavailable due to multiple possible recouplings. In such cases, direct construction of channel states can be accomplished through alternative methods, including algebraic or numerical diagonalization of the relevant Casimir operators associated with the harmonic oscillator algebra.

The full cluster channel is described as a linear combinations of the basis states we defined in \eqref{eq:wfcouple}
\begin{equation}
|\Xi_{\ell}\rangle=  \sum_n 
\chi_n
|\Phi_{n \ell}\rangle.
\label{eq:rgmx}
\end{equation}
This effectively expresses the physical relative motion of two fragments (cluster plus core) in harmonic oscillator basis. The amplitudes \(\chi_n\) are obtained variationally within the Resonating Group Method (RGM) ~\cite{Wildermuth:1979,Quaglioni2009initio},  by solving a generalized eigenvalue 
problem involving Hamiltonian (\(\mathcal{H}\)) and norm kernel (\(\mathcal{N}\)):
\begin{equation}
 \sum_{n'}\mathcal{H}_{nn'} \chi_{n'}  = E\sum_{n'}\mathcal{N}_{nn'}\chi_{n'},
 \quad
 \mathcal{H}_{nn'} = \langle \Phi_{n\ell} | H | \Phi_{n'\ell} \rangle,
 \quad
 \mathcal{N}_{nn'} = \langle \Phi_{n\ell} |  \Phi_{n'\ell} \rangle.
 \label{eq:kernels}
\end{equation}
Each solution $\{\chi_n\}$ is normalized via
\begin{equation}
\sum_{n,n'}\,
\chi_n^\star\,\mathcal{N}_{nn'}\,\chi_{n'}
\;=\; 1.
\label{eq:normalization}
\end{equation}
Since the norm kernel $\mathcal{N}$ is positive definite, it admits a unique positive-definite square root. Thus, defining
\begin{equation}
\tilde{\chi}_n
\;\equiv\;
\bigl(\sqrt{\mathcal{N}}\,\chi\bigr)_{n}
\;=\;
\sum_{n'} 
\bigl(\sqrt{\mathcal{N}}\bigr)_{n n'}\,\chi_{n'},
\label{eq:DefineChiTilde}
\end{equation}
we obtain the normalized channel wave function
\begin{equation}
|\tilde{\Xi}_{\ell}\rangle 
\;=\;
\sum_{n}\,\tilde{\chi}_n\,|\\,\Phi_{n\ell}\rangle.
\label{eq:XiTilde}
\end{equation}
Here, absorbing $\sqrt{\mathcal{N}}$ into $\chi$ converts Eq.~\eqref{eq:kernels} into a standard Hermitian eigenvalue problem. Consequently, the solutions $|\tilde{\Xi}_{\ell}\rangle$ become orthonormal in the usual sense:
\begin{equation}
\langle\,\tilde{\Xi}_{\ell'}\mid \tilde{\Xi}_{\ell}\rangle
\;= \delta_{\ell' \ell}
\label{eq:XiNorm}
\end{equation}
the orthogonality among different eigenstates similarly follows from the Hermiticity of $\mathcal{N}^{-1/2}\,\mathcal{H}\,\mathcal{N}^{-1/2}$.

\subsection{Applications}

\subsubsection{Shell model studies of clustering spectroscopic factors}
From the viewpoint of nuclear structure physics the level of clustering in any particular state can be assessed
by a spectroscopic factor (SF), defined as the overlap between cluster channels \eqref{eq:rgmx} and 
shell-model states for a parent nucleus $\Psi^{(A)}$ 
\begin{equation}
S_{\ell} 
\;\equiv\;
\bigl|\!\langle \Psi^{(A)} \,\bigl|\, \Xi_{\ell} \rangle\bigr|^{2}.
\label{eq:rgm1}
\end{equation}
Since the RGM equations in \eqref{eq:kernels} have in general multiple solutions we can have SF into effectively
different channels, which in traditional particle and potential problem would be characterized by a principal quantum number which represents the number of nodes in the wave function. 

Historically, many studies of $p$- and $sd$-shell nuclei have been carried out within the traditional shell model with a core~\cite{Draayer:1975,Volya2015,Chung:1978}. 
Despite its empirical nature, the traditional shell model remains one of the most successful and powerful predictive tools for describing a broad range of nuclear properties~\cite{Brown2022}. Even in the era of advanced computing and expanding \textit{ab initio} calculations, it continues to play a critical role in bridging fundamental theory and experimental observations. By simplifying the complex nuclear many-body problem, the shell model helps pinpoint and explain a variety of emergent phenomena, including quartet and clustering correlations~\cite{Sandulescu2024}.

Multiple recent studies, fueled by experimental efforts, have been carried out in the mass region bridging the $p$ and $sd$ shells. In particular, advancements in TTIK methods have invigorated this field by systematically investigating clustering above the $\alpha$-decay threshold~\cite{Avila2014,Barbui2022,Goldberg2022,Nauruzbayev2017,Nauruzbayev2017,Nurmukhanbetova2019,Nurmukhanbetova2023,Nurmukhanbetova2024,Nurmukhanbetova2024b,Upadhyayula2020}. Recent systematic studies and comparisons with experiment affirm the applicability of the presented approach. However, owing to the inherent limitations of the traditional shell model and its effective nature, many results remain qualitative. In particular, understanding the effective operators involved in four-nucleon (i.e., $\alpha$-particle) removal requires additional phenomenological adjustments.

Next, we consider several shell-model applications. Let us first focus on low-lying states, where the configuration space is restricted to a single oscillator shell. As a result, the entire cluster spectroscopy is limited to an effective operator within that shell. In an HO basis, this translates to a single $\Phi_{n\ell}$ contributing a non-zero overlap in evaluating the spectroscopic factor:
\begin{equation}
S_{\ell} 
\;\approx\;
|\chi_n|^2\,\bigl|\!\langle \Psi^{(A)} \,\bigl|\, \Phi_{n\ell} \rangle\bigr|^{2}.
\label{eq:SMSF}
\end{equation}
Within this limit, it is natural to assume all basis states are decoupled, leading to diagonal RGM equations and hence ${\chi}_n = 1$. Conservation of oscillator quanta enforces this condition, as exploited in analytic studies~\cite{filippov2003,filippov2004,lashko2008,lashko2019}.

\begin{figure}
\centering
\includegraphics[width=0.99\linewidth]{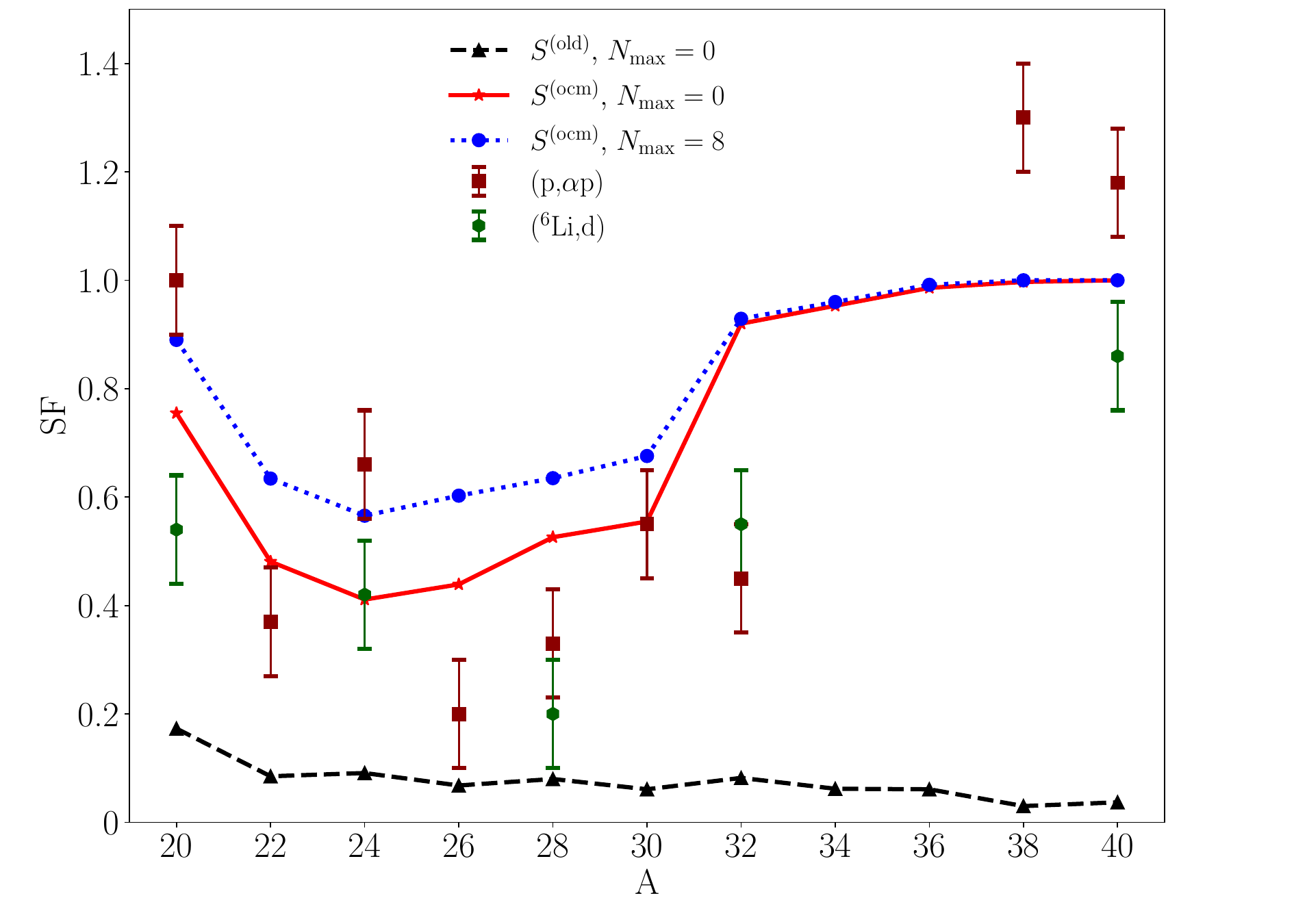}
\caption{\label{fig:sdSF}Spectroscopic factors for ground-state $\alpha$ transitions, $A \to (A-4) + \alpha$, in $sd$-valence $N=Z$ nuclei. Scattered points are experimental data from knockout and pickup reactions~\cite{Carey:1981,Anantaraman:1975}. Connected points show theoretical results using the USDB Hamiltonian~\cite{Brown2006New}, with the $\alpha$-particle wave function taken from an NCSM calculation (JISP16, $\hbar\omega = 14$ MeV) at the indicated $N_{\rm max}$. The dashed (black) line depicts the traditional shell-model spectroscopic factors [Eq.~\eqref{eq:SMSF}] with $\chi=1$, while the solid (red) and dotted (blue) lines show OCM spectroscopic factors. See also ref. 
\cite{Kravvaris2019}
}
\end{figure}

Figure~\ref{fig:sdSF} shows the evaluation of $\alpha$ clustering in $N=Z$, $sd$-shell nuclei for ground-state-to-ground-state transitions, where $\ell=0$ and $n=4$. The dashed black line presents spectroscopic factors calculated via Eq.~\eqref{eq:SMSF} with $\chi_n=1$ (labeled as $S^{(\text{old})}$). These values lie an order of magnitude below experimental data and fail to reproduce the observed trend of peaking at the beginning and end of the shell while dipping in the middle. This discrepancy is unsurprising, since an $\alpha$ particle boosted by eight quanta into the $sd$ shell contains only a 4\% $(sd)^4$ component, as indicated in Table~\ref{tab:ncsm-config}. The issue has prompted significant discussion and highlights the need to renormalize the spectroscopic factors.

A frequently used approach is to adopt the Orthogonality Conditions Model (OCM) spectroscopic factors~\cite{Fliessbach:1976,Fliessbach:1977,Volya2015}. Given that the channel wave function’s normalization in the limited shell-model space is so small, renormalizing it to unity is a natural choice, thereby establishing a sum rule for the SF in a given channel. Formally, one can argue that the proper RGM solution in an orthonormal basis corresponds to $\tilde{\chi}_n = 1$, and in a diagonal scenario this defines the SM cluster channel as
\begin{equation}
|\Xi_{\ell}\rangle 
= \frac{1}{\sqrt{\langle \Phi_{n\ell} \mid \Phi_{n\ell} \rangle}}\,|\Phi_{n\ell}\rangle.
\label{eq:SMRGM}
\end{equation}
It is important to emphasize that while Pauli blocking and the projection onto a limited valence space are related, they are distinct reasons for renormalization. Consequently, in OCM the projection onto the valence space is included in the normalization overlap. The spectroscopic factors computed with this renormalization are referred to as $S^{(\text{ocm})}_{\ell}$ in Fig.~\ref{fig:sdSF}, and they reproduce both the experimental data and its trend more accurately.

Recent experimental studies~\cite{Nurmukhanbetova2024} have tentatively identified the first limitations of the OCM method, noting that the procedure fails when the normalization due to Pauli blocking becomes extremely small, although further analysis is needed.

These results confirm strong clustering in the ground states of well-bound light nuclei. However, in these cases, the connection between a free $\alpha$ particle and one embedded in the nuclear medium is complex. This is evident from the direct overlaps between the $\alpha$ particle and the state being very small ($S^{(\text{old})}$). Yet, when the reaction channel is properly reconstructed, the resulting measured OCM spectroscopic factor is large (see Fig.~\ref{fig:sdSF}).
We will further focus on this complex interplay between the traditional nucleon–nucleon shell-model picture and cluster configurations in the discussion of $^{12}$C and $^{8}$Be in this and the following section.

\subsubsection{Alpha cluster resonances}
\begin{figure}[h]
\centering
\includegraphics[width=0.99\linewidth]{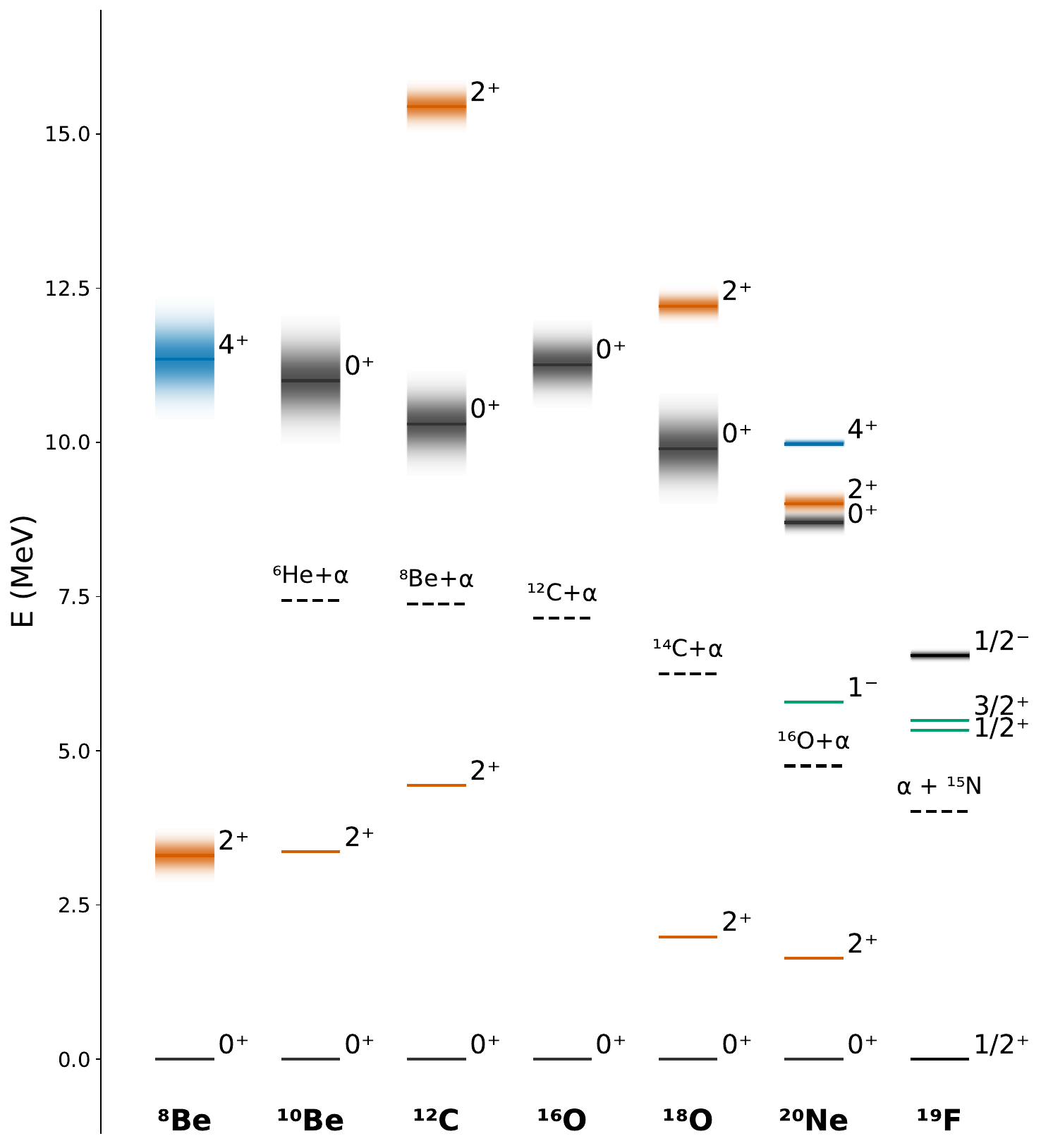}
\caption{\label{fig:level_scheme}Selected $\alpha$-clustered states in several light nuclei, with $\alpha$-decay thresholds indicated by dashed lines. Above the thresholds the resonances are broadened reflecting their width.   The case of $^{20}$Ne and $^{19}$F, including spin-orbit partners for the $\ell=1$ channel, is also shown.}
\end{figure}

While some clustered states, such as those in $^8$Be, have been known for decades, recent advances in experimental techniques and detailed analyses of scattering spectra have revealed numerous strongly clustered states in many nuclei (see reviews in Refs.~\cite{Freer2007,freer_rmp} and systematic investigations in Refs.~\cite{Avila2014,Barbui2022,Goldberg2022,Nauruzbayev2017,Nauruzbayev2017,Nurmukhanbetova2019,Nurmukhanbetova2023,Nurmukhanbetova2024,Nurmukhanbetova2024b,Upadhyayula2020}). 
The strength of clustering is often gauged by comparing the observed $\alpha$-decay width to the so-called Wigner limit, which represents the maximum possible width for an $\alpha$ particle in a potential model at the experimentally observed energy. Figure~\ref{fig:level_scheme} illustrates a selection of these strongly clustered states across various light nuclei.

Discussions of $\alpha$ clustering that now includes $N \neq Z$ nuclei, and the influence of valence particles or holes on high-lying cluster states, have been guided by experimental data and theoretical SF evaluations [Eqs.~\eqref{eq:rgm1} and \eqref{eq:SMRGM}]. In contrast to older, more restrictive approaches, novel cross-shell effective Hamiltonians now enable significant progress toward a microscopic understanding of clustering in highly excited states and distribution of clustering strength. In particular, the recently developed FSU shell-model Hamiltonian~\cite{Lubna2019Structure,Lubna2020Evolution}, designed for cross-shell particle-hole excitations, has proven invaluable for explaining seemingly excessive experimental clustering strength.

As an example, we highlight the $\ell=0$ and $\ell=1$ channels in $^{20}$Ne and $^{19}$F, which are also included in Fig. \ref{fig:level_scheme} as analyzed in Ref.~\cite{Volya2022a}. Selected results appear in Tables~\ref{tablL3} and~\ref{tablL4}, although many other non-clustered states—accurately reproduced by the shell model—are omitted here for brevity. 
Many additonal studies, including those of rotational bands, connection with already mentioned SU(3) symmetry \cite{Manakos1983Spectroscopic} as well as the 
emergence of rotational bands such as one seen in 20Ne can be found in Ref. \cite{Kravvaris2019}.

Beyond general theory-experiment comparisons, these studies emphasize the impact of an extra nucleon degree of freedom, as seen by comparing $^{15}$N+$\alpha$ and $^{16}$O+$\alpha$ decay channels in $^{19}$F and $^{20}$Ne. In $^{19}$F, lower-lying states couple to an $n=3$ channel that is unavailable in $^{20}$Ne due to Pauli blocking; these are low-lying, below-threshold states, and are hence inaccessible to $\alpha$-scattering experiments.
Meanwhile, in the $\ell=0$ channel, the 6.540~MeV $1/2^-$ state in $^{19}$F mirrors the 6.725~MeV $0^+$ state in $^{20}$Ne, both involving an $\alpha$ particle in an $n=4$ radial mode. For $\ell=1$, the 5.79~MeV $1^-$ resonance in $^{20}$Ne couples to the $n=4$ channel, and higher-lying $n=4$ clustered states are likewise identified in $^{19}$F as the spin-orbit partners $1/2^+$ (5.333~MeV) and $3/2^+$ (5.488~MeV), see Fig.~\ref{fig:level_scheme}.

Although the shell model accurately reproduces many observed features, certain highly excited states remain challenging. For instance, the very broad $0^+_4$ state in $^{20}$Ne—likely involving an $n=6$ channel—has a small calculated SF, raising questions about its apparent lack of collectivity in theory. In general, strong clustering appears to emerge from the “collectivization” of cross-shell excitations, in the region of energy where states of a new $\hbar \omega$ structure first appear and exhibit enhanced near-threshold strength for each new RGM channel. The mixing across different particle-hole configurations seems to be suppressed in clustered states. 
However, this phenomenon remains under active investigation.
Moreover, the super-radiance effect~\cite{Auerbach2011}, which seems to drive the collectivization of states deeply embedded in the reaction continuum, has been experimentally observed to play an important role in realigning the wave functions~\cite{Volya2022a,Barbui2022}.

\begin{table}[htb]
%\label{tab:3}
\centering
\begin{tabular}{@{}|c|c|c|c|c|c|@{}}
\hline
$J_{i}^\pi$ & $E$(MeV) & $n$ & SF$_\alpha$ & $E$(MeV) & $\gamma_\alpha$ \\
th & th & th & th & exp & exp \\
\hline
$0^{+}_{1}$ & 0 & 4 & 0.755 & 0 & \\
$0^{+}_{2}$ & 6.698 & 4 & 0.143 & 6.725 & 0.47 \\
$0^{+}_{3}$ & 7.547 & 5 & 0.007 & 7.191 & 0.017 \\
$0^{+}_{4}$ & 10.121 & 6 & 0 & 8.7 & broad \\
$0^{+}_{8}$ & 13.521 & 5 & 0.246 &  & \\
\hline
$1^{-}_{1}$ & 6.982 & 4 & 0.381 & 5.79 & 1.4 \\
$1^{-}_{2}$ & 7.918 & 4 & 0.379 & 8.708 &  \\
\hline

\end{tabular}

\caption{Lowest states coupled to  $\ell=0$ and $\ell=1$ clustering channels in $^{20}$Ne for the  $^{16}$O$+\alpha$ Columns identify state, theoretical excitation energy, number of nodes in the alpha channel, experimental energy, experimental alpha reduced width. The labels in the second row ``th" or ``exp" refer to results coming from theory and experiment, respectively. Correspondence between data from theory and experiment represent a suggested assignment. The data is from ref. \cite{Volya2022a} \label{tablL3}}
\end{table}

\begin{table}[htb]
%\resizebox{5cm}{!}

\begin{tabular}{@{}|c|c|c|c|c|c|@{}}
\hline
$J_{i}^\pi$ & $E$(MeV) & $n$ & SF$_\alpha$ & $E$(MeV) & $\gamma_\alpha$ \\
th & th & th & th & exp & exp \\
$1/2^{-}_{1}$ & 0.468 & 4 & 0.706 & 0.110 &  \\
$1/2^{-}_{2}$ & 6.900 & 4 & 0.020 & (6.095) & \\
$1/2^{-}_{3}$ & 7.092 & 4 & 0.041 & 7.048 & 0.12 \\
$1/2^{-}_{5}$ & 7.856 & 4 & 0.101 & 6.540$^*$ & 0.53 \\
\hline
$1/2^{+}_{1}$ & 0.000 & 3 & 0.874 & 0.000 & \\
$1/2^{+}_{2}$ & 6.060 & 4 & 0.311 & 5.333$^*$ & 1.16 \\
\hline
$3/2^{+}_{1}$ & 1.770 & 3 & 0.672 & 1.554 & \\
$3/2^{+}_{4}$ & 6.937 & 4 & 0.633 & 5.488$^*$ & 0.98 \\
\hline
\end{tabular}

\caption{Lowest $\ell=0$ and $\ell=1$ states in $^{19}$F, viewed as $^{15}$N+$\alpha$ channels. Columns are analogous to Table~\ref{tablL3}.
\label{tablL4}}

\end{table}

\subsubsection{Cluster Nucleon Configuraiton Interaction}
Going beyond the analysis of spectroscopic factors discussed in the previous subsections, we next apply the Cluster Nucleon Configuration Interaction (CNCIM) approach to \(^{21}\)Ne as a simple illustration, demonstrating that a drastically reduced channel basis can still reproduce key features of a full shell-model (SM) calculation. This offers a powerful extension to the traditional shell model, especially in cases where full diagonalization is not feasible.
We treat $^{20}$Ne as a $^{16}$O core plus $\alpha$ system, then add a neutron in the $d_{5/2}$ orbital. Concretely, we place an $\alpha$ particle with relative angular momentum $L = 0,2,4,6$ onto $^{16}$O and couple with the extra neutron, yielding 18 channel basis states---compared with 1935 many-body states in the full $sd$ SM space.

\begin{figure}[h]
\centering
\includegraphics[width=0.8\textwidth]{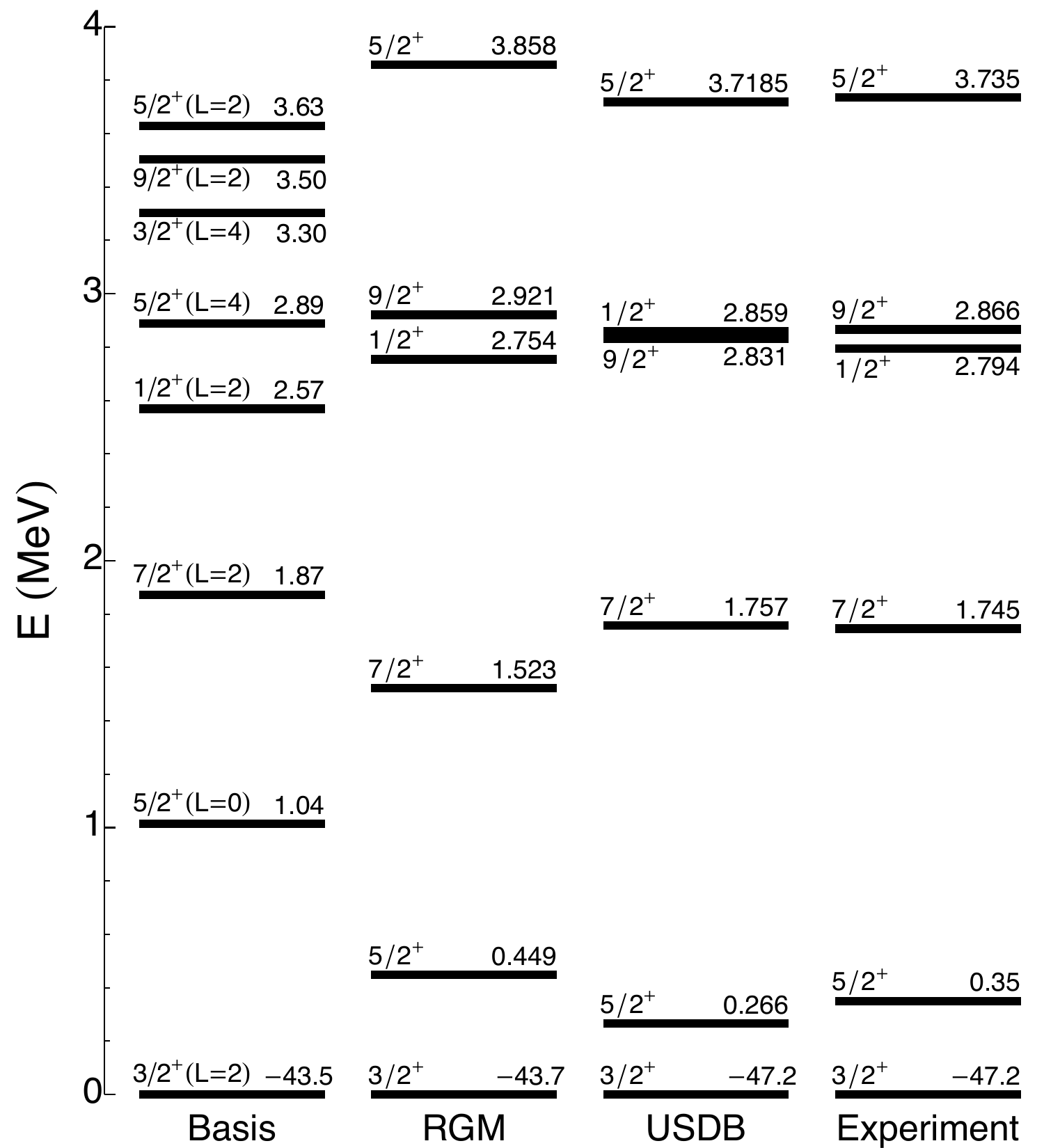}
\caption{Diagonal energies in the channel basis, low-lying states in RGM solution, full USDB and experimental spectra of $^{21}$Ne, see Ref. \cite{Kravvaris2019} }
\label{fig:Ne21RGM}
\end{figure}

In this reduced, non-orthogonal basis, we compute the Hamiltonian kernel using the USDB interaction \cite{Brown2006New} and solve it via the Resonating Group Method (RGM). Figure~\ref{fig:Ne21RGM} shows the resulting low-lying spectrum, alongside the full USDB calculation and experimental data. Ground states indicate total binding; energies of other levels are shown relative to the ground state.  
The main spectral features and excitation energies are well reproduced, although we note that simple RGM model underbinds by about 3.5~MeV compared to the full diagonalization.

Figure~\ref{fig:Ne21RGM} also includes a first column that lists the diagonal energies of the RGM basis states, 
$\langle \Xi_{J L}|H|\Xi_{J L}\rangle$, where total angular momentum $J$ couples $L$ with the unpaired neutron ($j=5/2$). The RGM solution clarifies the mixing of partial waves; for example, two $5/2^+$ states with $L=0$ and $L=4$ at 1.04 and 2.89~MeV in the first column undergo two-state mixing and repel each other. This implies that the first excited $5/2^+$ state in $^{21}$Ne is a mixture of $L=0$ and $L=4$, excluding $L=2$---a fact confirmed experimentally~\cite{Anantaraman:1978}. Further details of this study
including matrix elements of the Hamiltonian and the norm kernel can be found in  \cite{Kravvaris2018Clustering}

\subsubsection{Cluster relative motion}

Next, we move our discussion to clustering aspects in \textit{ab initio} no-core applications of shell-model (SM) methods. We begin with the nucleus \(^{8}\)Be, which, with its \(\alpha + \alpha\) structure, has long served as a benchmark for nuclear clustering studies~\cite{Ichimura1973Alphaparticle,Tanabe:1975,Suzuki1986Symplectic,Lovas1998Microscopic,wiringa_2000,Elhatisari:2015,Varga1992Clusterconfiguration}.

Table~\ref{tab:8Be} presents both experimental and theoretical energies/widths for the $0^+$, $2^+$, and $4^+$ cluster resonances, each interpreted as two $\alpha$ particles with relative angular momenta $\ell = 0, 2, 4$. A treatment based on the Resonating Group Method (RGM), which effectively employs a cluster-configuration-interaction (CI) approach with up to 12 basis channel states, in contrast to a full many-body expansion that would be prohibitively large. Here JISP16 \cite{shirokov_2007,Maris:2013} interaction with $\hbar\omega=25$ MeV is used.

\begin{figure}[h]
\centering
\includegraphics[width=0.99\linewidth]{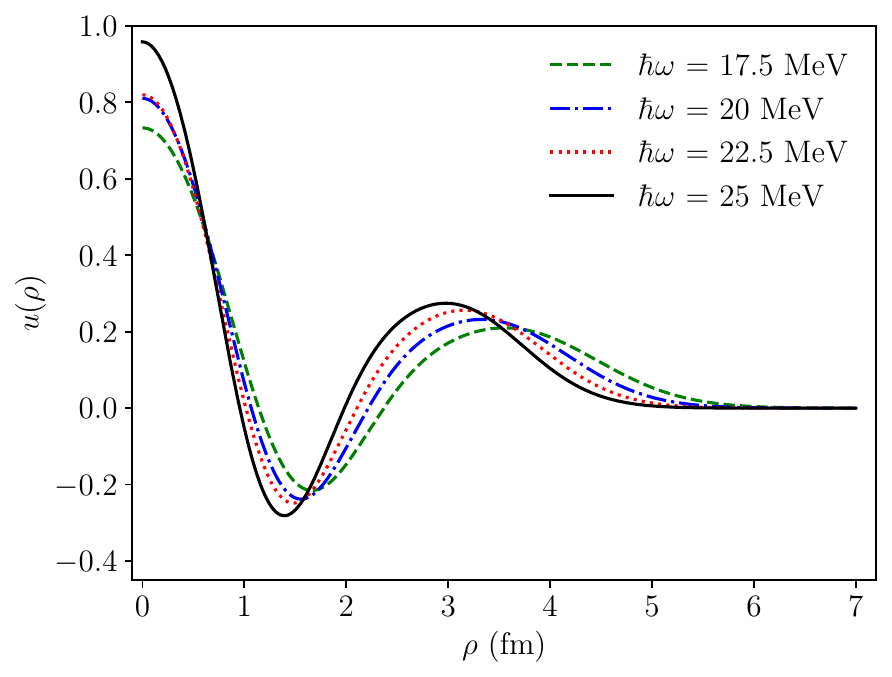}
\caption{\label{fig:alphaalphawf}Relative $\alpha+\alpha$ wave function for the $0^+$ RGM channel in $^8$Be, computed using the JISP16 interaction. Different curves correspond to various values of $\hbar\omega$.}
\end{figure}

A key signature of rotational two-alpha dynamics is the ratio $\mathrm{R}_{42}$ of $4^+$ to $2^+$ excitation energies, 
which is found to be about 3.5—close to the $\mathrm{R}_{42} = 3.3$ expected from a rotational band.  
This result, arising from a very different scattering perspective, is in quantitative agreement with our earlier rotation-based formulation in Sec.~2.4.
Figure~\ref{fig:alphaalphawf} shows the radial wave function for the $0^+$ channel,
\begin{equation}
u_{\ell}(\rho)=\sum_n \chi_n\,\phi_{n\ell}(\rho),
\label{eq:horad}
\end{equation}
illustrating the effect of different harmonic-oscillator parameters $\hbar\omega$. Beyond the spatial size of each $\alpha$ cluster, $u_{\ell}(\rho)$ describes the relative motion of the two $\alpha$ particles and can be matched to external Coulomb wave functions allowing to study asymptotic normalization and decay widths. 

The decay widths in Table~\ref{tab:8Be} are extracted using the standard \( R \)-matrix approach~\cite{Thomas1954formulation},
\begin{equation}
\Gamma_\ell 
= 
\frac{\hbar^2 k}{\mu}\,
\frac{\rho_c^{2}\,u_\ell^{2}(\rho_c)}{F_{\ell}^{2}(\eta,k\rho_c)\;+\;G_{\ell}^{2}(\eta,k\rho_c)},
\end{equation}
by matching the interior wave function to the asymptotic solution at \(\rho_c = 3.6\,\mathrm{fm}\). Beyond the range of the nuclear potential, the result is not sensitive to the exact matching point; however, in a restricted HO basis, this issue must be treated with care. The chosen matching location maximizes the outgoing flux while minimizing sensitivity to the matching point.
This study simultaneously reproduces the nearly-bound $0^+$ ground state ($\Gamma = 5.6\,\text{eV}$) and the broad $4^+$ resonance, highlighting the effectiveness of the cluster-based RGM approach. 

To conclude this discussion, we pause on Fig.~\ref{fig:alphaalphawf}, which highlights that the $^{8}$Be ground state is a long-lived state of two $\alpha$ particles, with a resonant wave function suggesting that the two $\alpha$ particles move around each other with an average separation of about 3.5~fm, as can be inferred from the figure. This result is also obtained from a different perspective in Sec.~\ref{sec:itagaki}.

\renewcommand{\arraystretch}{1.2}
\begin{table}[h]
\centering
\begin{tabular}{cccccc}
\hline\hline
$\ell$ & $E_{\rm ex}$ & $\Gamma$ & $E^{(\mathrm{RGM})}_{\rm ex}$ & $\Gamma^{(\mathrm{RGM})}$ & ${S}_\ell$ \\[5pt]
\hline
0 & 0.0 & 5.6$^*$ & 0.0 & 8.9$^*$ & 0.69 \\
2 & 3.0 & 1.5 & 4.6 & 1.4 & 0.66 \\
4 & 11.4 & 3.5 & 16.0 & 2.7 & 0.51 \\
\hline\hline
\end{tabular}
\caption{\label{tab:8Be}Experimental and RGM results for $^8$Be with the JISP16 interaction, allowing up to 12 HO quanta in each $\ell$ channel. Energies and widths are given in MeV except for those marked with $*$ (in eV).}
\end{table}

\subsubsection{Hoyle state}

One important future direction of the method is its extension to multi cluster problems. In this context, $^{12}$C, including both its ground state and the Hoyle state, occupies a central role in this review, as it is examined across this section and Secs.~2 and 4, consistently revealing the same underlying physics from complementary perspectives. A classic example is the Hoyle state in $^{12}$C, which has been studied in various approaches~\cite{bijker_2002,Chernykh2007Structure,Epelbaum2011initio}.

The Hoyle state, an excited $0^+$ state just 285~keV above the triple-alpha threshold, decays almost exclusively via an intermediate $^8$Be resonance, itself only 93~keV above the two-alpha threshold~\cite{Raduta:2011,Itoh:2014,Smith:2017,DellAquila2017Highprecision}. This sequential decay path, in addition to being favored by Coulomb dynamics, highlights important structural features of the three-alpha configuration. A first application of the method to this problem is presented in Ref.~\cite{Kravvaris2019}, and we summarize it below.

While no direct analytic method analogous to Moshinsky coefficients exists for the three-cluster case, the channels can be  constructed through a sequential coupling of Jacobi coordinates. In the present calculation, up to $N_{\rm max} = 12$ oscillator quanta are distributed among the two relative coordinates. The RGM framework is applied to a system of three identical $\alpha$ particles, each in an $s^4$ configuration, using the JISP16 interaction with $\hbar\omega = 25$~MeV. The minimal configuration allowed by the Pauli principle corresponds to $N = 8$, representing a filled $0s$ shell with eight nucleons occupying the $0p$ shell.

\begin{table}[h]
\renewcommand{\arraystretch}{1.1} % Adjust this value to make rows tighter or looser
\centering
\begin{tabular}{@{}c c c c r@{}}
$J^\pi$  & $E^{\rm(exp)}_{\rm ex}$ & $E^{\rm(NCSM)}_{\rm ex}$ & $E^{\rm(RGM)}_{\rm ex}$ & $S_\ell$ \\
\hline\hline
$0^+$ &  0    & 0     & 0     & 0.42 \\
$2^+$ &  4.4  & 6.06  & 3.61  & 0.49 \\
$4^+$ & 14.1  & 19.8  & 13.6  & 0.60 \\
\hline\hline
\end{tabular}
\caption{Excitation energies (in MeV) for rotational band members in $^{12}$C: experimental data, NCSM results, and RGM results (both with $\hbar\omega = 25$~MeV), along with spectroscopic factors $S_\ell$ for triple-alpha decay.}
\label{tab:C12}
\end{table}

% \begin{table}[h]
% \centering
% \begin{tabular}{ c   c   c   c  r  r}
% $J^\pi$  & $E^{\rm(exp)}_{\rm ex}$ & $E^{\rm( NCSM)}_{\rm ex}$ & $E^{\rm(rgm)}_{\rm ex}$ & ${\cal S}^{\rm(rgm)}_\ell $ & Channels \\[0.3em]
% \hline\hline
% $0^+$ &  0    & 0     & 0     & 0.42 & 8  \\
% $2^+$ &  4.4  & 6.06  & 3.61  & 0.49 & 16 \\
% $4^+$ & 14.1  & 19.8  & 13.6  & 0.60 & 16 \\
% \hline\hline
% \end{tabular}
% \caption{Excitation energies (in MeV) for rotational band members in $^{12}$C: experiment, NCSM, and RGM results (both with $\hbar\omega = 25$ MeV), spectroscopic factors for triple-alpha decay, and number of basis channels included.}
% \label{tab:C12}
% \end{table}

As seen in Table~\ref{tab:C12}, the RGM and NCSM results both suggest significant clustering in the lowest $0^+$, $2^+$, and $4^+$ states. While the NCSM is not ideally suited to describe the Hoyle state, the triple-alpha spectroscopic factor 
${S}(0^+_2) = 0.257$ is reasonable. Furthermore, the squared overlap of the triple-alpha channel ($J^\pi = 0^+$) with a two-fragment channel consisting of the $^8$Be ground state ($N_{\rm max}=4$) and an $\alpha$ in relative motion with $n=2$, $\ell=0$, is 0.51. This large overlap supports the dominance of the sequential decay mechanism through $^8$Be.

These results essentially reiterate the findings from the MCSM in Sec.~2. The states listed in Table~\ref{tab:C12} are members of the ground-state band seen in Fig.~4. While here we do not project onto the intrinsic frame to pictorially display the density profile of a rotating three-$\alpha$ system as in Fig.~5, we instead infer this structure from the energies of states that form a rotational band, from quadrupole moments that can be shown to follow rotor-model systematics, and from the consistency of the spectroscopic factors $S_\ell$ in the last column of Table~\ref{tab:C12}. 
The gradual increase of $S_\ell$ with angular momentum highlights the role of rotational motion, which slightly deforms the two-$\alpha$ configuration and increases their separation, thereby explaining the growth of $S_\ell$.

Our numerical results for the spectroscopic factors discussed above show that both the ground state and the Hoyle state couple to the same lowest asymptotic three-$\alpha$ channel. This channel is open for the Hoyle state but closed for the ground state, highlighting the mixing discussed in Fig.~3. Furthermore, this channel couples strongly to the $^{8}$Be$+\alpha$ configuration, placing the qualitative picture shown in Fig.~3 on a quantitative footing.

\subsubsection{Scattering problem}
Having established a cluster-based framework for spectroscopy of bound and weakly bound states, we now illustrate its extension to scattering. 
In particular, $\alpha+\alpha$ scattering serves as an instructive example for studying resonances and continuum dynamics in light nuclei. 

% In the context of our approach, it is natural to proceed in the HO basis and continue to rely on the expansion of the radial motion in HO functions.

Unlike the case of deeply bound states, the HO expansion is a poor choice for weakly bound and scattering states because of spatially extended nature of the wave functions. However, the analytic form of the basis functions remedies this issue. The J-matrix method, Ref.~\cite{Alhaidari2008Jmatrix}, also known as the Harmonic Oscillator Representation of Scattering Equations (HORSE)~\cite{Bang2000PMatrix,Shirokov:2016}, has been extensively discussed in the literature~\cite{Alhaidari2008Jmatrix}. In its traditional form, which we discuss here, the method is limited to the case where remotely the Hamiltonian matrix is represented just by the kinetic energy operator. The Coulomb problem presents a significant challenge for the standard J-matrix/HORSE method, but the method can be appropriately modified, as discussed in Refs.~\cite{Bang2000PMatrix,Shirokov:2016,Kravvaris2018Clustering}.

The integer $n$ in Eq.~\eqref{eq:horad}, which enumerates the basis states, coincides with the number of nodes in the radial part of the wave function. The method relies on the asymptotic limit $r\rightarrow \infty$ in coordinate space being equivalent to the configuration-space limit $n\rightarrow \infty$, with the approximate correspondence
\begin{equation}
r=\sqrt{\frac{\hbar}{m\omega} (2n+\ell+3/2)}.
\label{eq:r0}
\end{equation}

Thus, the RGM solution \eqref{eq:rgmx}, expressed in radial form \eqref{eq:horad} needs to be matched with the asymptotic solution for the free-space Hamiltonian, such that the asymptotic behavior is expressed as
\begin{equation}
\chi_n \simeq \alpha F_{n\ell} + \beta G_{n\ell},
\label{eq:nsym}
\end{equation}
where $F_{n\ell}$ and $G_{n\ell}$ represent the regular and irregular solutions for the free-space Hamiltonian (which are Coulomb or Bessel functions), expanded in the HO basis; see Ref.~\cite{Yamani1975matrix}.

The central idea, describing the J-matrix (or HORSE) method~\cite{Alhaidari2008Jmatrix,Bang2000PMatrix,Shirokov:2016}, is represented by the following equation, showing the structure of the RGM matrix:
\begin{equation}
\scalebox{0.92}{$
\left(
\begin{NiceArray}{cccccc}
\cline{1-3}
\multicolumn{1}{|c}{\mathcal{H}_{00}} & \cdots & \multicolumn{1}{c|}{\mathcal{H}_{0n}} & 0 & \cdots & 0 \\
\multicolumn{1}{|c}{\vdots}          & \ddots & \multicolumn{1}{c|}{\vdots}          & 0 & 0      & \vdots \\
\multicolumn{1}{|c}{\mathcal{H}_{n0}} & \cdots & \multicolumn{1}{c|}{\mathcal{H}_{nn}} & T_{n\,n+1} & 0 & 0 \\
\cline{1-3} \cline{4-6}
0      & 0      & T_{n+1\,n} & \multicolumn{1}{|c}{T_{n+1\,n+1}} & T_{\cdots} & \multicolumn{1}{c|}{0} \\
\vdots & 0      & 0          & \multicolumn{1}{|c}{T_{\cdots}}   & T_{\cdots} & \multicolumn{1}{c|}{\ddots} \\
0      & \cdots & 0          & \multicolumn{1}{|c}{0}            & \ddots     & \multicolumn{1}{c|}{\ddots} \\
\cline{4-6}
\end{NiceArray}
\right)
\left(
\begin{NiceArray}{c}
\chi_0 \\ \vdots \\ \chi_n \\ \chi_{n+1} \\ \chi_{n+2} \\ \vdots
\end{NiceArray}
\right)
=
E
\left(
\begin{NiceArray}{cccccc}
\cline{1-3}
\multicolumn{1}{|c}{\mathcal{N}_{00}} & \cdots & \multicolumn{1}{c|}{\mathcal{N}_{0n}} & 0 & \cdots & 0 \\
\multicolumn{1}{|c}{\vdots}          & \ddots & \multicolumn{1}{c|}{\cdots}          & 0 & 0      & \vdots \\
\multicolumn{1}{|c}{\mathcal{N}_{n0}} & \cdots & \multicolumn{1}{c|}{\mathcal{N}_{nn}} & 0 & 0 & 0 \\
\cline{1-3} \cline{4-6}
0      & 0      & 0      & \multicolumn{1}{|c}{1}      & 0 & \multicolumn{1}{c|}{\vdots} \\
\vdots & 0      & 0      & \multicolumn{1}{|c}{0}      & 1 & \multicolumn{1}{c|}{0} \\
0      & \cdots & 0      & \multicolumn{1}{|c}{\cdots} & 0 & \multicolumn{1}{c|}{\ddots} \\
\cline{4-6}
\end{NiceArray}
\right)
\left(
\begin{NiceArray}{c}
\chi_0 \\ \vdots \\ \chi_n \\ \chi_{n+1} \\ \chi_{n+2} \\ \vdots
\end{NiceArray}
\right)
$}
\label{eq:blocks}
\end{equation}

%%%%%%%%%%%%%%%%%%%%%%%%%%%%%%%%%%%%%%%%%%%%%%%%%%%%%%%%%

% At the core of the method is an approximation that assumes the Hamiltonian and norm kernels in Eq.~\eqref{eq:kernels} are range-limited in configuration space up to some maximum value $n$, related to distance $r$ as described \eqref{eq:r0}. The upper-left block, spanning from $0$ to $n$ nodes, is computed exactly. 
% Beyond this, an asymptotic form is assumed: the norm kernel becomes the identity matrix, and the Hamiltonian kernel is given by the kinetic energy operator, represented by a lower right block. The solution, starting from $\xi_n$, is matched to Eq.~\eqref{eq:nsym}.

% Here, $E$ denotes the continuous scattering energy. In the asymptotic solution, the kinetic energy in the harmonic oscillator basis is represented by a tridiagonal matrix. Consequently, only a single matrix element, $T_{n\,n+1}$, connects the two blocks (spaces).
% Relating $\xi_n$ and $\xi_{n+1}$ we obtain
% \begin{equation}
% \alpha F_{n\ell} + \beta G_{n\ell} =  \left(\frac{1}{E \mathcal{N}_{\mathcal{P}\mathcal{P}} - \mathcal{H}_{\mathcal{P}\mathcal{P}}}\right)_{nn} T_{n\,n+1} (\alpha F_{n+1\ell} + \beta G_{n+1\ell}),
% \label{eq:horse}
% \end{equation}
% where first term on the left is the diagonal $n$-element of the inverse matrix. 
% This relation allows one to determine the ratio $\beta/\alpha$ which determines the scattering  phase shift at the energy $E$.

At the core of the method is an approximation that assumes the Hamiltonian and norm kernels in Eq.~\eqref{eq:kernels} are range-limited in configuration space up to some maximum value $n$, related to the radial distance $r$ as described in Eq.~\eqref{eq:r0}. The upper-left blocks in \eqref{eq:blocks}, spanning from $0$ to $n$ nodes, is computed exactly, let us call this space 
${\cal P}$ and denote matrices and vectors restricted to this subspace with superscript ${\cal P}$. 
Beyond this point, an asymptotic form is assumed: the norm kernel becomes the identity matrix, and the Hamiltonian kernel is given by the kinetic energy operator, represented by the lower-right blocks on both sides of equation \eqref{eq:blocks}. The solution, starting from $\xi_n$, is matched to the asymptotic form given in Eq.~\eqref{eq:nsym}.

Here, $E$ denotes the continuous scattering energy. In the asymptotic region, the kinetic energy Hamiltonian in the harmonic oscillator basis is represented by a tridiagonal matrix. Consequently, only a single matrix element, $T_{n\,n+1}$, connects the two blocks (or spaces). 
\begin{equation}
\sum_{m=0}^{m=n}\left (E \mathcal{N}_{n m} - \mathcal{H}_{nm}\right ) \chi_m=   T_{n\,n+1} \chi_{n+1},
\label{eq:horse1}
\end{equation}

Relating $\xi_n$ and $\xi_{n+1}$ and writing them both in the asymptotic form, we obtain
\begin{equation}
\alpha F_{n\ell} + \beta G_{n\ell} =  \left(\frac{1}{E \mathcal{N}^{(\mathcal{P})} - \mathcal{H}^{(\mathcal{P})}} \right)_{nn} T_{n\,n+1} (\alpha F_{n+1\ell} + \beta G_{n+1\ell}),
\label{eq:horse}
\end{equation}
where the first term on the right is the $n$th diagonal element of the inverse matrix.  
This relation allows one to determine the ratio $\beta/\alpha$, which in turn determines the scattering phase shift at energy $E$.

\begin{figure}[t]
\centering
\includegraphics[width=0.8\textwidth]{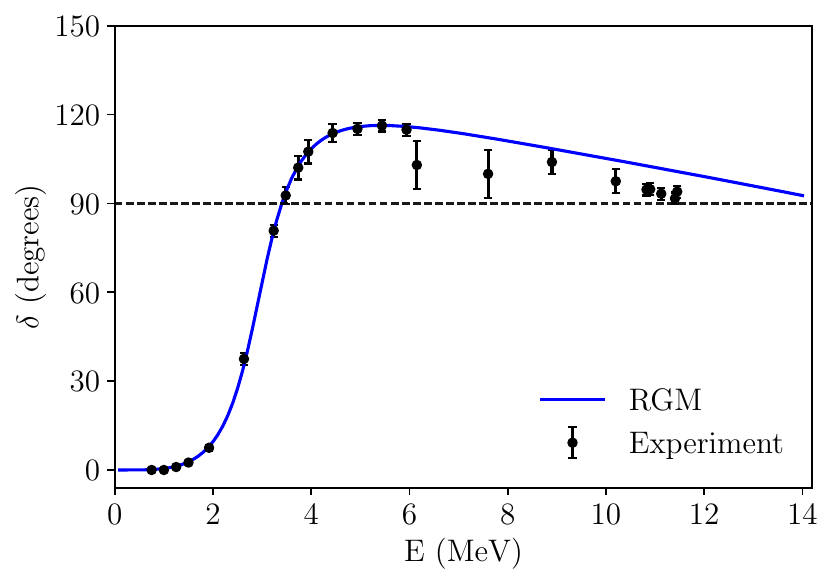}
\caption{(Color online) Phase shifts for $\alpha+\alpha$ scattering in the $\ell=2$ channel. Experimental data are from Ref.~\cite{Afzal1969Systematic}, while the line shows RGM theoretical results, for further details see Ref. \cite{Kravvaris2019}}
\label{fig:AlphaAlphaPshifts}
\end{figure}

Figure~\ref{fig:AlphaAlphaPshifts} shows the \(\ell = 2\) (\(D\)-wave) phase shifts for \(\alpha + \alpha\) scattering. The solid line corresponds to the phase shift obtained using this method with HO truncation up to \(N = 12\) quanta. Once the threshold is adjusted to the experimental \(Q\) value, the agreement with data~\cite{Afzal1969Systematic} is quite good. 
 
The scattering phase shift presented here constitutes a direct experimental observable associated with the $2^+$ state in $^{8}$Be, which is also discussed in Table~\ref{tab:8Be}. This provides strong experimental support for the validity of these results and for their interpretation in terms of clustering, as developed in this section and in Secs.~\ref{sec:otsuka} and \ref{sec:itagaki}.

%%%%%%%%%%%%%%%%%%%%%%%%%%%%%%%%%%%%%%%%%%%%%%%%%%%%%%%%%%%%
%%%%  From Itagaki

\section{Cluster-shell competition and its modeling \label{sec:itagaki}}

One of the most intriguing features of nuclear structure physics is the interplay between shell and cluster structures~\cite{PhysRevC70054307}. 
This is mainly caused by the effect of the spin-orbit interaction, which strengthens the symmetry of the $jj$-coupling shell model.
This interaction is well known to be vital in explaining the observed magic numbers of 28, 50, 82 and 126.
The spin-orbit interaction also breaks clusters, where some of the strongly correlated nucleons are spatially localized.

Nevertheless, as we discussed in this article, the $\alpha$ cluster structure is important in the light mass region. 
Be isotopes are known to have a robust $\alpha$-$\alpha$ cluster structure: $^8$Be decays into two $\alpha$ clusters, 
and the molecular orbital structure of valence neutrons appears in neutron-rich Be isotopes~\cite{PhysRevC61044306, PhysRevC62034301}, 
which has been confirmed by the {\it ab initio} calculation as we have seen.
The persistence of the $\alpha$-$\alpha$ cluster structure is due to the relative %alpha-alpha 
distance, which is about 3–4~fm and large compared to the range of the spin-orbit interaction.

In light nuclei, it is considered that these two different models (shell and cluster) coexist and compete with each other.
Although the $\alpha$-$\alpha$ cluster structure persists in $^8$Be, when one more $\alpha$ cluster is added in $^{12}$C the interaction among 
$\alpha$ clusters becomes stronger and the system has a shorter $\alpha$-$\alpha$ distance.
In this case, the $\alpha$ clusters are trapped within the interaction range of the spin-orbit interaction.
In $^{12}$C, although the three $\alpha$ cluster structure remains in the ground state, the $jj$-coupling shlell model components mix in.

We can model this transition from the cluster state to the shell state owing to the spin-orbit interaction with clear perspective.
The $\alpha$ clusters are spin-zero systems, 
so the spin-orbit interaction — a rank-one non-central interaction — does not contribute for the systems consisting of the $\alpha$ clusters only.
However, we have developed the antisymmetrized quasi cluster model (AQCM) ~\cite{PhysRevC71064307, PhysRevC87054334, PhysRevC103044303}.
This method enables us to smoothly transform the wave functions of the $\alpha$-cluster model to those of the $jj$-coupling shell model. We refer to the clusters that experience the effects of the spin-orbit interaction due to this model as quasi-clusters, which alows us to discuss the intermediate state between this transition from the cluster state to the shell state.
We previously introduced AQCM to ¹²C and discussed the competition between cluster states and the $jj$-coupling shell model state.

Here, we summarize the basic concept of AQCM, which allows the smooth transformation of cluster model wave functions 
to $jj$-coupling shell model ones.
In AQCM, as in many other cluster models including the Brink model, each single particle is described by a Gaussian form. 
\begin{equation}	
  \phi^{\tau, \sigma} \left( {\bf r} \right)
  =
  \left(  \frac{2\nu}{\pi} \right)^{\frac{3}{4}} 
  \exp \left[- \nu \left({\bf r} - {\bf \zeta} \right)^{2} \right] \chi^{\tau,\sigma}, 
  \label{spwf} 
\end{equation}
where the Gaussian center parameter ${\bf \zeta}$
is related to the expectation 
value of the position of the nucleon,
and $\chi^{\tau,\sigma}$ is the spin-isospin part of the wave function.
The Slater determinant is constructed from 
these single-particle wave functions by antisymmetrizing them.
For the Gaussian center parameters
$\left\{ {\bf \zeta}_i \right\}$,
ihere four single-particle 
wave functions with different spin and isospin
sharing a common 
${\bf \zeta}$ value correspond to an $\alpha$ cluster.
This cluster wave function is transformed into
$jj$-coupling shell model based on the AQCM.
When the original value of the Gaussian center parameter ${\bf \zeta}$
is ${\bf R}$,
which is 
real and
related to the spatial position of this nucleon, 
it is transformed 
by adding the imaginary part as
\begin{equation}
  {\bf \zeta} = {\bf R} + i \Lambda {\bf e}^{\text{spin}} \times {\bf R}, 
  \label{AQCM}
\end{equation}
where ${\bf e}^{\text{spin}}$ is a unit vector for the intrinsic-spin orientation of this
nucleon. 
The control parameter, labelled as $\Lambda$, is associated with the breaking of the $\alpha$ cluster. 
With a finite value of $\Lambda$, the two nucleons with opposite spin orientations have complex 
conjugate ${\bf \zeta}$ values.
This situation corresponds to the time-reversal motion of the two nucleons.

Here, we explain the intuitive meaning of this procedure.
Including the imaginary part allows us to connect the single-particle wave function 
directly to the spherical harmonics of the $jj$-coupling shell model.  
Suppose the Gaussian center parameter, represented by the vector symbolized as 
${\bf \zeta}$ has an $x$-component and the spin direction is defined along the $z$-axis
 (i.e., a spin-up nucleon). 
According to Eq.~(\ref{AQCM}), the imaginary part of the $x$ component of the Gaussian center parameter is given to the $y$ component. 
When we expand $-\nu \left({\bf r} - {\bf \zeta} \right)^{2}$
in the exponent of Eq. (\ref{spwf}), a factor corresponding to the cross term of this expansion appears:
$\exp \left[ 2 \nu {\bf \zeta} \cdot {\bf r} \right]$.
The factor 
$\exp \left[ 2 \nu {\bf \zeta} \cdot {\bf r} \right]$
contains all the information related to the angular momentum of this single particle.
The Taylor expansion allows us to show that
the $p$ wave component of $\exp \left[ 2 \nu {\bf \zeta} \cdot {\bf r} \right]$ 
is $2 \nu {\bf \zeta} \cdot {\bf r}$,
which is proportional to $\left( x + i \Lambda y \right)$. 
At the limit of $\Lambda = 1$, this is proportional to $Y_{11}$ of the spherical harmonics.
The nucleon is introduced as spin-up, and thus
the coupling with the spin part gives the stretched state of the angular momentum,
$ \left| 3/2\ 3/2 \right\rangle $ of the $jj$-coupling shell model, 
where the spin-orbit interaction acts attractively. 
For the spin-down nucleon, we introduce the complex conjugate ${\bf \zeta}$ value,
which gives $ \left| 3/2\ -3/2 \right\rangle $. 

This transformation is quite general, and we can easily generate 
the $jj$-coupling shell model wave functions corresponding to the magic numbers 28, 50, and 82
starting with the cluster wave functions.
In the case of $^{12}$C, we prepare three quasi clusters.
The next two nucleons are generated by rotating the ${\bf \zeta}$ values and spin-directions of these two nucleons by $2\pi/3$. 
The last two nucleons are generated by changing the rotation angle to $4\pi/3$. 
Eventually, all the six nucleons have spin-stretched states, 
and after the antisymmetrization, the configuration becomes the subclosure configuration of 
$ \left( s1/2 \right)^2 \left( p3/2 \right)^4$.
This procedure is applied for both proton and neutron parts.

\begin{figure}[tb]
  \centering
    \includegraphics[width=5.5cm]{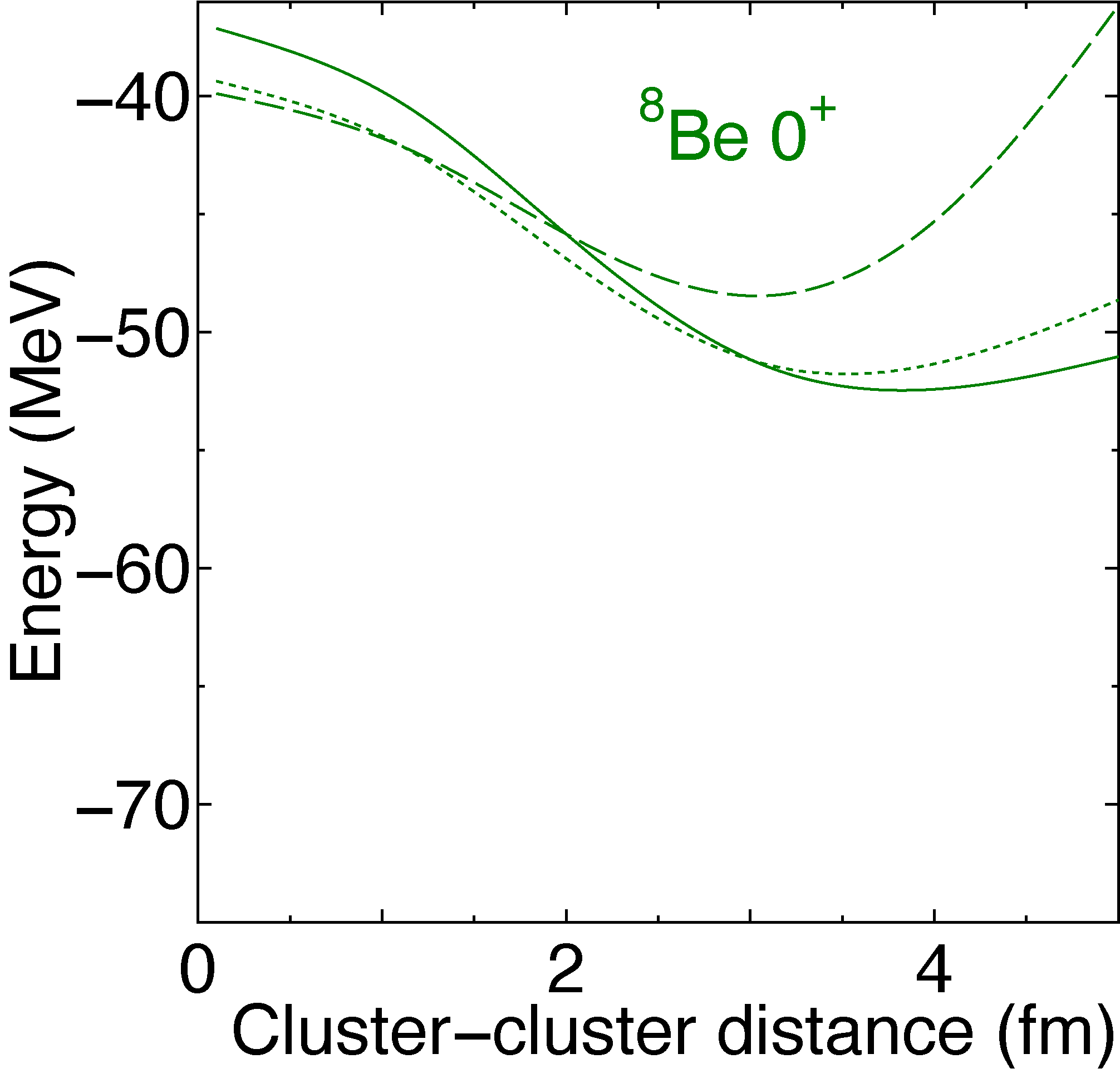} 
 
  \caption{
Energy curves of $0^+$ state of 
$^8$Be as a function of the distance between two $^4$He clusters. Solid line is for $\Lambda = 0$ (pure two $\alpha$'s)
and dotted and dashed lines are for two quasi-clusters with  $\Lambda=0.1$ and 0.2, respectively.
See the details in Ref.~\cite{PhysRevC107024309}.
    }
  \label{be8}
\end{figure}

We start the discussion with $^8$Be.
Our Hamiltonian gives the energy of $-27.57$~MeV for the $\alpha$ cluster, and thus,
$-55.1$~MeV is the two-$\alpha$ threshold energy (experimentally $-56.6$~MeV, to which our theoretical value does not contradict).
Figure~\ref{be8} shows the
energy curves of the $0^+$ state of 
$^{8}$Be as a function of the distance between two $^4$He clusters. The solid line is for $\Lambda = 0$ (pure two $\alpha$'s),
and the dotted and dashed line are for two quasi-clusters with $\Lambda=0.1$ and 0.2, respectively. 
The energy minimum point appears around the relative distance of $\sim$3.5~fm. 
This distance is quite large, and this is outside of the interaction range of the spin-orbit interaction.
Therefore, the $\Lambda$ value that gives the minimum energy is zero (solid line),
which means that the  $\alpha$ clusters are not broken.
The $\alpha$ breaking effect can be seen in more inner regions, where the energies of dotted and dashed lines
are lower than the solid line.
The $\alpha$ clusters are surely broken there. 
However, at short relative distances,
the energy  itself is high enough, and the spin-orbit interaction only plays
a role in reducing the increase of the excitation energy to some extent
when two clusters get closer.

%%%%%%%%%%%%%%%%%%%%%%%%%%%%%%%%%%%%%%%%%%%%%%%% 
%%%%%%%%%%%%%%%%%%%%%%%%%%%%%%%%%%%%%%%%%%%%%%%%

%%%%%%%%%%%%%%%%%%%%%%%%%%%%%%%%%%%%%%%%%%%%%%%% 
\begin{figure}[tb]
  \centering
    \includegraphics[width=5.5cm]{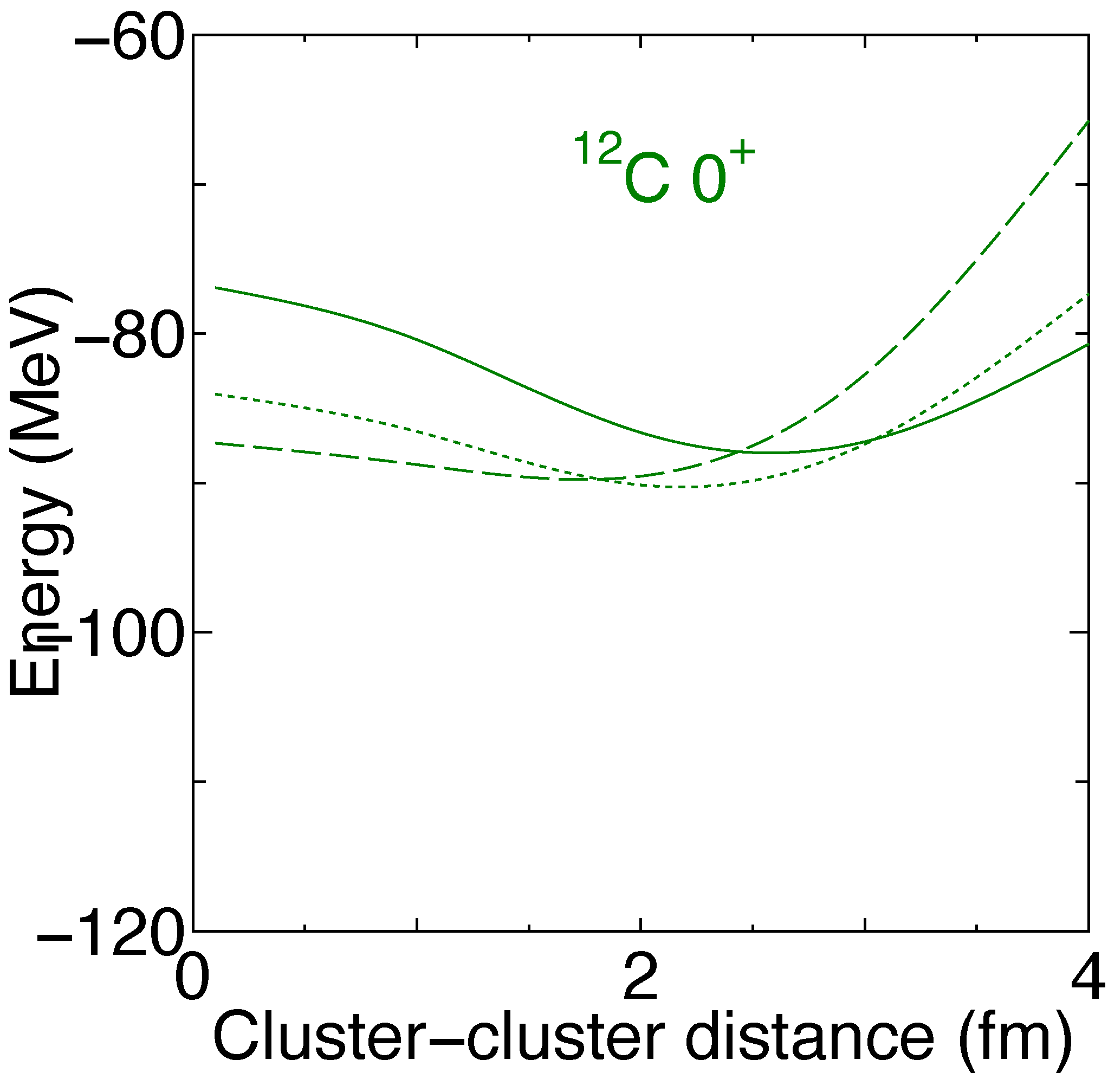} 
\caption{
Energy curves of $0^+$ state of 
$^{12}$C as a function of the distance between three $^4$He clusters
with equilateral triangular configuration.
Solid line is for $\Lambda = 0$ (pure three $\alpha$'s)
and dotted and dashed lines are for two quasi-clusters with  $\Lambda=0.1$ and 0.2, respectively.
See the details in Ref.~\cite{PhysRevC107024309}.
    }
  \label{c12}
\end{figure}
%%%%%%%%%%%%%%%%%%%%%%%%%%%%%%%%%%%%%%%%%%%%%%%% 

Next we discuss $^{12}$C.
The three-$\alpha$ threshold energy is $-82.7$~MeV in our calculation compared with the 
experimental value of $-84.9$~MeV.
Figure~\ref{c12} shows the
energy curves of $0^+$ state of 
$^{12}$C with an equilateral triangular configuration
as a function of the distance between two $^{4}$He clusters. The solid line is for $\Lambda = 0$ (pure three $\alpha$'s).
Since one $^4$He is added to $^8$Be,
the energy minimum point appears around the relative distance of 2.5--3.0~fm, shorter by
1~fm than the previous $^8$Be case before allowing the breaking of $\alpha$ clusters.
Therefore, it is considered that the three $\alpha$ clusters step in the interaction range of the spin-orbit interaction.
The dotted line ($\Lambda=0.1$) and dashed line ($\Lambda =0.2$) almost degenerate at the
region of the lowest energy (the relative cluster-cluster distance shrinks to 2~fm there).

It can be summarized that
the cluster breaking effect is negligibly small in $^8$Be, where $\alpha$--$\alpha$ cluster structure keeps
enough distance; they stay out of the interaction range of the spin-orbit interaction, 
which breaks the $\alpha$ clusters.
The situation is completely different in the $^{12}$C case since the additional $\alpha$ cluster
shrinks the cluster-cluster distance, and clusters are in the interaction range of the spin-orbit interaction.
The ground state of $^{12}$C contains the component of the $jj$-coupling shell model.

%%%%%%%%%%%%%%%%%%%%%%%%%%%%%%%%%%%%%%%%%%%%%%%%%%%%%%%%%%%%

\section{Remarks and prospects transcending sections}
\label{sec:end_summary}

We here present some remarks and prospects over multiple sections, keep aside subjects presented in individual sections such as rotational features in Sect.~\ref{sec:otsuka}.

The three sections, Sects.~\ref{sec:otsuka}-\ref{sec:itagaki}, of this article approach $\alpha$ clustering in light nuclei from distinct theoretical perspectives, yet they converge on an identical physical picture. It is worth making this convergence explicit, as it constitutes one of the unique messages of this work.
At the level of methodology, all three approaches are rooted in the same conceptual foundation: a configuration-interaction (CI) and variational description based on nucleonic degrees of freedom and realistic nucleon-nucleon interactions. What differs is the strategy of configuration selection and representation. Section~\ref{sec:otsuka} employs a very large-scale no-core shell model diagonalization (MCSM), where cluster correlations are not assumed but emerge organically from the superposition of a vast number of many-body basis states. Section~\ref{sec:volya} adopts a complementary strategy, the Cluster-Nucleon Configuration Interaction (CNCIM),  in which traditional shell-model-like Slater determinant configurations are combined with microscopically constructed cluster channel configurations, enriching the basis in a physically targeted way. Section-\ref{sec:itagaki} uses the antisymmetrized quasi-cluster model (AQCM), where cluster wave functions built from Gaussian single-particle states with complex center parameters are smoothly connected to $jj$-coupling shell-model wave functions through a continuous deformation parameter, thereby modeling the intermediate regime between pure cluster and pure shell structure. Together, the three approaches thus bridge the two structural limits from complementary view points.

All three approaches are applied to the same nuclei, and the degree of agreement across methods is instructive. The ground state of $^8$Be, a {\it di}-$\alpha$ resonance just above the two-$\alpha$ threshold, provides the clearest example. Figures 5, 18, and 20, drawn from Sects. 2, 3, and 4 respectively, independently arrive at the same result: the two $\alpha$ clusters orbit each other with a center-to-center separation of approximately 3.5–3.6 fm. This value emerges from a two-dimensional nucleon density profile in the intrinsic frame, from the structure of the RGM relative-motion wave function, and from the minimum of the AQCM energy surface as a function of cluster-cluster distance. The agreement across such different theoretical languages is not a coincidence: this separation reflects a balance between kinetic energy and nuclear attraction that is robust across model choices, and it places the two clusters well outside the range of the spin-orbit interaction, explaining why the $\alpha$ structure in $^8$Be is essentially unperturbed.  The two $\alpha$ clusters rotate at this nearly fixed separation, giving rise to a rotational band consistently identified across approaches.

$^{12}$C presents a richer case where the three perspectives are genuinely complementary rather than merely confirmatory. All three sections find that the ground state is not a pure cluster state but contains a significant admixture of shell-model-type structure, yet each illuminates a different facet of this competition: Sect. 2 quantifies the mixing and its energetic consequence, Sect. 3 addresses its manifestation in spectroscopic factors, and Sect. 4 provides the dynamical mechanism: the contraction of the inter-cluster distance to approximately 2.5–3.0 fm upon adding a third alpha cluster brings the system within the range of the spin-orbit interaction, generating shell-model admixtures that have no counterpart in $^8$Be. 
For the Hoyle state, the approaches confirm that alpha clustering is the dominant structural feature, with shell-model components present as a secondary admixture from the ground state.

Experimental evidence for $\alpha$ clustering extends well beyond the light nuclei discussed here, and the question of how clustering strength evolves across the nuclear chart remains open. Near-threshold clustering phenomena, only briefly touched upon in this article, represent a particularly rich subject: the restructuring of many-body wave functions in the vicinity of cluster-decay thresholds, the role of superradiance in organizing broad doorway and narrow trapped states embedded in the continuum, all remain areas of active investigation.

A broader perspective connects $\alpha$ clustering to the more general question of four-nucleon correlations and their evolution across the nuclear chart. The connection between alpha particles, proton-neutron quartets, and pairing correlations in heavier nuclei is well motivated theoretically but not yet fully explored in a unified framework. The crossover from a regime dominated by pairing, through quartet condensation, to explicit spatial clustering as a function of nuclear size or proton-neutron asymmetry represents a yet unsolved problem in nuclear many-body physics. 

The methods presented and developed here offer promising tools for addressing these open questions in nuclear physics and beyond.

%%%%%%%%%%%%

\subsection*{\label{sec:acknowledgements}Acknowledgments}

All authors thank Dr. Takashi Nakamura for arranging the excellent opportunity to publish this article.
Regarding the work presented in Sect.~2, 
TO acknowledges Drs. T. Yoshida, T. Abe, Y. Tsunoda, N. Shimizu, N. Itagaki, Y. Utsuno, H. Ueno,  J. Vary and P. Maris for very fruitful collaborative works shown primarily in Refs.~\cite{otsuka_2022,otsuka_2025}.   TO thanks Dr. K. C. W. Li and Dr. M. Kimura for useful discussions on the clustering, Dr. S. Kuma and Dr. T. Azuma for various information about atomic molecules, and Dr. Y. Aritomo for updating on nuclear fission studies.  TO is grateful to Dr. T. Kobori for encouragements crucially contributing to this work.  
TO and AV acknowledge the visitor program of GANIL, and TO thanks the Alexander von Humboldt Foundation for the Research Award, as some parts of their works were made under these supports.  The MCSM (incl. QVSM)  calculations quoted in Sect.~2 were performed on the supercomputers K and Fugaku at RIKEN AICS  K and Fugaku at RIKEN AICS (hp190160, hp200130, hp210165, hp220174, hp230207, hp240213, hp250224).
The work presented in Sect.~2 are supported in part by JSPS KAKENHI Grant Number
JP19H05145 and JP21H00117 as well as JP25K00998. 
It was also supported in part by MEXT as ``Program for Promoting Researches on the Supercomputer Fugaku'' (Simulation for basic science: from fundamental laws of particles to creation of nuclei, JPMXP1020200105, Simulation for basic science: approaching the new quantum era, JPMXP1020230411), and by JICFuS.  
Regarding the work presented in Sect.~3, 
AV acknowledges partial support by the US Department of Energy (DOE), Office of Science, Office of Nuclear Physics grant DE-SC0009883. 
Regarding the work presented in Sect.~4, NI acknowledges partial support by JSPS KAKENHI Grant Numbers JP22K03618 and JP25K01005.

\bibliographystyle{sn-mathphys}  % or use another style, e.g., plain, unsrt, etc.  apsrev4-2
\bibliography{revised_arxiv}   %epj_cluster_14

\end{document}